\documentclass[twocolumn]{aastex61}
\pdfoutput=1 
\usepackage{amsmath,amstext}
\usepackage[T1]{fontenc}
\usepackage{apjfonts} 
\usepackage[figure,figure*]{hypcap}
\usepackage{scrextend}
\usepackage{footnote}
\usepackage{hyperref}
\usepackage{subfigure}
\usepackage{array}

\usepackage{epsfig}
\usepackage{natbib}

\shorttitle{CMZoom Overview}
\shortauthors{CMZoom team}

\begin{document}

\def\Msun{\hbox{M$_{\odot}$}}
\def\Lsun{\hbox{L$_{\odot}$}}
\def\kms{km~s$^{\rm -1}$}
\def\micron{$\mu$m}
\def\deg{$^{\circ}$}
\def\arcmin{$^{\prime}$}
\def\Vlsr{\hbox{V$_{LSR}$}}
\def\arraystretch{0.7}

\hyphenation{kruijs-sen}

\title{CMZoom: Survey Overview and First Data Release}

\author{Cara Battersby}
\affiliation{University of Connecticut, Department of Physics, 196 Auditorium Road, Unit 3046, Storrs, CT 06269 USA}
\affiliation{Center for Astrophysics | Harvard \& Smithsonian, 60 Garden St., Cambridge, MA 02138 USA}

\author{Eric Keto}
\affiliation{Center for Astrophysics | Harvard \& Smithsonian, 60 Garden St., Cambridge, MA 02138 USA}

\author{Daniel Walker}
\affiliation{Joint ALMA Observatory, Alonso de C\'ordova 3107, Vitacura, Santiago, Chile}
\affiliation{National Astronomical Observatory of Japan, 2-21-1 Osawa, Mitaka, Tokyo, 181-8588, Japan}

\author{Ashley Barnes}
\affiliation{Argelander-Institut f\"{u}r Astronomie, Universit\"{a}t Bonn, Auf dem H\"{u}gel 71, 53121, Bonn, Germany}

\author{Daniel Callanan}
\affiliation{Astrophysics Research Institute, Liverpool John Moores University, 146 Brownlow Hill, Liverpool L3 5RF, UK}
\affiliation{Center for Astrophysics | Harvard \& Smithsonian, 60 Garden St., Cambridge, MA 02138 USA}

\author{Adam Ginsburg}
\affiliation{University of Florida Department of Astronomy, Bryant Space Science Center, Gainesville, FL, 32611, USA}

\author{H Perry Hatchfield}
\affiliation{University of Connecticut, Department of Physics, 196 Auditorium Road, Unit 3046, Storrs, CT 06269 USA}

\author{Jonathan Henshaw}
\affiliation{Max-Planck-Institute for Astronomy, Koenigstuhl 17, 69117 Heidelberg, Germany}

\author{Jens Kauffmann}
\affiliation{Haystack Observatory, Massachusetts Institute of Technology, 99 Millstone Road, Westford, MA 01886, USA}

\author{J.~M.~Diederik Kruijssen}
\affiliation{Astronomisches Rechen-Institut, Zentrum f{\"u}r Astronomie der Universit{\"a}t Heidelberg, M{\"o}nchhofstra{\ss}e 12-14, D-69120 Heidelberg, Germany}

\author{Steven N.~Longmore}
\affiliation{Astrophysics Research Institute, Liverpool John Moores University, 146 Brownlow Hill, Liverpool L3 5RF, UK}

\author{Xing Lu}
\affiliation{National Astronomical Observatory of Japan, 2-21-1 Osawa, Mitaka, Tokyo, 181-8588, Japan}

\author{Elisabeth A.~C.~Mills}
\affiliation{Department of Physics and Astronomy, University of Kansas, 1251 Wescoe Hall Drive, Lawrence, KS 66045, USA}

\author{Thushara Pillai}
\affiliation{Boston University Astronomy Department, 725 Commonwealth Avenue, Boston, MA 02215, USA}

\author{Qizhou Zhang}
\affiliation{Center for Astrophysics | Harvard \& Smithsonian, 60 Garden St., Cambridge, MA 02138 USA}

\author{John Bally}
\affiliation{CASA, University of Colorado, 389-UCB, Boulder, CO 80309}

\author{Natalie Butterfield}
\affiliation{Green Bank Observatory, 155 Observatory Rd, PO Box 2, Green Bank, WV 24944, USA}

\author{Yanett A.~Contreras}
\affiliation{Leiden Observatory, Leiden University, PO Box 9513, NL 2300 RA Leiden, the Netherlands}

\author{Luis C.~Ho}
\affiliation{Kavli Institute for Astronomy and Astrophysics, Peking University, Beijing 100871, China}
\affiliation{Department of Astronomy, School of Physics, Peking University, Beijing 100871, China}



\author{J{\"u}rgen Ott}
\affiliation{National Radio Astronomy Observatory, 1003 Lopezville Rd., Socorro, NM 87801, USA}

\author{Nimesh Patel}
\affiliation{Center for Astrophysics | Harvard \& Smithsonian, 60 Garden St., Cambridge, MA 02138 USA}

\author{Volker Tolls}
\affiliation{Center for Astrophysics | Harvard \& Smithsonian, 60 Garden St., Cambridge, MA 02138 USA}


\begin{abstract}
We present an overview of the \textit{CMZoom} survey and its first data release. \textit{CMZoom} is the first blind, high-resolution survey of the Central Molecular Zone (CMZ; the inner 500 pc of the Milky Way) at wavelengths sensitive to the pre-cursors of high-mass stars. \textit{CMZoom} is a 550-hour Large Program on the Submillimeter Array (SMA) that mapped at 1.3 mm all of the gas and dust in the CMZ above a molecular hydrogen column density of 10$^{23}$\,cm$^{-2}$ at a resolution of $\sim$3$\arcsec$~(0.1 pc). In this paper, we focus on the 1.3 mm dust continuum and its data release, but also describe \textit{CMZoom} spectral line data which will be released in a forthcoming publication.
While \textit{CMZoom} detected many regions with rich and complex substructure, its key result is an overall deficit in compact substructures on 0.1 - 2 pc scales (the compact dense gas fraction: CDGF). In comparison with clouds in the Galactic disk, the CDGF in the CMZ is substantially lower, despite having much higher average column densities. CMZ clouds with high CDGFs are well-known sites of active star formation. The inability of most gas in the CMZ to form compact substructures is likely responsible for the dearth of star formation in the CMZ, surprising considering its high density. The factors responsible for the low CDGF are not yet understood but are plausibly due to the extreme environment of the CMZ, having far-reaching ramifications for our understanding of the star formation process across the cosmos.
\end{abstract}


\section{Introduction}

The inner $\sim$ 500 pc of the Milky Way (the Central Molecular Zone or CMZ) is an ideal testbed for probing a possible environmental dependence of the processes that govern star formation. The CMZ is close enough to study star formation in detail, while also hosting extreme conditions that provide a strong lever arm on tests for star formation as a function of environment. The extreme conditions of the CMZ are a result of the unique physical properties of its molecular gas and its location within the Galaxy's nucleus. Gas in the CMZ experiences an intense UV background field \citep[$G_0\sim10^3-10^4$;][]{lis01,goi04,clark13}, elevated cosmic ray ionization rates \citep[$\xi\sim10^{-15}-10^{-14}$;][]{oka05,got13,har15,padovani20}, X-ray flares \citep{ter10,ter18}, and dynamical stresses like shearing and compression due to the bar potential \citep{gus80,lon13b,kru17,kru19}. Compared to gas in the disk of the Galaxy, gas in the CMZ has higher temperatures \citep{gus85,mil13,gin16,kri17}, greater densities \citep{gus83,wal86,mil18a}, elevated turbulence \citep[]{she12,kau17a,henshaw19}, richer chemistry \citep{req06,req08,arm15,zen18}, and stronger magnetic fields \citep{cru96,pil15}. While such gas conditions may be uncommon in the present day (seen only in other galaxy centers), they may be more typical of gas conditions of galaxies in the early universe, at the peak of cosmic star formation \citep{kru13}. The CMZ serves as an important local analog for these systems as its relative proximity \citep[about 8.1 kpc;][]{rei19, Gravity18, Gravity19} enables detailed study of the influence these gas conditions have on the star formation process.

Past star formation events in the CMZ have built up a nuclear star cluster with a mass in excess of that of the black hole, SgrA*, \citep{schodel09} and a nuclear stellar disk with R$\sim$ 200 pc and a mass of $10^9$ M$_\odot$, about 2\% of the total stellar mass in the Milky Way \citep{lau02,sch15,mcm17}. The Fermi lobes, fossils of an outflow centered on the CMZ \citep{su10}, suggest that the most recent starburst in the CMZ may have occurred within the past 10 Myr \citep{bor17}, if the Fermi lobes are due to star formation activity. Currently, the CMZ hosts three young massive clusters with ages of 2-6 Myr \citep[the Young Nuclear, Arches, and Quintuplet clusters; ][] {lu13,cla18a,cla18b}, possibly from the tail end of this starburst \citep{kru17}, as well as a comparable population of isolated massive stars, which may have been tidally stripped from these or other clusters \citep{mau10,hab14}. These clusters show tantalizing evidence for a top-heavy initial mass function \citep{sto05,maness07,lie12,lu13,hosek19}, which could be an indication that the unique environment of the CMZ significantly modifies the star formation process. Furthermore, the initial mass function of these clusters may be representative of a large fraction of the star formation in the CMZ, given recent measurements of a large cluster formation efficiency in this region \citep{gin18b}.

Though star formation in the CMZ may have been active millions of years ago, there is a surprising lack of present day star formation given the dense gas reservoir in this region \citep{imm12, lon13a}. While the fraction of the Galactic star formation rate (SFR) taking place in the CMZ \citep[0.05-0.1 M$_\odot$ yr$^{-1}$, 3-6\% of the total rate in the Milky Way;][]{cro12,lon13a,rob10,cho11,lic15, koe15, bar17} is roughly the same as the fraction of the Milky Way's molecular gas in this region \citep[$3\times10^7$ M$_\odot$, $\sim$4\% of all the molecular gas in the Milky Way;][]{dah98}, it is far below what would be expected \citep{lad10} considering the high density of this gas \citep{lon13a}. This discrepancy is likely not due to missed star formation, since more sensitive surveys of star formation tracers are not substantially revising the amount of star formation \citep[]{yus09, imm12,mil15,lu15,bar17,ric18,lu19a,lu19b}. Even considering that previous studies may have overestimated the fraction of CMZ gas with densities $\gtrsim 10^4$ cm$^{-3}$ (\citealt{mil18a} show that rather than being 100\% this fraction is only $\sim$15\% in a representative CMZ cloud sample), the inefficiency by which the CMZ produces stars remains problematic. The currently favored explanation for this discrepancy is tied to the gas dynamics and particularly the turbulence of CMZ gas \citep{kru14, rat14}, either its large magnitude, which can provide additional support against gravitational collapse \citep[e.g.][]{kru05, pad11}, or its solenoidal nature \citep[e.g.][]{fed16, bar17,kru19}. 

Recent studies have begun searching for additional links between the dynamics of CMZ gas, unique to its location in the nuclear potential of the Milky Way, and the current star formation rate. Almost all active sites of star formation in the CMZ are confined to the central 200\,pc, along or interior to an eccentric stream of gas clouds orbiting Sgr A* \citep{mol11, lon13b, lon13a, kdl15, hen16, bar17}. It has been suggested that star formation in this stream may be dynamically triggered by the pericenter passage of gas clouds near the global gravitational potential well coincident with SgrA* \citep{lon13a,jeffreson18}. This idea is supported by simulations of the gas \citep{dal19} and observations of the evolution of gas properties along portions of the stream \citep{kri17}. However, this model is likely not sufficient to describe all of the star formation observed in the CMZ \citep{ken13, sim18b}; \citet{jeffreson18} estimate that 20\% of CMZ clouds may be tidally forced into star formation. 

Ultimately, addressing these open questions requires making a direct connection between the environmental properties of the gas (both physical and dynamical) and the resulting star formation. This requires long-wavelength observations of the dust continuum and spectral lines on sub-pc scales sufficient to resolve individual star-forming clumps and cores across the entire CMZ. However, until recently, large-area surveys of dense molecular gas have been limited to low-resolution single-dish studies, or interferometric surveys probing gas on cloud ($\geq$10$\arcsec$, pc) scales \citep[e.g.][]{jon12, gin16, hen16, lon17, kri17, pou18}. Interferometric studies required to probe gas at arc-second scales are observationally much more expensive to cover large areas at sufficient sensitivity, so have been limited to focused studies on individual clouds \citep{vog87,mon09,bal14, jon14, rat14, rat15, lu15,mil15,mil18b,gin18,barnes2019} or samples of a small number of clouds \citep{kau17a, kau17b, lu19a, lu19b}.

With the \textit{CMZoom} survey \citep[this work and][]{bat17iau}, we have conducted the first large-scale, high-resolution survey of the CMZ at submm wavelengths, producing the first unbiased census of sites of high-mass star formation and the physical and kinematic properties at sub-pc scales across the whole CMZ. The large area ($\sim$ 350 arcmin$^2$) and high ($\sim$ 0.1 pc) resolution of the survey are chosen to enable addressing key open questions about the nature of star formation in the Galactic center environment.

In this paper, we give an overview of the \textit{CMZoom} survey and present the full dust continuum maps. These data are made publicly available on the Harvard Dataverse here (\href{https://dataverse.harvard.edu/dataverse/cmzoom}{https://dataverse.harvard.edu/dataverse/cmzoom}). Subsequent papers release our source catalog (Hatchfield et al. in prep.), spectral line data (Callanan et al. in prep.), and association of our sources with star formation tracers (Hatchfield et al. in prep.). This paper is organized as follows: Section 2 details the source selection, spectral setup, antenna configurations, and observing strategy. Section 3 outlines the data calibration and imaging process, including combination with single-dish data. Section 4 describes the data, including the astrometric accuracy, beam size, noise, and comparison with previous datasets and outlines the data release. Section 5 details the compact substructure revealed in the survey, the compact dense gas fraction (CDGF). Section 6 presents a summary of the paper.

\section{Submillimeter Array CMZoom Data}
The \textit{CMZoom} survey was one of the first large-scale projects undertaken at the Submillimeter Array (SMA)\footnote{The Submillimeter Array is a joint project between the Smithsonian Astrophysical Observatory and the Academia Sinica Institute of Astronomy and Astrophysics and is funded by the Smithsonian Institution and the Academia Sinica.}. The survey (Project ID: 2013B-S091) took about 550 hours (61 nights, see Section \ref{sec:obs_strategy} for more details) on the SMA, in compact and subcompact configuration, at 230 GHz covering wideband (8+ GHz) dust continuum and a number of key spectral lines. The resulting images have an angular resolution of about 3$\arcsec$~(0.1 pc), and a spectral resolution of about 1 \kms, over all of the highest column density gas (above a Herschel column density threshold of 10$^{23}$ cm$^{-2}$) in the inner 5\deg~$\times$ 1\deg~of the Galaxy. In total, the \textit{CMZoom} mosaic covered about $5 \times 10^6~\Msun$~of dense gas in the CMZ (measured from the Herschel column density map). With a total CMZ mass of about $2-6 \times 10^7~\Msun$~\citep{mor96}, this corresponds to covering about 10$-$25\% by mass of the CMZ, selected to be of the highest column density. It is important to note that throughout the text, we assume that most, if not all, of the 1.3 mm continuum emission is due to the thermal dust continuum. For most parts of the CMZ, this is likely correct, however, for some regions with very strong free-free or synchrotron emission, there might be a substantial contribution to the continuum emission, which should be considered in a detailed analysis of these data.

\begin{figure*}
\begin{center}
\includegraphics[trim=0 6.5cm 0 5cm, clip, width=1\textwidth]{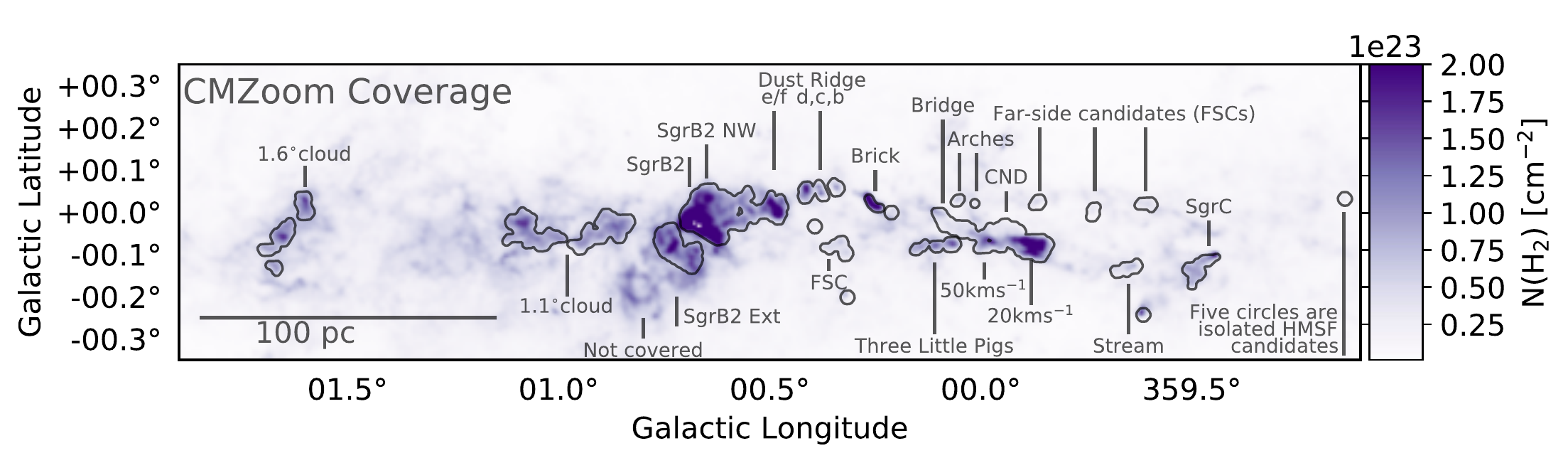}\\
 \singlespace\caption{The Central Molecular Zone as seen in N(H$_2$) derived from the Herschel cold dust continuum \citep[][Battersby et al., in prep.]{mol10} in units of cm$^{-2}$ in the colorscale with the \textit{CMZoom} coverage is shown as gray contours. The 
 figure shows colloquial names or notes on each observed region, as they are referenced to in Table \ref{table-sources} and throughout the text.
Within the inner 5\deg~longitude $\times$ 1\deg~latitude of the Galaxy, \textit{CMZoom} is complete above a column density threshold of 10$^{23}$ cm$^{-2}$, with the exception of the cloud to the SE of Sgr B2 and isolated bright pixels, and with the addition of a few clouds as noted in Section \ref{sec:coverage}. \textit{CMZoom} covered 974 individual mosaic pointings over about 550 hours on the SMA.  }
\label{fig:coverage}
\end{center}
\end{figure*}

\subsection{Source Selection}
\label{sec:coverage}
We expect that the highest column density structures in the CMZ are the most relevant for understanding high-mass star formation, and such regions are well-suited to observation with the SMA. Therefore, the \textit{CMZoom} survey was designed to map all of the highest column density gas, above a Herschel column density threshold of 10$^{23}$ cm$^{-2}$, in the inner 5\deg~longitude $\times$ 1\deg~latitude. This is  \citep[about 700 $\times$ 150 pc based on Galactic Center distance of 8.15 kpc from][]{rei19} of the Galactic Center, which is the distance adopted for the remainder of the text. The only exception that meets these criteria, but was not observed, is one isolated cloud to the Southeast of Sgr B2 at a much lower latitude than the main part of the CMZ. The Herschel column density map was derived using data from the Hi-GAL survey \citep{mol10} as described in \citep[][and Battersby et al., in prep.]{bat11}.

In addition to the nearly complete coverage above this column density threshold, a number of regions of interest were also mapped (see Figure \ref{fig:coverage}), for a total area of approximately 350 square arcminutes, or at a distance of 8.15 kpc a square of 45 pc on a side. These additional regions include `far side cloud candidates' (G0.326$-$0.085, G359.734$+$0.002,  G359.611$+$0.018, G359.865$+$0.02, and G359.65$-$0.13), the circumnuclear disk (CND: G359.948$-$0.052), pointings toward the `Arches' ionized filaments (G0.014$+$0.021 and G0.054$+$0.027), isolated high-mass star forming (HMSF) candidates (G0.393$-$0.034, G0.316$-$0.201, G0.212$-$0.001, G359.615$-$0.243, and G359.137$+$0.031),  
and a bridge of emission, connecting the Dust Ridge and the 50 \kms~cloud (G0.070$-$0.035) detected to have strong H$_2$CO features in the APEX-CMZ survey \citep{gin16}. A full region file of mosaic pointings is released in the Dataverse  (\href{https://dataverse.harvard.edu/dataverse/cmzoom}{https://dataverse.harvard.edu/dataverse/cmzoom}), and the full coverage can be seen in Figure \ref{fig:coverage}.

Our column density cutoff is based on smoothed column density contours, in an effort to produce maps of mostly contiguous regions, therefore individual bright pixels above this threshold are not included. Similarly, some lower-level emission at the edges of clouds, or in between bright emission is included. With the inclusion of select regions of interest with lower column densities, the full column density distribution of the mapped regions is not a clean cutoff at 10$^{23}$ cm$^{-2}$. 

The regions were observed in a series of mosaics, with pointings separated by a half-beam for complete Nyquist sampling. To develop an optimized grid of mosaic pointings for our irregularly-shaped clouds, a script was developed to take an arbitrary, irregular polygon, denoted by SAOImage DS9 Regions\footnote{\href{http://ds9.si.edu/site/Home.html}{http://ds9.si.edu/site/Home.html}} using the `polygons' shape option, and sample that region with a regularly-spaced hexagonal grid of circular mosaic pointings, separated by a half-beam. This code is made publicly available on the \textit{CMZoom} GitHub page\footnote{\href{https://github.com/CMZoom}{https://github.com/CMZoom}} .

\begin{figure*}
\begin{center}
\includegraphics[width=1\textwidth]{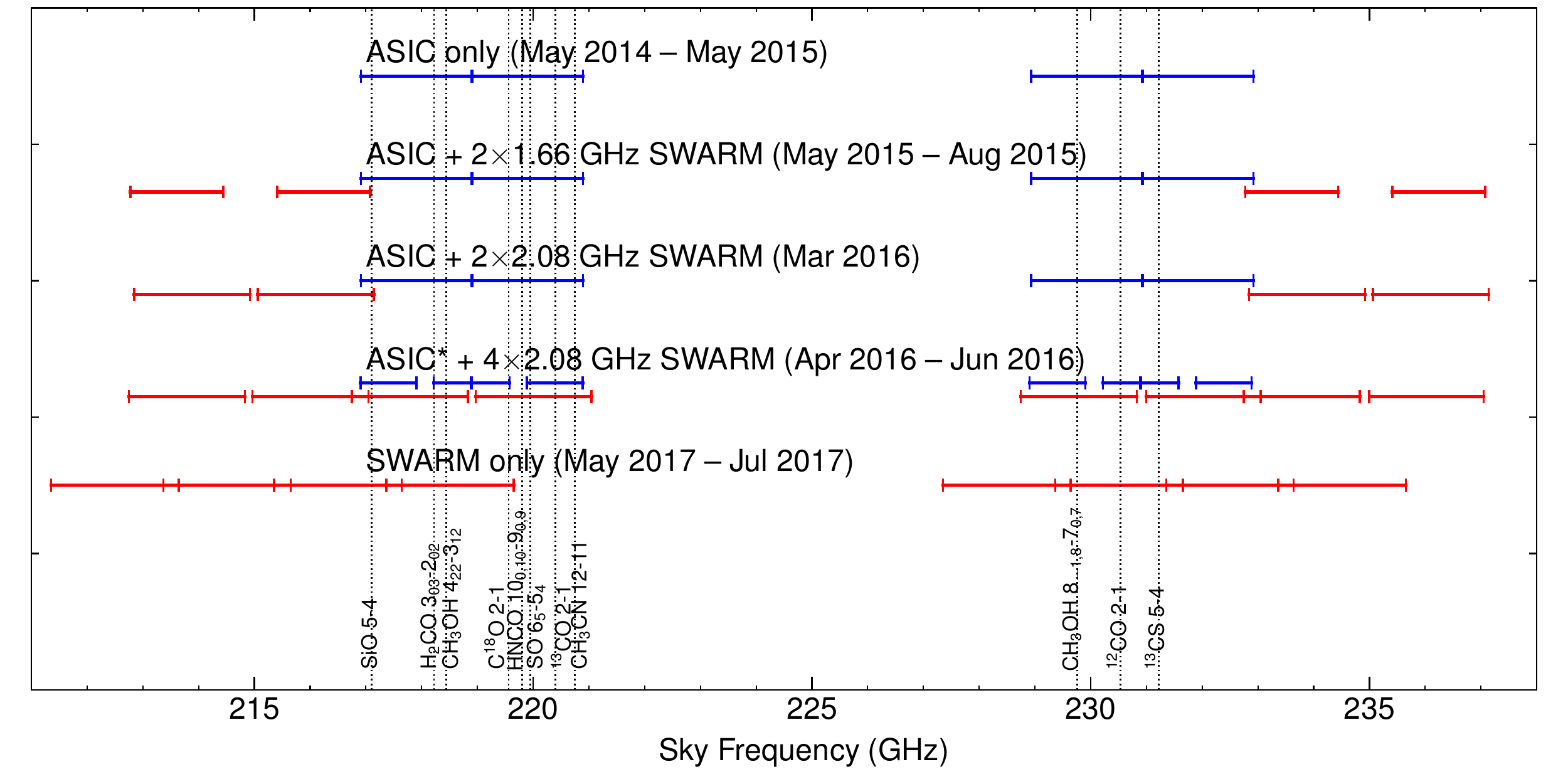}
\vspace{-5mm}
\singlespace\caption{Spectral coverage of the \textit{CMZoom} survey over time. Observations for the \textit{CMZoom} survey took place over the period of May 2014 to July 2017. During this time, the SMA transitioned from the {\sc asic} to the {\sc swarm} correlator, starting with {\sc asic} only (spectral coverage shown in blue) and ending with {\sc swarm} (spectral coverage shown in red) only, with varying degrees of overlap in between. The ASIC* from April to June 2016 indicates a transitionary period where ASIC was operating differently than its standard mode. 
Our key target spectral lines are shown as dashed lines in the plot. Coverage of these lines was maintained over the lifetime of the survey, except for a few tertiary lines in the three tracks observed in 2017 and a few other oddities that will be discussed in the spectral line data release paper.}
\label{fig:spec_coverage}
\end{center}
\end{figure*}

\subsection{Spectral Setup}
Observations for the \textit{CMZoom} survey were completed using the 230 GHz receiver at the Submillimeter Array (SMA) over the course of four years (May 2014 to July 2017), during which time the SMA was gradually upgraded from the {\sc asic} correlator to the wideband {\sc swarm} correlator \citep{pri16} in phases. Therefore, the \textit{CMZoom} survey mirrors this variable bandwidth coverage over time, with the first observations being limited to the 8 GHz {\sc asic} correlator and the final observations containing the full 16 GHz {\sc swarm} coverage (Figure \ref{fig:spec_coverage}). The early observations covered 216.9 GHz to 220.9 GHz in the lower sideband, and 228.9 GHz to 232.9 GHz in the upper sideband. The most extended coverage with {\sc swarm} was 211.5 - 219.5 GHz in the lower sideband and 227.5 - 235.5 GHz in the upper sideband, while observations in between are bookended by these extremes (Figure \ref{fig:spec_coverage}). The spectral resolution is consistently about 0.812 MHz (1.1 km s$^{-1}$) over the entire bandwidth across the published datasets.We note that the newer raw {\sc swarm} data are of substantially higher spectral resolution (a factor of 8), but were spectrally smoothed to 1.1 \kms~to maintain consistency with previous {\sc asic} data and to maintain manageable file sizes for image processing and analysis. Due to high turbulence, spectral lines in the CMZ are generally wider than this \citep[e.g.][]{she12, kau17a}, therefore, this smoothing should not substantially affect the results. This will be discussed further in the forthcoming publication releasing the spectral line data.

In addition to the 230 GHz dust continuum, which is the focus of this paper, the following spectral lines were targeted and consistently included in all observations. In the lower sideband, \textit{CMZoom} observed the triplet of para-H$_2$CO lines of 3$_{0,3}$--2$_{0,2}$, 3$_{2,2}$--2$_{2,1}$, and 3$_{2,1}$--2$_{2,0}$ at 218.222192, 218.475632, and 218.760066 GHz \citep[all frequencies are rest frequencies from CDMS][as compiled on splatalogue\footnote{\href{https://www.cv.nrao.edu/php/splat/index.php}{https://www.cv.nrao.edu/php/splat/index.php}}]{mueller2005}, respectively, $^{13}$CO and C$^{18}$O J=2--1 at 220.39868420 and 219.56035410 GHz, respectively, SiO 5--4 at 217.104980 GHz, and a number of CH$_3$OH and CH$_3$CN lines. In the upper sideband, \textit{CMZoom} observed the $^{12}$CO J=2--1 line at 230.538000 GHz, the H30$\alpha$ recombination line at 231.90092784 GHz, as well as a number of CH$_3$OH transitions. We note that the extended bandwidth observations cover substantially more spectral lines than those listed here. Preliminary analysis suggest incredibly rich spectra toward Sgr B2 and other Galactic Center regions, which clearly illustrate the benefits of the extended spectral coverage.

\begin{figure*}
\begin{center}
\includegraphics[width=1\textwidth]{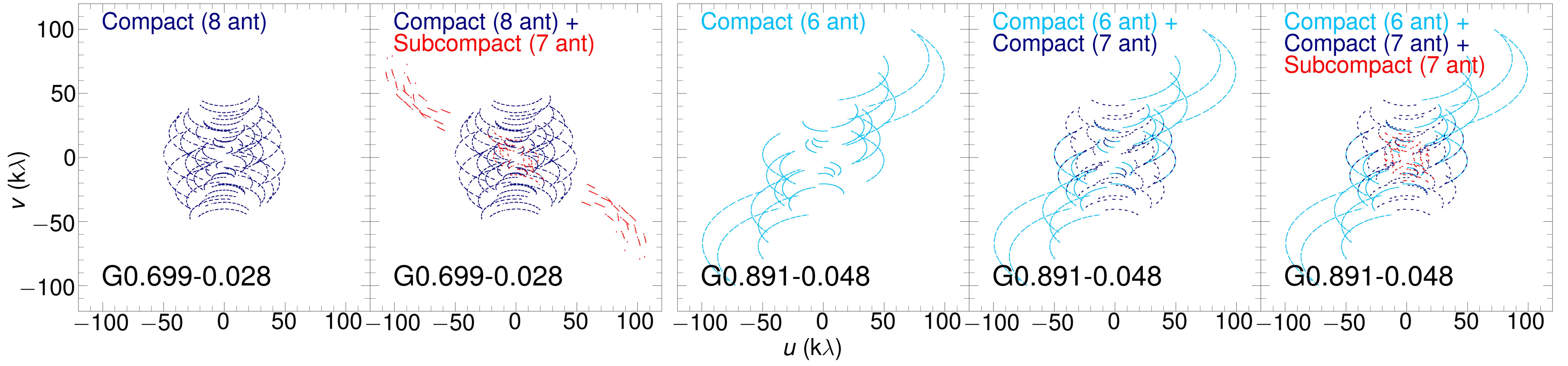}
\singlespace\caption {Two example regions of the overall UV coverage for the \textit{CMZoom} survey, demonstrating that the observing strategy was sometimes unusual with different numbers of antennas, dates, and length of time per pointing, but that the overall coverage is sufficient in all survey data. The left two panels show the compact and compact + subcompact UV coverage for one track toward source G$0.699$-$0.028$ observed with 8 antennas in compact configuration on June 10, 2014 (track 7), as well as in subcompact configuration in March 27, 2014 (track Pilot9). Note that one antenna was out of place in the subcompact configurations here (panel 2) and compact in the following panels (panels 3-5), but it did not adversely affect the imaging, so its baselines were included. The right three panels show one track toward G$0.891$-$0.048$. The region was first observed with incomplete UV coverage (only 6 antennas) in compact configuration on May 11, 2015 (track 20), then followed up with 7 antennas on May 4, 2016 (track 20) in compact configuration, and also in subcompact configuration (track 35) on June 7, 2015.}
\label{fig:uv_coverage}
\end{center}
\end{figure*}

\subsection{Array Configurations}
\label{sec:config}
In order to achieve good sensitivity over a range of spatial scales, the \textit{CMZoom} mosaic pointings were observed in both `compact' and `subcompact' configurations, standard SMA array configurations with maximum baseline lengths of 70m and 30m, respectively. The resulting maps, including the observations in both configurations, are sensitive to structures of about 3$\arcsec$~on the smallest scales and about 45$\arcsec$~on larger scales, corresponding to physical scales of 0.12-1.8 pc at a Galactic Center distance of 8.15 kpc. We increase our sensitivity to large-scale structures by combining with single-dish data, as described in Section \ref{sec:SD}. As the observations for the \textit{CMZoom} survey were carried out over 4 years, there were slight variations in the exact compact array configurations. The array configurations used are noted in Table \ref{table-obs} and the antenna positions for these arrays are included in Appendix \ref{appendix-configs} and Table \ref{table-config}. These configurations are similar enough that we anticipate no substantive change in the data properties. 

Figure \ref{fig:uv_coverage} demonstrates two examples of the UV coverage of our survey data, one track towards G0.699$-$0.028 and the other towards G0.891$-$0.048. Both UV coverage plots show some strange features, such as the out-of-place antenna creating two streams on the UV plot for G0.699$-$0.028 or the lopsided coverage of the 6 antenna compact track for G0.891$-$0.048, but were overall considered satisfactory. A number of special circumstances, and therefore, unusual features, are expected for our multi-configuration large 4-year survey, with partial observing nights, antenna moves, variable weather conditions, and missing antennas. Much of the data for the \textit{CMZoom} survey show similar strange features, however, we have ensured that all observations meet a minimum threshold for satisfactory UV coverage, as determined by the overall beam shape and size (see Section \ref{sec:rms}). 

In general, observation of a region was performed over one or multiple `tracks:' an interferometry term for a full night's worth of observations, relating to the UV tracks that the observation makes over time (see, for example, Figure \ref{fig:uv_coverage}). In some circumstances observation of a region was partly complete, but not satisfactory (either lacking sensitivity and/or UV coverage due to weather, missing antennas, or short observing nights). In some cases, such as tracks 17 and 18, we combined the pointings for the two tracks into one (so observed $\sim$40 pointings in a night instead of 20) for re-observation. In other more severe cases, the tracks were completely re-observed. The details for each observation are included in Table \ref{table-obs}.

\begin{table*}
\begin{center}
{
\begin{tabular}{cccc}
\hline\hline
Track & Observation & \# of & Array \\
Number & Date & Antennas & Config. \\
\hline
J2014\tablenotemark{*} & 6/09/2012 & 7 & compact \\ 
Pilot1\tablenotemark{*} & 5/21/2013 & 5 & subcompact \\ 
Pilot2\tablenotemark{*} & 8/23/2013 & 5 & subcompact \\ 
Pilot3\tablenotemark{*} & 7/24/2013 & 6 & compact \\ 
      & 8/03/2013 & 5 & compact \\ 
      & 8/09/2013 & 5 & compact \\
Pilot4\tablenotemark{*} & 7/25/2013 & 6 & compact \\ 
Pilot5\tablenotemark{*} & 8/01/2013 & 6 & compact \\ 
      & 8/02/2013 & 6 & compact \\ 
Pilot6\tablenotemark{*} & 3/10/2014 & 7 & subcompact \\ 
      & 3/21/2014 & 7 & subcompact \\ 
Pilot7\tablenotemark{*} & 3/19/2014 & 7 & subcompact \\ 
Pilot8\tablenotemark{*} & 3/22/2014 & 7 & subcompact \\ 
Pilot9\tablenotemark{*} & 3/27/2014 & 7 & subcompact \\ 
1 & 5/25/2014 & 7 & compact-1\\
2 & 5/24/2014 & 7 & compact-1 \\
3 & 5/30/2014 & 7 & compact-1 \\
4 & 6/02/2014 & 7 & compact-1 \\ 
5 & 6/04/2014 & 7 & compact-1 \\
6 & 6/07/2014 & 8 & compact-1 \\
7 & 6/10/2014 & 8 & compact-1 \\
8 & 6/13/2014 & 8 & compact-1 \\
9 & 6/14/2014 & 8 & compact-1 \\
10 & 6/15/2014 & 7 & compact-1 \\
   & 6/20/2014 & 7 & compact-1 \\
11 & 6/16/2014 & 7 & compact-1 \\
12 & 6/22/2014 & 8 & compact-1 \\
   & 7/15/2017\tablenotemark{a} & 8 & compact-6 \\
13 & 6/24/2014 & 8 & compact-1 \\
14 & 6/27/2014 & 8 & compact-1 \\
   & 5/30/2016\tablenotemark{b} & 7 & compact-5 \\
15 & 7/09/2014 & 8 & compact-1 \\
16 & 7/10/2014 & 8 & compact-1 \\
   & 4/14/2015 & 7 & compact-2 \\
17 & 4/16/2015 & 7 & compact-2 \\
   & 5/03/2016\tablenotemark{b}  & 7 & compact-5 \\
   & 7/15/2017\tablenotemark{a} & 8 & compact-6 \\
18 & 5/09/2015 & 6 & compact-2 \\
   & 5/03/2016\tablenotemark{b}  & 7 & compact-5 \\ 
19 & 5/10/2015 & 6 & compact-2 \\
   & 5/04/2016\tablenotemark{c} & 7 & compact-5 \\
   & 7/31/2017\tablenotemark{d} & 8 & compact-6 \\
20 & 5/11/2015 & 6 & compact-2 \\
   & 5/04/2016\tablenotemark{c} & 7 & compact-5 \\
   & 7/31/2017\tablenotemark{d} & 8 & compact-6 \\
21 & 7/25/2014 & 7 & subcompact \\
22 & 7/27/2014 & 7 & subcompact \\
23 & 7/28/2014 & 7 & subcompact \\
\hline\hline
\end{tabular}
\quad
\begin{tabular}{cccc}
\hline \hline
Track & Observation & \# of & Array \\
Number & Date & Antennas & Config. \\
\hline
24 & 7/29/2014 & 7 & subcompact \\
25\tablenotemark{$\dagger$} & 8/04/2014 & 7 & subcompact\\
26 & 5/22/2015 & 7 & compact-2 \\
   & 5/30/2016\tablenotemark{b} & 7 & compact-5 \\
27 & 5/23/2015 & 7 & compact-2 \\
   & 6/01/2016\tablenotemark{e} & 7 & compact-5 \\
28 & 5/24/2015 & 7 & compact-2 \\
   & 6/01/2016\tablenotemark{e} & 7 & compact-5 \\
29 & 5/26/2015 & 7 & compact-2 \\
   & 6/04/2016\tablenotemark{f} & 7 & compact-5 \\
30 & 6/02/2015 & 7 & compact-2 \\ 
   & 6/04/2016\tablenotemark{f} & 7 & compact-5 \\
31 & 3/25/2016 & 8 & compact-4 \\
32 & 7/10/2015 & 6 & compact-3 \\
   & 3/16/2016 & 7 & compact-4 \\
33 & 7/22/2015 & 6 & compact-3 \\
   & 3/29/2016 & 8 & compact-4 \\
34 & 6/05/2015 & 6 & subcompact \\
35 & 6/07/2015 & 7 & subcompact \\
36 & 6/06/2015 & 7 & subcompact \\
37 & 6/09/2015 & 7 & subcompact \\
38 & 6/10/2015 & 7 & subcompact \\
39 & 6/13/2015 & 7 & subcompact \\
40 & 6/15/2015 & 7 & subcompact \\
41 & 6/17/2015 & 6 & subcompact \\
42 & 6/18/2015 & 7 & subcompact \\
43 & 6/22/2015 & 7 & subcompact \\
44 & 5/31/2017 & 7 & subcompact \\ 
45 & 7/23/2015 & 6 & compact-3 \\
   & 3/28/2016 & 8 & compact-4 \\
46 & 7/27/2015 & 6 & compact-3 \\
   & 4/30/2016 & 7 & compact-5 \\
47 & 7/28/2015 & 6 & compact-3 \\
   & 5/01/2016 & 7 & compact-5\\ 
48 & 5/07/2016 & 7 & compact-5 \\
49 & 5/02/2016 & 7 & compact-5 \\
50 & 5/08/2016 & 7 & compact-5 \\
51 & 5/10/2016 & 6 & compact-5 \\
52 & 5/14/2016 & 7 & compact-5 \\
   & 5/17/2016 & 7 & compact-5 \\
53 & 5/28/2016 & 7 & compact-5 \\
54 & 5/29/2016 & 7 & compact-5 \\
55 & 5/21/2016 & 7 & compact-5 \\
56 & 5/22/2016 & 7 & compact-5 \\
57 & 5/23/2016 & 7 & compact-5 \\
58 & 6/05/2016 & 7 & compact-5 \\
59 & 6/07/2016 & 7 & compact-5 \\
60 & 6/11/2016 & 7 & compact-5 \\
61 & 6/17/2016 & 8 & compact-5 \\
\hline\hline
\end{tabular}
}

{\singlespace\caption{A summary of \textit{CMZoom} observations, including prior pilot observations marked with a \tablenotemark{*} from \citet{lu19a} and that from \citet{jon14}, labeled J2014. All those tracks marked with \tablenotemark{a-f} were repeated observations which included pointings from multiple different tracks, superscript letters match which tracks were combined. \tablenotemark{$\dagger$We note that track 25 was exploratory in nature, and was used to inform the planning of future observations. As such, this track is not used for this paper, and will not be featured in the first data release.}}}
\end{center}
\label{table-obs}
\end{table*}

\begin{table*}
\begin{center}
{
\centering
\begin{tabular}{|l||l|l|c|c|c|c|c|}
\hline\hline
Region Name & Colloquial Name & Track Numbers & $N_\mathrm{pointings}$ & Mask & Median RMS & $\theta_\mathrm{maj}$ &  $\theta_\mathrm{min}$ \\
 &  &  & (\#) & (\#) & (mJy beam$^{-1}$) & (\arcsec) & (\arcsec) \\
\hline
G1.683-0.089 & 1.6\deg~cloud & 41, 53 & 8 & 1 & 4.5 & 5.0 & 2.5 \\
G1.670-0.130 & 1.6\deg~cloud & 41, 53 & 6 & 2 & 3.9 & 4.9 & 2.5 \\
G1.651-0.050 & 1.6\deg~cloud & 11, 23 & 24 & 3 & 3.4 & 3.1 & 3.0 \\
G1.602+0.018 & 1.6\deg~cloud & 10, 23 & 21 & 4 &2.1 & 4.6 & 2.5  \\
G1.085-0.027 & 1.1\deg~cloud & 15, 16, 24, 37 & 34 & 5 &2.9 & 3.2 & 3.0  \\
G1.038-0.074 & 1.1\deg~cloud & 42, 53, 54, 55 & 46 & 6 & 4.3 & 3.7 & 2.5  \\
G0.891-0.048 & 1.1\deg~cloud & 17, 18, 19, 20, &  & & &  \\
 & & 34, 35 & 82 & 7 &3.4 & 3.7 & 2.7  \\
G0.714-0.100 & Sgr B2 extended & 43, 44, 57, 58, &  & & & \\
 & & 59, 60, 61 & 94 & 8 & 4.8 & 3.4 & 2.8  \\
G0.699-0.028 & Sgr B2 & 7, 8, 9, Pilot7, &  & & & \\
 & & Pilot8, Pilot9 & 74 & 9 & 13. & 3.1 & 3.0 \\
G0.619+0.012 & Sgr B2 NW & 39, 40, 41, 45, 46, 47, & & & & \\
 &  & 48, 49, 50, 51, 52 & 175 & 10 & 2.4 & 6.5 & 3.2  \\
G0.489+0.010 & Dust Ridge Clouds e+f /Sgr B1 & 3, 4, 5, 21, 22  & 44 & 11 & 2.8 & 4.0 & 3.0  \\
-- & Dust Ridge Clouds e+f /Sgr B1 & Pilot2, Pilot5 & 6 & -- &-- & -- & -- \\
G0.412+0.052 & Dust Ridge Cloud d & 3, 21 & 13 & 12 & 3.0 & 3.0 & 2.9  \\
G0.393-0.034 & (isolated HMSF candidate) & 13, 24 & 7 & 13 & 2.8 & 3.4 & 3.1\\
G0.380+0.050 & Dust Ridge: Cloud c & 2, 21 & 9 & 14 & 3.8 & 3.0 & 3.0 \\
G0.340+0.055 & Dust Ridge: Cloud b  & 2, 21 & 9 & 15 & 2.8 & 3.0 & 2.9 \\
G0.326-0.085 & (far-side stream candidate - FSC) & 1, 21 & 20 & 16 & 3.6 & 3.1 & 3.0 \\
G0.316-0.201 & (isolated HMSF candidate) & 13, 24 & 7 & 17 & 6.5 & 3.5 & 3.1 \\
G0.253+0.016 & Brick & KJ2012, Pilot6 & 6 & 18 & 3.7 & 4.3 & 3.0 \\
G0.212-0.001 & (isolated HMSF candidate) & 12, 23 & 7 & 19 & 3.1 & 3.3 & 3.0\\
G0.145-0.086 & Three Little Pigs: Straw Cloud & 14, 24 & 6 & 20 & 3.5 & 3.4 & 2.9\\
G0.106-0.082 & Three Little Pigs: Sticks Cloud & 14, 24 & 5 & 21 & 3.3 & 3.4 & 2.8\\
G0.070-0.035 & (APEX H$_2$CO bridge) & 32, 33, 37 & 39  & 22 & 2.7 & 4.2 & 2.9\\
G0.068-0.075 & Three Little Pigs: Stone Cloud & 14, 24 & 10 & 23 & 3.2 & 3.3 & 3.0\\
G0.054+0.027 & Arches w1 & 43, 57 & 4 & 24 & 8.3 & 3.6 & 3.0 \\
G0.014+0.021 & Arches e1 & 43, 52 & 1 & 25 & 17. & 3.4 & 2.9 \\
G0.001-0.058 & 50 \kms~Cloud & 29, 35 & 24 & 26 & 4.2 & 4.3 & 1.6 \\
-- & 50 \kms~Cloud & Pilot2, Pilot4 & 4 & -- & -- & -- & --\\
G359.948-0.052 & Circumnuclear Disk & 42, 43, 55, 56, 57 & 40 & 27 & 5.2 & 3.7 & 2.9 \\
G359.889-0.093 & 20 \kms~cloud & 26, 27, 28, 36, 38 & 67 & 28 & 4.4 & 4.0 & 1.6 \\
-- & 20 \kms~Cloud & Pilot1, Pilot3 & 8 & -- & -- & -- & --\\
G359.865+0.022 & (far-side stream candidate - FSC) & 30, 37 & 8 & 29 & 2.9 & 4.1 & 2.7 \\
G359.734+0.002 & (far-side stream candidate - FSC) & 6, 22 & 8 & 30 & 3.1 & 3.3 & 3.0 \\
G359.648-0.133 & (stream: Sgr C to 20 \kms cloud) & 30, 38 & 16 & 31 & 2.7 & 4.2 & 2.6 \\
G359.611+0.018 & (far-side stream candidate - FSC) & 6, 22 & 10 & 32 & 3.0 & 3.3 & 3.0 \\
G359.615-0.243 & (isolated HMSF candidate) & 12, 23 & 7 & 33 & 5.6 & 3.3 & 3.0  \\
G359.484-0.132 & Sgr C & 31, 32, 38 & 28 & 34 & 2.4 & 4.1 & 2.9 \\
-- & Sgr C & Pilot1, Pilot4 & 35 & -- & -- & -- & --\\
G359.137+0.031 & (isolated HMSF candidate) & 16, 36 & 7 & 35 & 5.3 & 3.7 & 2.8  \\
\hline
\end{tabular}}

{\singlespace\caption{Regions observed with the \textit{CMZoom} survey as a mosaic of SMA pointings, in order of decreasing Galactic longitudes. The region names are approximately the central coordinates of each mosaic in Galactic coordinates. Many of these regions covered have more common colloqiual names and for those that do not, we include source notes in parentheses. In some cases, for example the 20 \kms~cloud, the full cloud is covered by multiple mosaic regions instead of just one. Details of the observations associated with each track are in Table \ref{table-obs}. The median RMS values refer to the dust continuum, the spectral line RMS values will be described in a forthcoming publication. }}
\end{center}
\label{table-sources}
\end{table*}

\subsection{Observing Strategy}
\label{sec:obs_strategy}
In total, the \textit{CMZoom} survey covered 61 tracks. The exact number of hours spent on the survey is nontrivial to calculate and somewhat ambiguous, since many tracks were repeated or partially repeated several times, the number of antennas varied, the overall observing time per night was not constant, and there were some significant pauses for technical issues. By looking at a few tracks in some detail across these various conditions, we estimate that, on average, each successful track corresponds to slightly greater than 9 hours of observation, adding up to a total of about 550 hours of observations for the survey, including overhead. This estimate is made assuming that each track had no major issues. Considering that a number of tracks needed additional executions, the total real on-sky observing time is even larger. 


The full \textit{CMZoom} survey consists of 974 individual mosaic pointings (see Figure \ref{fig:coverage}). \textit{CMZoom} observed 45 compact configuration tracks, with an average of 21.5 pointings per night, and 16 subcompact configuration tracks, with an average of about 60 pointings per night. These pointings were repeated throughout the night to ensure good UV coverage. The observations were done in a loop, wherein each pointing was observed for ten 7.7 second scans at a time, followed by gain calibration observations, then this sequence was repeated. The scans were the minimal length allowed, in order to avoid losses when scans were corrupted. 

Bandpass and flux calibrations were performed at the beginning and/or end of the night depending on source availability. Most often, 2-3 sources were observed for bandpass calibration (3C454.3, 3C279, and Saturn), and 2-4 for flux calibration (Callisto, Titan, Neptune, Uranus, Mars). The system temperature was regularly corrected throughout the observation with a chopper wheel. We performed gain calibrations approximately every 15 minutes on 1733-130 and secondary gain calibrations on 1744-312 approximately every 30 minutes. Pointing was done at the beginning of the night and about an hour after sunrise or sunset when the dishes were subject to slight pointing changes due to warming from the sun. Data from the secondary gain calibrator were reduced like a normal science source and imaged, which was used to test the success of the data reduction. Pointing corrections were performed regularly throughout each track (at the very least at the start and an hour after sunset or sunrise). Typical pointing accuracy with the SMA is about 2\arcsec \citep{Ho04}.

The \textit{CMZoom} survey aimed to achieve a uniform RMS noise, which is a challenge, given varying observational conditions, as well as changes to the backend over the course of the 4-year survey. These variations include normal weather variations, total `track' time variations due to length of time that the Galactic Center is visible above the horizon over the course of the season or technical issues, the number of antennas included in each observation, receiver bandwidth (spanning the range of 8-16 GHz over the survey), as well as the fundamental dynamic range limit imposed by very bright targets. Throughout the survey, we have carefully monitored the noise levels in each region of the survey and requested track repetitions or partial repetitions (such as joint re-observations, as explained in Section \ref{sec:config}) where required to achieve moderate uniformity. Many regions had sufficiently good RMS noise on the first time, while others required many repetitions (see Table \ref{table-obs}). In these cases, if the data for some tracks were truly unusable, they were discarded, otherwise, we included all data that made a positive contribution to the overall RMS noise. Only tracks included in the final images are included in Table \ref{table-obs}. In this manner, we were able to achieve a median RMS noise of 13 mJy Sr$^{-1}$, discussed in more detail in Section \ref{sec:rms}. We note that the dust continuum data for clouds G0.054$+$0.027 and G0.014$+$0.021 (Arches w1 and e1) are amongst the noisiest in the entire survey due to peculiarities of their observation and these data are of overall poor quality.


\begin{figure*}
\begin{center}
\includegraphics[width=1\textwidth]{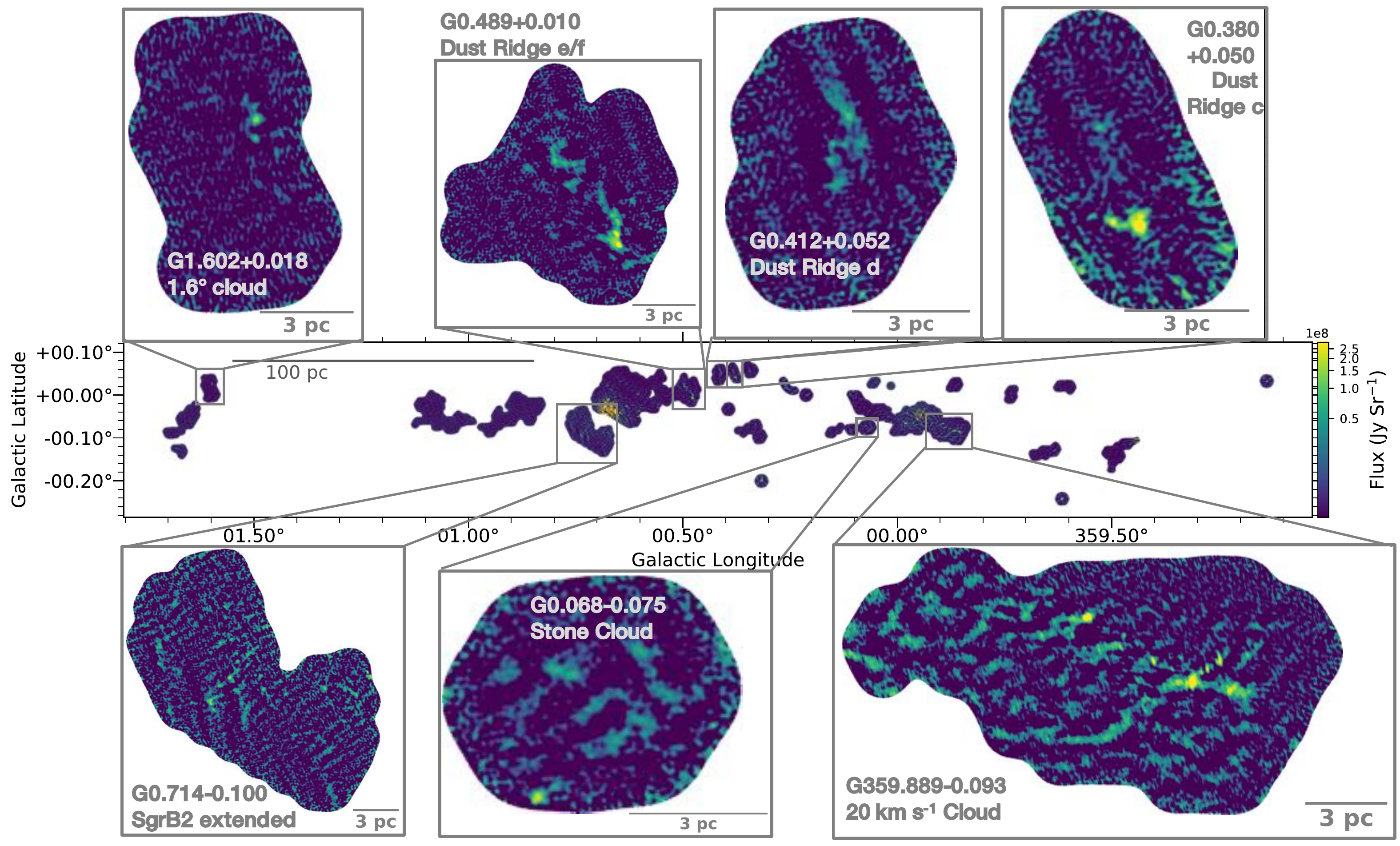}
\singlespace\caption{A mosaic of the full coverage of the \textit{CMZoom} survey in 1.3 mm dust continuum, with zoom-ins toward various regions of interest. The images are all on the same color scale. The full image gallery is in Appendix \ref{appendix-images}. Locally higher noise is seen in the vicinity of the strong continuum sources Sgr B2 ($l\sim0\degr.7$) and Sgr A* ($l\sim359\degr.9$)}
\label{fig:mosaic}
\end{center}
\end{figure*}

\section{Data Calibration and Imaging}
\subsection{SMA Data Calibration}
All datasets, independent of their correlator setups ({\sc asic}-only, {\sc asic}$+${\sc swarm} and {\sc swarm}-only), were calibrated using the {\sc mir idl} software package following standard SMA calibration procedures\footnote{\label{ft:keto}Eric Keto's {\sc mir idl} webpage: \href{https://www.cfa.harvard.edu/sma/mir/}{https://www.cfa.harvard.edu/sma/mir/}}. Before importing into {\sc mir idl}, {\sc swarm} data was ``rechunked" (divided into larger bins) by a smoothing factor of 8, in order to match the previous {\sc asic} spectral resolution of 1.1 \kms. Some tracks also required an updated baseline calibration, as noted in the SMA observer logs, which was the first step in the calibration process. For the most part, poor weather visibility datasets with system temperatures higher than 400 K were discarded, however, this was not a strict rule. Generally, any data of sufficient quality to improve the overall region RMS was included. See the full list of tracks included in the final data in Table \ref{table-obs}.

Once the above tasks were completed, the first calibration step was to calibrate the system temperature over the course of the night. The next step was to perform a bandpass calibration using observations of either 3C454.3, 3C279, Saturn, or a combination of these sources when available. Bandpass data were inspected for noise spikes in every baseline, which were subsequently removed by averaging the adjacent channels. When multiple correlators ({\sc asic}$+${\sc swarm}) were used for the observations, they were bandpass calibrated independently. Gain calibration is the next step, both phase and amplitude, performed with standard SMA routines. Both phase and amplitude of our phase calibrators on each baseline were inspected to identify  ``bad'' data and phase jumps. Phase jumps required some data to be flagged, split into separate time intervals and calibrated independently.  

Flux calibration was performed based on comparison with the brightness of planets and their satellites, based on models of brightness temperature adopted from the Atacama Large Millimeter/submillimeter Array's (ALMA's)
{\sc CASA} software as outlined in Eric Keto's {\sc mir idl} webpage (see footnote~\ref{ft:keto}). Flux calibrations were checked against the standard SMA calibrator list\footnote{\href{http://sma1.sma.hawaii.edu/callist/callist.html}{http://sma1.sma.hawaii.edu/callist/callist.html}} with reasonable agreement. The uncertainty in the absolute flux calibration was estimated to be $\sim 10-20$\%. Next, Doppler corrections were performed on all science targets. The final step of the data calibration was a careful inspection of the data as a function of time and frequency. At the end of data reduction, we imaged our secondary gain calibrator (1744-312) for every track to verify the quality of phase transfer that was based on our primary calibrator (1733-130). The imaging and deconvolution were accomplished using the {\sc MIRIAD} and {\sc CASA} software packages, as explained in further detail in the following section.

\subsection{Imaging Pipeline}
The fully-calibrated SMA compact and subcompact data are merged, deconvolved (cleaned), and imaged in {\sc casa}. First, however, the calibrated {\sc mir} data files produced by {\sc mir idl} must be processed and prepared for imaging. Due to the large number of data files we opted to develop an imaging pipeline, such that the full survey data products could be generated in a fully-automated, uniform, and repeatable manner. In the following subsections, we describe this process in detail, and the complete scripts have been made publicly available on the CMZoom GitHub page$^3$.

The first step of the pipeline is to extract the relevant data for the given science target. The script takes the given name of the science target and the path(s) to the corresponding calibrated SMA data files. In general, each source will have at least two calibrated data files, corresponding to the compact and subcompact data. However, many sources were observed over multiple nights and therefore can have many associated data files. This is typically either because the source is large, and therefore required multiple tracks to complete the pointing mosaic, or because the track was of marginal quality and had to be repeated to achieve satisfactory quality when combined. 

Each file is successively loaded into {\sc mir}, where the meta-data are inspected to determine whether the data were taken with the {\sc asic} or {\sc swarm} correlator, or some combination of the two. It is necessary to make such a distinction, as the data from the two correlators must be exported and processed separately prior to imaging. This is due to the fact that the {\sc swarm} correlator provides many more channels per spectral window, and therefore requires a greater number of channels to be flagged on the edges of the windows. Having determined the correlator information, the script then uses the {\sc idl2miriad} routine to export the source data in {\sc miriad} format. At the time of our analysis, there was a known bug when exporting entire sidebands and converting to {\sc casa} measurements sets, where the frequency information of data cubes is offset and gaps are introduced between each spectral window. To circumvent this, we export each spectral window individually, process it separately, and recombine all windows again before imaging.

All spectral window data associated with the given science target are loaded into {\sc miriad} to be processed prior to imaging. In general, there are 48 spectral windows for the {\sc asic} data and 2--4 for the {\sc swarm} data. The noisy edge channels for each spectral window are flagged using the \emph{uvflag} command. The number of edge channels flagged are 10 and 100 for the {\sc asic} and {\sc swarm} windows, respectively, which corresponds to approximately 10\% of the full bandwidth. Each spectral window is then exported as a \emph{uvfits} file using the \emph{fits} command, and then imported into {\sc casa} as a MeasurementSet. For a given source, all corresponding tracks are concatenated per sideband using the \emph{concat} command. This ultimately results in two or four measurement sets per source, depending on the correlator used, either {\sc asic} or {\sc swarm}, which each have two sidebands.

\begin{figure*}
\begin{center}
\subfigure{
\includegraphics[width=1.0\textwidth]{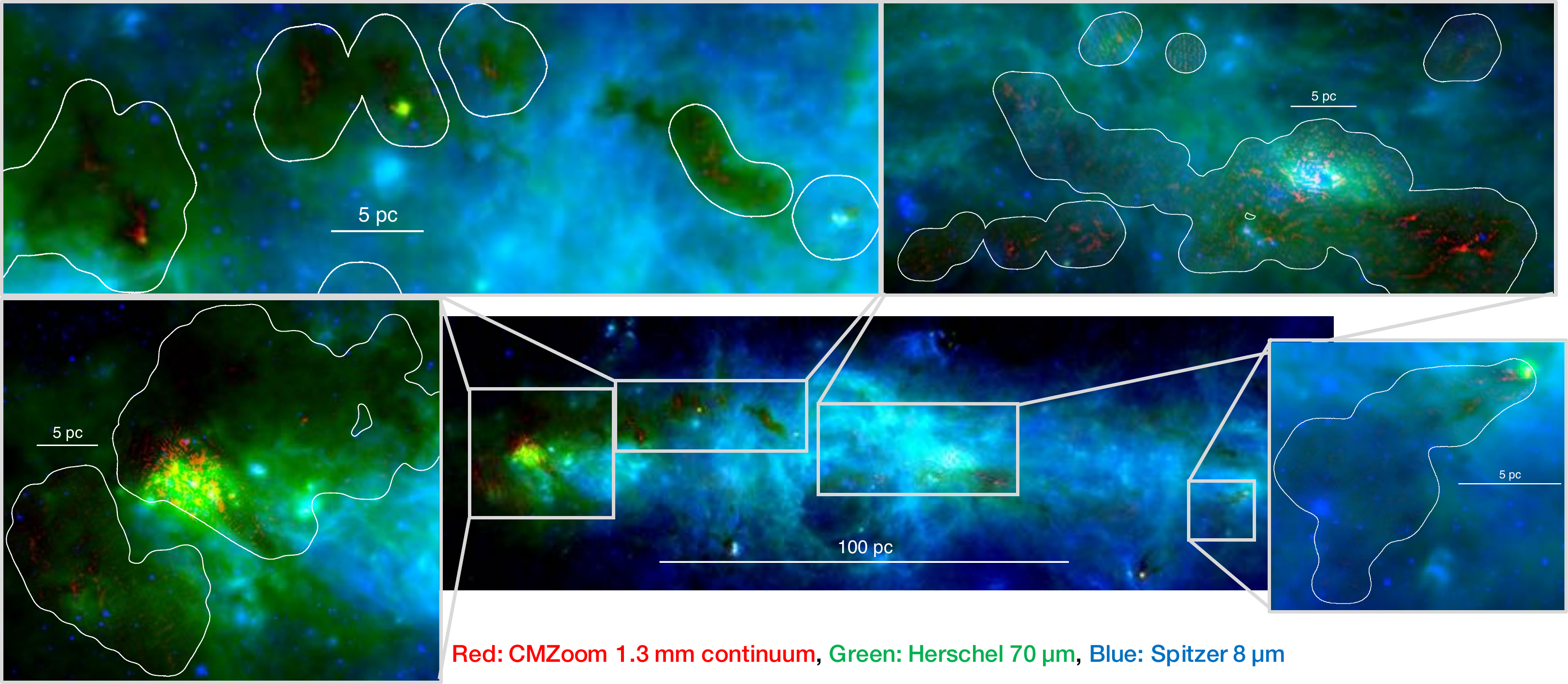}}
\subfigure{
\includegraphics[width=1.0\textwidth]{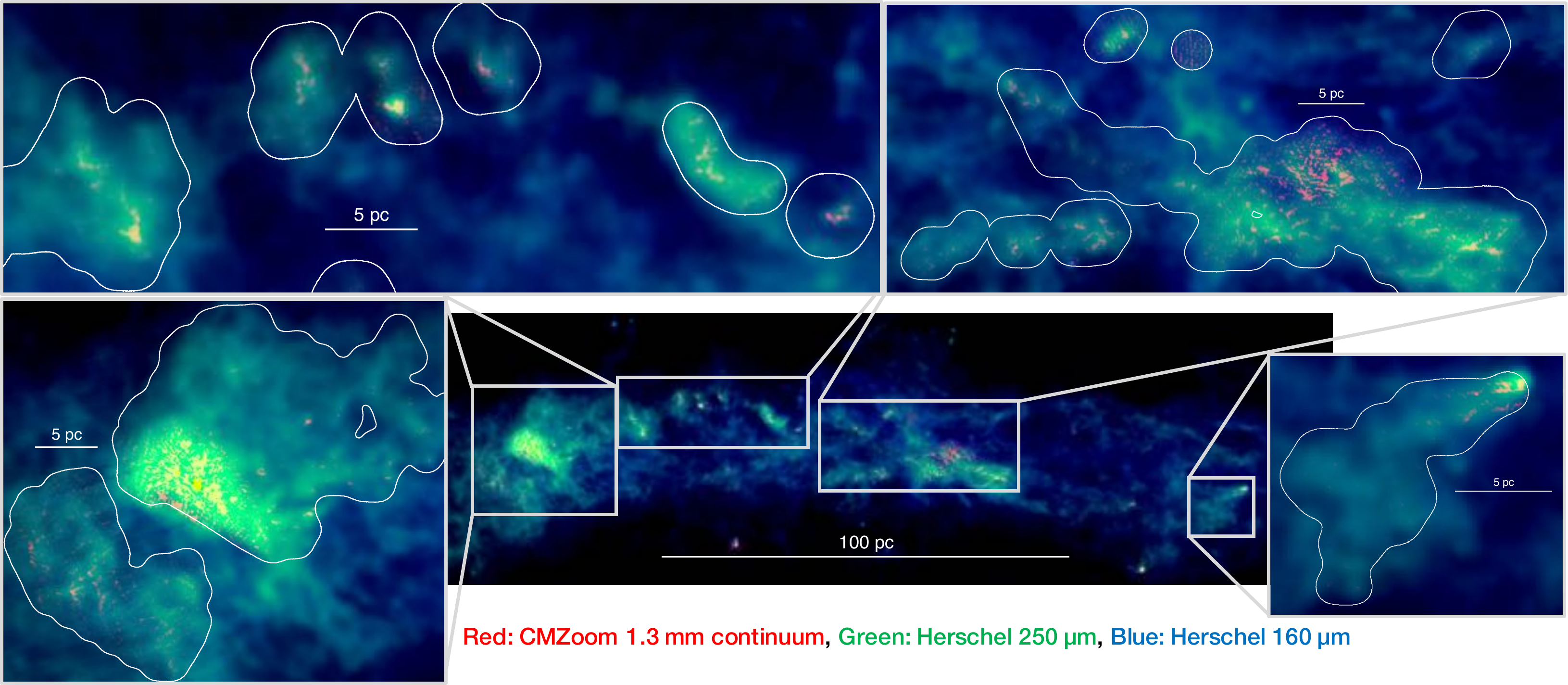}}
\singlespace\caption{Three color images of the inner $\sim$200 pc of the Galaxy, highlighting new \textit{CMZoom} data in red, with various zoom-in views towards regions of interest. This figure shows only part of the full \textit{CMZoom} coverage which extends in longitude in both directions (see Figures \ref{fig:coverage} and \ref{fig:mosaic}). All figures show \textit{CMZoom} 1.3 mm dust continuum in red (not primary beam corrected) with survey coverage contours in white. The \textit{top} figures show Herschel Hi-GAL 70 \micron~\citep{mol10} in green, and Spitzer GLIMPSE 8 \micron~\citep{ben03} in blue. The \textit{bottom} figures show Herschel Hi-GAL 250 and 160 \micron~\citep{mol10} in green and blue, respectively. }
\label{fig:3color_wow}
\end{center}
\end{figure*}

\subsection{Continuum imaging}

Prior to imaging the dust continuum emission, the continuum component of the emission must be subtracted from the spectral line data by identifying the line free channels. To do this in an automated way, we utilized the \emph{findContinuum} function of the hif\_findcont task of the ALMA Cycle 7 pipeline version 5.6.1 \citep{Humphreys16}.\footnote{\href{https://almascience.nrao.edu/documents-and-tools/alma-science-pipeline-users-guide-casa-5-6.1}{https://almascience.nrao.edu/documents-and-tools/alma-science-pipeline-users-guide-casa-5-6.1}} This is done in the image plane and is used to inspect data cubes and determine the uncontaminated continuum-only channels. First we use the \emph{tclean} command with zero iterations to generate dirty cubes for all measurement sets for a given source. The \emph{findContinuum} routine then takes each dirty cube, creates an averaged spectrum, and searches for any emission that is greater than some user-defined threshold. We choose a threshold of 5-$\sigma$ as this is a standard choice in the literature. We find this threshold provides a good compromise between false positives and completeness. Anything fainter than this threshold is not likely to contribute substantially to the continuum flux. The program then determines the range of channels that do not have emission above this limit (i.e. the line-free, continuum-only channels). This routine outputs the identified line-free channels in plain-text format such that they can be fed directly into the \emph{tclean} command in {\sc casa} to generate a continuum image from the data cubes.

\begin{figure*}
\begin{center}
\includegraphics[width=1.0\textwidth]{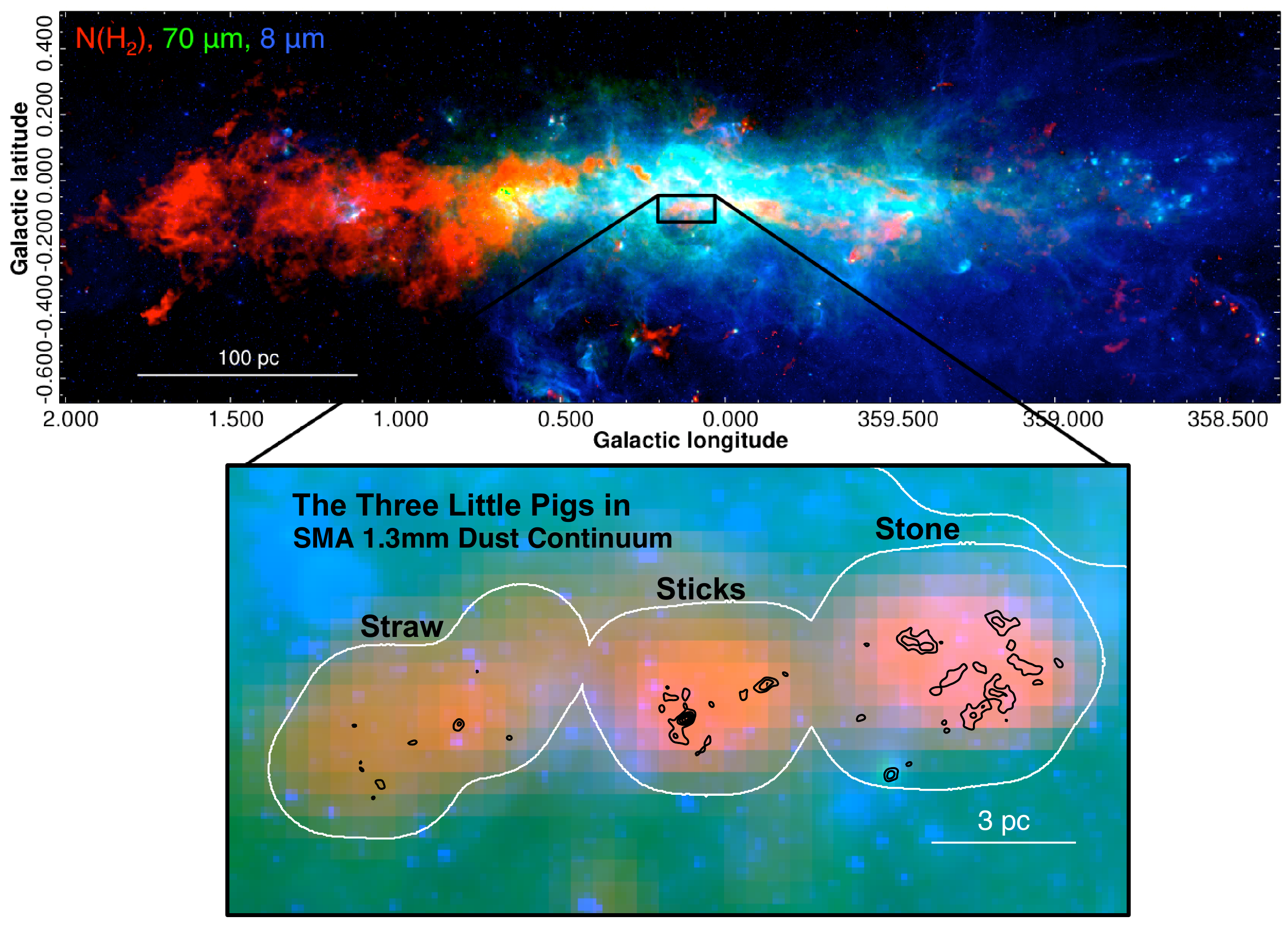}
\singlespace\caption{A three-color image of the CMZ, as seen with Herschel in N(H$_2$) \citep[from][Battersby et al. in prep.]{mol10} in red, 70 \micron~\citep{mol10} in green, and Spitzer GLIMPSE 8 \micron~\citep{ben03} in blue. The lower-panel is a zoom-in toward three clouds dubbed the ``Three Little Pigs.'' In previous single-dish data, the clouds have similar column densities, masses, temperatures, and sizes, however, the detailed structure seen with the SMA dust continuum (shown as the four black contour levels at 2, 4, 6, and 8 $\sigma$) tells a different story. From left to right, the clouds increase in their levels of substructure, from the `Straw Cloud' on the left, with very little substructure, to the `Sticks Cloud' in the middle with moderate substructure, and finally to the `Stone Cloud' on the right, with the most substructure.}
\label{fig:tlp}
\end{center}
\end{figure*}

To generate images of the 1.3~mm dust continuum emission, all measurement sets for a given source are imaged together using \emph{tclean} using the multi-frequency synthesis gridder. The \emph{spw} parameter is used to specify the continuum-only channels for each measurement set that were determined in the previous step using \emph{findContinuum}. A range of input parameters were explored for \emph{tclean} to determine how they affected the resultant images. We decided to use the \emph{multiscale} parameter with scales of [0, 3, 9, 27], to better recover both the large- and small-scale structures within the images. We use the \emph{Briggs} weighting scheme with a \emph{robust} parameter of 0.5, as this yields a fair compromise between the angular resolution and the noise properties of the resulting image. We also set the pixel scale to 0.5$^{\prime}$$^{\prime}$, which equates to 6-8 pixels per beam major axis given typical synthesized beams of approximately 3-4$^{\prime}$$^{\prime}$. To apply clean masks during the cleaning, we used the \emph{auto-multithresh}\footnote{\href{https://casa.nrao.edu/casadocs/casa-5.3.0/synthesis-imaging/masks-for-deconvolution}{https://casa.nrao.edu/casadocs/casa-5.3.0/synthesis-imaging/masks-for-deconvolution}} parameter in \emph{tclean} \citep{automultithresh20}. This auto-masking algorithm is implemented to iteratively generate and grow masks in a way that is similar to how a user would manually create masks. This requires several user-defined input parameters. We use the recommended parameter values for ALMA 7~m (ACA) observations, as the array is reasonably similar to the SMA. These parameters are: \emph{sidelobethreshold}  =  1.25, \emph{noisethreshold} = 5.0, \emph{lownoisethreshold} = 2.0, \emph{minbeamfrac} = 0.1, and \emph{growiterations} = 75. To determine the appropriate cleaning threshold for each region in the survey, we first make rough continuum maps using a uniform cleaning threshold of 5 mJy beam$^{-1}$, which corresponds to $\sim$2 $\sigma$ RMS of the survey. We then take the residual maps for each region, and measure the RMS using a number of rectangular regions of various sizes that are placed randomly within the confines of each mosaic. We then take the median of the RMS values, which is used as the final cleaning threshold per region, which we  set to 2 $\sigma$. We set the clean iterations arbitrarily high such that the algorithm reached the threshold value and was not limited by the number of iterations. The images used in the remainder of the paper have been corrected for the primary beam using \emph{pbcor} (with the exception of Figures \ref{fig:tlp}, \ref{fig:3color_wow}, \ref{fig:carma}), but we also release the uncorrected version of the data.

The final images are then exported from {\sc casa} as {\sc fits} files. In addition to {\sc fits} files of the individual source dust continuum emission, we also produce a full \text{CMZoom} survey dust continuum emission mosaic {\sc fits} file. We do this via a combination of different Python packages. First, each individual survey image is transformed from units of Jy beam$^{-1}$ to MJy Sr$^{-1}$ using {\sc radio\_beam}\footnote{\href{https://github.com/radio-astro-tools/radio-beam}{https://github.com/radio-astro-tools/radio-beam}} to extract the beam information, which is then used with {\sc astropy}\footnote{\href{http://www.astropy.org/}{http://www.astropy.org/}} to account for the beam and transform the units. This conversion is performed as the different survey regions have differing beam properties (see Section \ref{sec:rms}), and it is therefore not appropriate to include the beam information in the units of the mosaic. The transformed images are then reprojected on to a large fits image using the {\sc reproject}\footnote{\href{https://reproject.readthedocs.io/en/stable/}{https://reproject.readthedocs.io/en/stable/}} package to obtain a full survey mosaic with consistent units of MJy/sr. The {\sc reproject} package is also used to transform the native images from J2000 to Galactic coordinates, due to a known bug at the time in the {\sc casa} transformation that has since been fixed\footnote{\href{https://casa.nrao.edu/casadocs/casa-5.4.0/introduction/release-notes-540}{https://casa.nrao.edu/casadocs/casa-5.4.0/introduction/release-notes-540}}. 

\subsection{Combination with Single-Dish Data}
\label{sec:SD}
We release SMA-only data products, including the combined SMA compact and subcompact configuration data for each region, as well as data products that have been combined with single-dish (zero- and small-spacing) data to achieve better recovery of structure at large spatial scales (see Section \ref{sec:release}). 

The Bolocam Galactic Plane Survey (BGPS) surveyed the Galactic center region at 1.1 mm (271.1 GHz) with a resolution of 33\arcsec~\citep{bal10, agu11, gin13}, and is currently the best data for combination, due to its proximity in frequency, and resolution being reasonably well-matched with the SMA primary beam (about 45$\arcsec$). For the dust continuum emission, we scale the BGPS data to the SMA-observed wavelength (about 1.3mm), assuming a spectral index of 1.75 \citep{bat11} and combine with the SMA and BGPS data using the CASA task {\sc feather}. The BGPS achieves a complete sensitivity to large spatial scales of 80$\arcsec$ or 3 pc and partial recovery of spatial scales up to 300$\arcsec$ or 12 pc \citep[for more details see][]{gin13}.


We have investigated other methods for single-dish combination, such as using the single-dish data as a model for the SMA cleaning, then combining. However, we find that the \textit{feather} task performs equally well and choose this method for this work. 

\subsection{Spectral Line Data}
Information about the \textit{CMZoom} spectral line data processing, overview, and release will be provided in a forthcoming publication. We have currently in hand APEX data from \citet{gin16} for single-dish combination with the lower-sideband SMA data (about 216-220 GHz). The resolution is about 30$\arcsec$~and will complement the SMA data with sensitivity to large spatial scales, which is especially important for recovering emission from extended and diffuse gaseous structures in the molecular clouds.

\begin{figure*}
\centering
\includegraphics[trim = 0mm 0mm 0mm 0mm, clip, angle=0, width=0.75\textwidth]{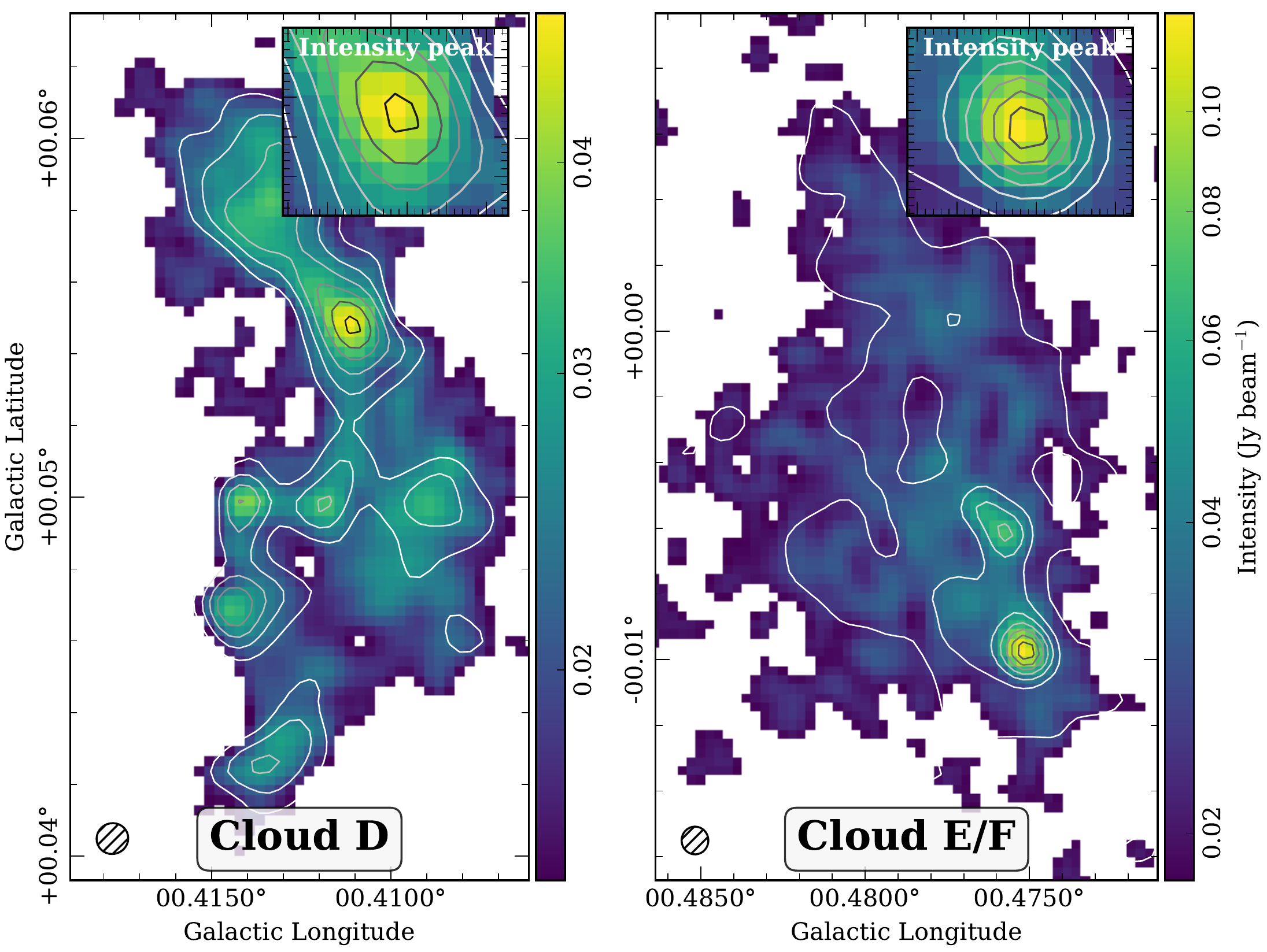}
\singlespace\caption{\textit{CMZoom} observations compare favorably with more recent ALMA observations \citep{barnes2019}. Shown in colorscale are the SMA observations towards the Dust Ridge Cloud D \textit{(left, also known as G0.412$+$0.052)} and Cloud E \textit{(right, also known as G0.489$+$0.010)} molecular clouds. These have been masked above a 3\,$\sigma$ level, where $\sigma\,=\,5$\,mJy~beam$^{-1}$ for Cloud D and $\sigma\,=\,6$\,mJy~beam$^{-1}$ \citep[upper limits from][]{wal18}. Overlaid are contours from similar frequency ALMA observations ($\sim 250$ GHz) towards these sources, which have been smoothed to the approximate resolution of the SMA observations for comparison (see beam in lower left of each panel). The contours have been plotted in colors of white to black with increasing intensity. Also shown for clarity is a zoom-in of the flux peak in the upper right of each panel, with an identical color scale and contours to the full image.}
\label{sma_alma_comp}
\end{figure*}

\section{Data Description}

The \textit{CMZoom} survey mapped about 350 square arcminutes of the highest column density gas in the inner 500 pc of the Galaxy in the 1.3~mm dust continuum and a variety of spectral line tracers. The approximate survey resolution is 3.25\arcsec~(about 0.1 pc), and median RMS noise level is 13 MJy Sr$^{-1}$ (see Section \ref{sec:rms}). The spatial recovery of features and astrometric accuracy are supported by comparison with recent ALMA data (see Section \ref{sec:pointing}). While every attempt has been made to ensure good recovery of structures on various spatial scales and to minimize imaging artifacts, we caution the reader to interpret these images with care. In particular, structures near the noise limit or the edge of the map should be interpreted with caution. Structures on scales smaller than the beam should not be considered reliable. In areas near very bright emission (e.g., Sgr B2 and Sgr A*), there are known imaging artifacts, as removal of the  sidelobes from bright sources is an imperfect and nonlinear process. 

 Figure \ref{fig:mosaic} shows the full \textit{CMZoom} survey mosaic as well as the 1.3~mm dust continuum images for a handful of clouds from the \textit{CMZoom} Survey. The full image gallery is in Appendix \ref{appendix-images}. Figure \ref{fig:3color_wow} shows the SMA \textit{CMZoom} 1.3 mm continuum data in three-color images along with Herschel \citep{mol10} and Spitzer \citep{ben03} infrared data.  The fully-processed SMA \textit{CMZoom} data are publicly released with this publication (see Section \ref{sec:release} for details). 

Figure \ref{fig:tlp} demonstrates the power of the \textit{CMZoom} survey. This figure highlights three clouds within the CMZ, which have similar column densities, temperatures, masses, and overall appearances in previous single-dish data. However, the \textit{CMZoom} data tell a different story of these three clouds, dubbed the ``Three Little Pigs." Going from left to right, these clouds show an increasing degree of substructure on small scales, from the scantly substructured `Straw Cloud' (G0.145$-$0.086) on the left (higher Galactic longitude), to the moderately substructured `Sticks Cloud' (G0.106$-$0.082) in the middle, to the highly substructured `Stone Cloud' (G0.068$-0.075$) on the right (lower Galactic longitude). Without the detailed look at these clouds with the SMA, their internal differences would not have been uncovered. The nature of the differing degree of substructure is still the subject of ongoing investigation. 

\begin{figure*}
\centering
\subfigure{
\includegraphics[ clip, angle=0, width=1.0\textwidth]{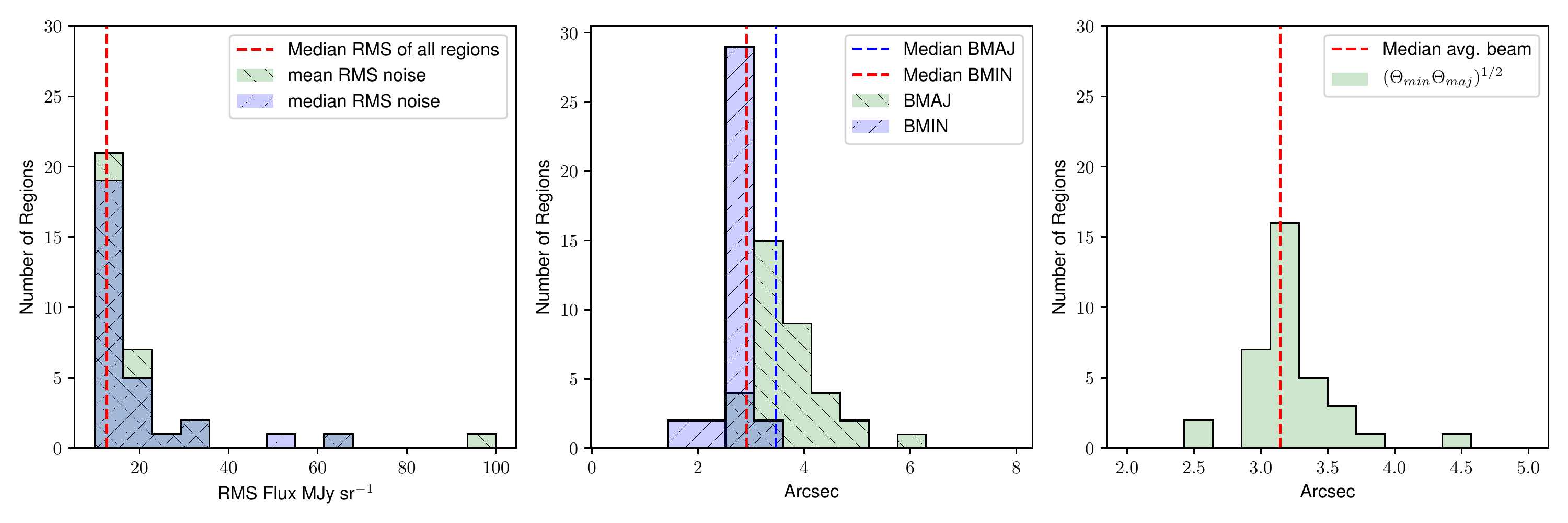}}\\
\singlespace\caption{The overall angular resolution of the \textit{CMZoom} survey is about $3\farcs{}2$ (0.13 pc) and the RMS noise level is about 13 MJy Sr$^{-1}$. There is variation across the survey, due to varying observing conditions and antenna configurations, as well as variations in the dynamic range based on the source emission in each field. Left: Histograms of the median and mean RMS noise value for each of the regions in the survey. Middle: Histograms of the variation of the beam major and minor axes over all surveyed regions. Right: A histogram of the geometrically averaged beam size $(\theta_{\rm min} \cdot \theta_{\rm maj})^{1/2}$ over all regions in the survey. }
\label{fig:rms}
\end{figure*}

\subsection{Astrometric Accuracy}
\label{sec:pointing}
To investigate the source structure recovery and astrometric accuracy of the SMA observations presented in this work, we compare to similar frequency ($\sim$259 GHz) ALMA observations that overlap with part of the {\it CMZoom} survey coverage. These observations focus on the Cloud D and Cloud E dust-ridge molecular clouds (i.e. G$0.412+0.052$ and G$0.489+0.010$ in Table \ref{table-sources}), and were taken using the Band 6 receiver ($\sim$\,250 GHz) with the C43-2 array configuration (max baseline of $\sim$\,300\,m) as part of ALMA Cycle 2 project (project ID: 2013.1.00617). These achieved an angular resolution of $\sim$\,1\arcsec, and form the basis of higher resolution observations to study sources of interest highlighted by the {\it CMZoom} survey \citep[full detail presented in][]{barnes2019}.

As a qualitative comparison, we first smooth the ALMA observations to the SMA beamsize in each map ($\sim 3\arcsec$) to match the angular resolution. These ALMA observations are then overlaid as contours on the SMA observations shown in color-scale. These maps are presented in Figure\,\ref{sma_alma_comp}, from which initial inspection shows no obvious differences between the SMA and ALMA observations in either the pointing centers or the structures recovered. Focusing in more detail on the peaks in continuum emission for both clouds, which are shown in the upper right of each panel, confirms that there is no spatial offset larger than the beam-size for both clouds. For a more quantitative comparison, we conduct a cross correlation analysis to analyze the difference between the two images. This was done using the {\sc image registration} \footnote{\href{https://image-registration.readthedocs.io/en/latest/}{https://image-registration.readthedocs.io/en/latest/}} python package, which was specifically designed to determine if any spatial offsets exist between two sets of observations. We use the {\sc cross correlation shifts} function in the package to conduct this analysis, after setting the mean values of both images to zero, as recommended such that zero values are not correlated between two datasets. We find that the offset for both clouds is significantly smaller than the size of the beam ($< 0.1$\arcsec), hence confirming the manual inspection. We can, therefore, conclude with confidence that the astrometric accuracy for the SMA observations is consistent to within a beam-size with independent ALMA observations. Another benefit of this comparison is the revelation that, qualitatively, the structures recovered with the SMA appear to be consistent with the structure seen in the ALMA maps.

\subsection{Beam Size and RMS Noise Level}
\label{sec:rms}

Due to a variety of observing conditions, the \textit{CMZoom} average beam size shows some variation. Here, and in the remainder of the text, we are referring to the `synthesized' interferometric beam size of the SMA.
Figure \ref{fig:rms} outlines the beam sizes as reported by the imaging procedures. The beam minor axis ranges from $1\farcs{}6$ to $3\farcs{}1$, with 75\% of the maps having a minor beam axis of $2\farcs{}7$ to $3\farcs{}0$. The major axis varies from $3\farcs{}0$ to $6\farcs{}3$, with 75\% of the maps having a major beam axis of $3\farcs{}3$ to $4\farcs{}1$. The elongation of the beams varies between maps. In most maps, the major--to--minor beam axis ratio is between 1 and 2. Three maps have much more elongated beams, with axis ratios between 2.5 and 3.0. As a general trend, the beam minor axis decreases when the beam major axis increases.

The resulting angular resolution in a map with a beam of size $\theta_{\rm min} \times \theta_{\rm maj}$ can be described by the geometrically averaged beam size, $(\theta_{\rm min} \cdot \theta_{\rm maj})^{1/2}$. This property is included in Figure \ref{fig:rms}, and it ranges from $2\farcs{}6$ to $4\farcs{}8$, with 75\% of the maps having a geometrically averaged beam size between $3\farcs{}0$ and $3\farcs{}3$. We thus summarize that the CMZoom maps have a typical angular resolution of $3\farcs{}2\pm{}0\farcs{}1$ (mean of the geometric combination of major and minor axes with standard deviation). At a Galactic Center distance of 8.15 kpc, this corresponds to a spatial resolution of 0.13 pc. 

The \textit{CMZoom} continuum maps are corrected for the primary beam pattern, and have a spatially intricate noise distribution. We estimate the noise using the residual map produced during the imaging process, ${R}_{x,y}$ and computing the local standard deviation, $\sigma$, from the definition 
\begin{equation}
    \sigma_{x,y} =\sqrt{ \left<(R_{x,y} - \left<R_{x,y}\right>)^2\right>}
\end{equation} 
where $\left<R_{x,y}\right>$ is the mean of the residual map evaluated over some region centered at pixel $x,y$.  We use a Gaussian kernel $G$ with FWHM=14 pixels as this region and perform a weighted average across the Gaussian.  The weighted average computed at each pixel is equivalent to the convolution, so  $\left<R_{x,y}\right> = R_{x,y} \ast G$.  We can therefore write 
\begin{equation}
        \sigma_{x,y} =\sqrt{ \left<\left(R_{x,y} - \left(R \ast G\right)_{x,y}\right)^2\right>}
\end{equation}
which, following the same logic, becomes
\begin{equation}
        \sigma_{x,y} =\sqrt{ \left(R_{x,y} - (R \ast G)_{x,y}\right)^2 \ast G_{x,y}}
\end{equation}
In words, we first smooth the residual map, calculate the difference between the smoothed residual map and the un-smoothed residual map, square the resulting difference map, smooth the difference map, and then take the square root of the map. The Gaussian kernel size for the smoothing has a FWHM of 14 pixels, chosen to be about twice the effective radius of typical resolved compact source emission in the continuum maps and is justified using simulated observations in the catalog paper (Hatchfield et al., in prep.). The resulting map a first-order estimate of the root-mean-squared (RMS) noise level in the maps, given that the maps are dominated by regions free of emission. The median value of the RMS noise over all maps is about 13 MJy Sr$^{-1}$.

We use this as our estimate of the local noise level on spatial scales relevant to the dense emission in the maps. We compare the standard deviation in a map to the 95.4th percentile of the negative and positive emission, respectively. In theory, if the noise is Gaussian and the noise level in an intensity map $I$ is $\sigma(I)$, the intensity where the 95.4th percentile is achieved, $I_{95.4\%}$, should be identical to $2\,\sigma(I)$. We indeed find that $|I_{95.4\%}|/\sigma(I)\approx{}2$ for negative pixels in all but three maps. Specifically, a higher ratio is found in the regions G0.253$+$0.016 and G0.699$-$0.028. Ratios $2\lesssim{}|I_{95.4\%}|/\sigma(I)\lesssim{}4$ prevail for positive pixels in most maps, except for the images of the aforementioned regions that also violate $|I_{95.4\%}|/\sigma(I)\approx{}2$ for negative pixels. This suggests that these particular maps have larger noise levels that are not well characterized by Gaussian noise, the most extreme example being the region around Sgr B2. We suspect this is the case due to the presence of significant imaging artifacts due to the extremely bright and complicated emission structure in that region. 

The conversion from MJy Sr$^{-1}$ to mJy beam$^{-1}$ varies with the beam properties for each region, so we report the typical noise properties and averages in units of MJy Sr$^{-1}$, such as in Figure \ref{fig:rms}. The median value of the RMS noise over all maps is about $13~\rm{}MJy\,Sr^{-1}$ and the median RMS noise level in individual maps varies between 7 and $64~\rm{}MJy\,Sr^{-1}$ (the high values are due dynamic range issues as a result of bright pixels in Sgr B2), as shown in Figure \ref{fig:rms}. The middle 75\% of the maps have median RMS noise values from 11 to $18~\rm{}MJy\,Sr^{-1}$. The RMS noise maps generated using this method are used for the construction  of the continuum catalog (Hatchfield et al., in prep.) and therefore strongly influence the completeness limit of the catalog. The local noise maps are used in the production of the catalog to ensure robust findings.

\subsection{Data Release}
\label{sec:release}
The fully-processed 1.3~mm dust continuum images, both SMA-only and combined with single-dish, are made publicly available on the \textit{CMZoom}~Dataverse\footnote{\textit{CMZoom}~Dataverse: \href{https://dataverse.harvard.edu/dataverse/cmzoom}{https://dataverse.harvard.edu/dataverse/cmzoom}} upon publication of this paper. Most users will likely want the full 1.3 mm dust continuum mosaic that is SMA-only and primary beam corrected. This dataset is marked on the front page of the Dataverse as the `most simple and complete map' for easy access. Some users may also want the full continuum mosaic mask, residual map, noise map,  mosaic feathered with single-dish, or non-primary beam corrected mosaic. These are available in the ``Full Survey Mosaics'' dataset. There are an additional two datasets in the Dataverse that contain images of individual regions,  primary beam corrected (most people will want this version) and non-primary beam corrected. Finally, the reduced visibility data for each track is also be released on the \textit{CMZoom}~Dataverse as a separate dataset. Observation details for each track can be found in Table \ref{table-obs} and the association of tracks with sources is in Table \ref{table-sources}. The complete raw SMA data are available in the SMA archive\footnote{SMA archive: \href{https://www.cfa.harvard.edu/cgi-bin/sma/smaarch.pl}{https://www.cfa.harvard.edu/cgi-bin/sma/smaarch.pl}}. 

\section{Compact Substructures in the CMZ}
\label{sec:compact}
The \textit{CMZoom} survey provides a uniquely comprehensive view into the paucity of star formation in the CMZ, allowing us to address questions of whether compact cores in the CMZ have difficulty collapsing into stars, or whether the gas in the CMZ inefficiently produces compact, dense cores. Work from \citet{wal18} and \citet{barnes2019} demonstrate that cores in the CMZ seem to be similar to cores in the disk, based on their masses, radii, and virial parameters. \citet{lu19b, lu19a} investigate the dense gas substructure and star formation efficiency of cores in a handful of clouds in the inner CMZ and find that $<10$\% (as low as 1\% for most) of the total gas in the clouds is substructured into bound, dense cores. In this section, we evaluate the insights from our unbiased large survey of the CMZ in relation to the CMZ's inefficiency at forming stars.

In this section, we show that the \textit{CMZoom} survey reveals that high column density gas in the CMZ is largely devoid of compact substructures on 0.1 - 2 pc scales. Figures \ref{fig:mosaic} and \ref{fig:3color_wow} demonstrate this qualitatively over the entire survey and in selected clouds, respectively, and in this section, we examine the level of compact substructures quantitatively. The \textit{CMZoom} survey region selected the highest column density gas in the CMZ, yet much of this gas appears to lack substantial compact substructure. This low-level of substructure is in contrast to what is expected based on observations of the Galactic disk at comparably high column densities \citep[e.g.][]{zha15, bat14b, gin17, Lu2018}. See Section \ref{sec:uncertainties} for a discussion of the uncertainties in this analysis, and the reasoning and conclusion of separating SgrC into a `dense' and `diffuse' component for this analysis. 

In order to quantify the lack of compact substructure in the CMZ, we define the term Compact Dense Gas Fraction (CDGF). The CDGF is a measurement of the fraction of a cloud that is contained in compact substructures consisting of dense gas. Using these SMA interferometer data, we are not sensitive to just one definition of dense gas, but rather, to gas that it is over-dense compared with its background and sub-structured on 0.1 - 2 pc scales. Typical number densities of \textit{CMZoom} compact substructures from the catalog are about 10$^{4-7}$ cm$^{-3}$ (Hatchfield et al., in prep.). The term CDGF differs from the term `dense gas fraction' in the literature in that a low CDGF does not mean a low fraction of the gas is at high density, but rather, that a small fraction of the total gas is clumped up into compact substructures. As will be discussed in the proceeding sections, the \textit{CMZoom} survey measured very low CDGFs toward most clouds in the CMZ, despite these clouds containing very high fractions of dense gas.

\subsection{Methods to Measure the Compact Dense Gas Fraction (CDGF)}
\label{sec:dgf_methods}

We quantitatively measure the compact dense gas fraction (CDGF) in the \textit{CMZoom} survey in two ways. For both methods we use the SMA-only version of our maps, not the feathered version combined with single dish data, since we are interested in measuring the compact sub-structure. The CDGFs presented here are more sensitive to quantifying the level of compact substructure within a region rather than to large-scale changes in the density structure.
We caution that the terms `dense gas fraction' or `compact dense gas fraction' are not generally well-defined in the literature and the derived numbers are highly dependent on the exact measurement being made and the assumptions that go into it. This is especially true of different datasets and tracers and for observations vs. simulations, but there can even be substantial variation in how this term is applied using the same datasets. For this reason, we implement two different measures of the CDGF, outlined below, and emphasize the importance of a careful apples-to-apples comparison with other datasets or simulations. 

\begin{figure*}[!ht]
\includegraphics[width=1\textwidth]{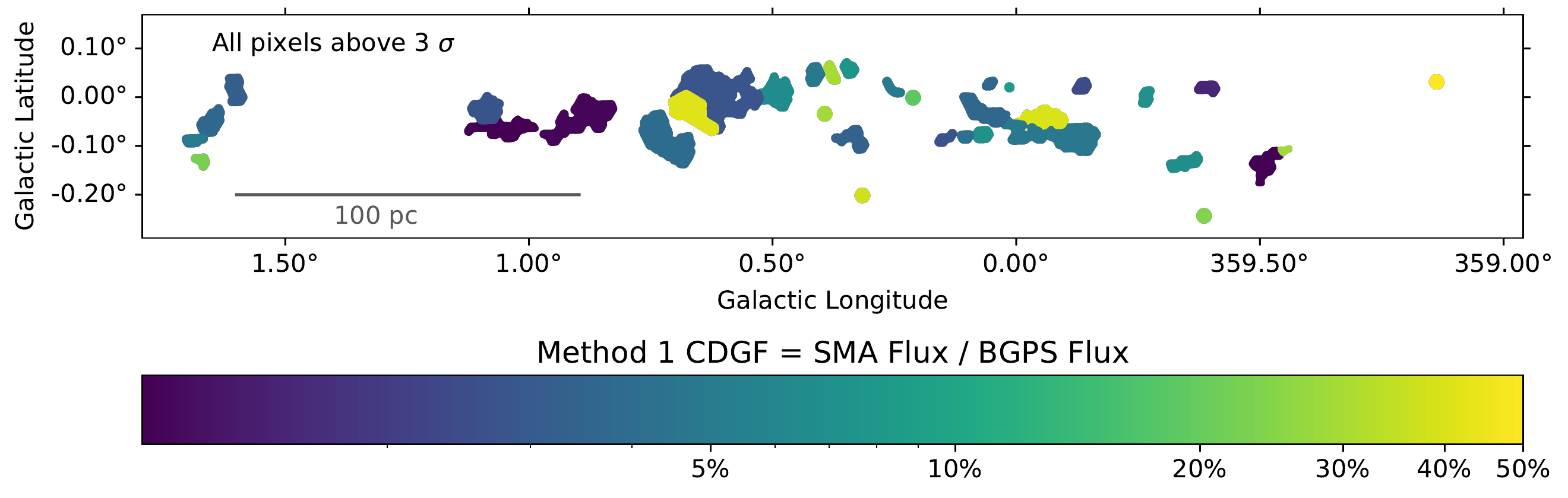}
\singlespace\caption{The method 1 compact dense gas fraction (CDGF), as measured by comparing the SMA-only (interferometer) and BGPS (single-dish) total flux in each cloud. This map shows the total SMA flux divided by the total BGPS flux in each cloud above 3 $\sigma$. This figure highlights that the vast majority of gas in the CMZ (as measured by BGPS dust emission) is \textbf{not} in compact, dense cores of a few tens of parsecs in size (as measured by the SMA dust emission). Most clouds in the CMZ have CDGFs below 10\%. Note that the apparently high CDGF seen toward the CND, located near $\ell$ = 0\deg, is likely due to contamination from the strong synchrotron source Sgr A* and resulting imaging artifacts (see Section \ref{sec:dgf_results} for more details). The CND is removed from the remaining CDGF plots.}
\label{fig:dgf_sma_bgps}
\end{figure*}

\begin{figure*}
\includegraphics[width=0.9\textwidth]{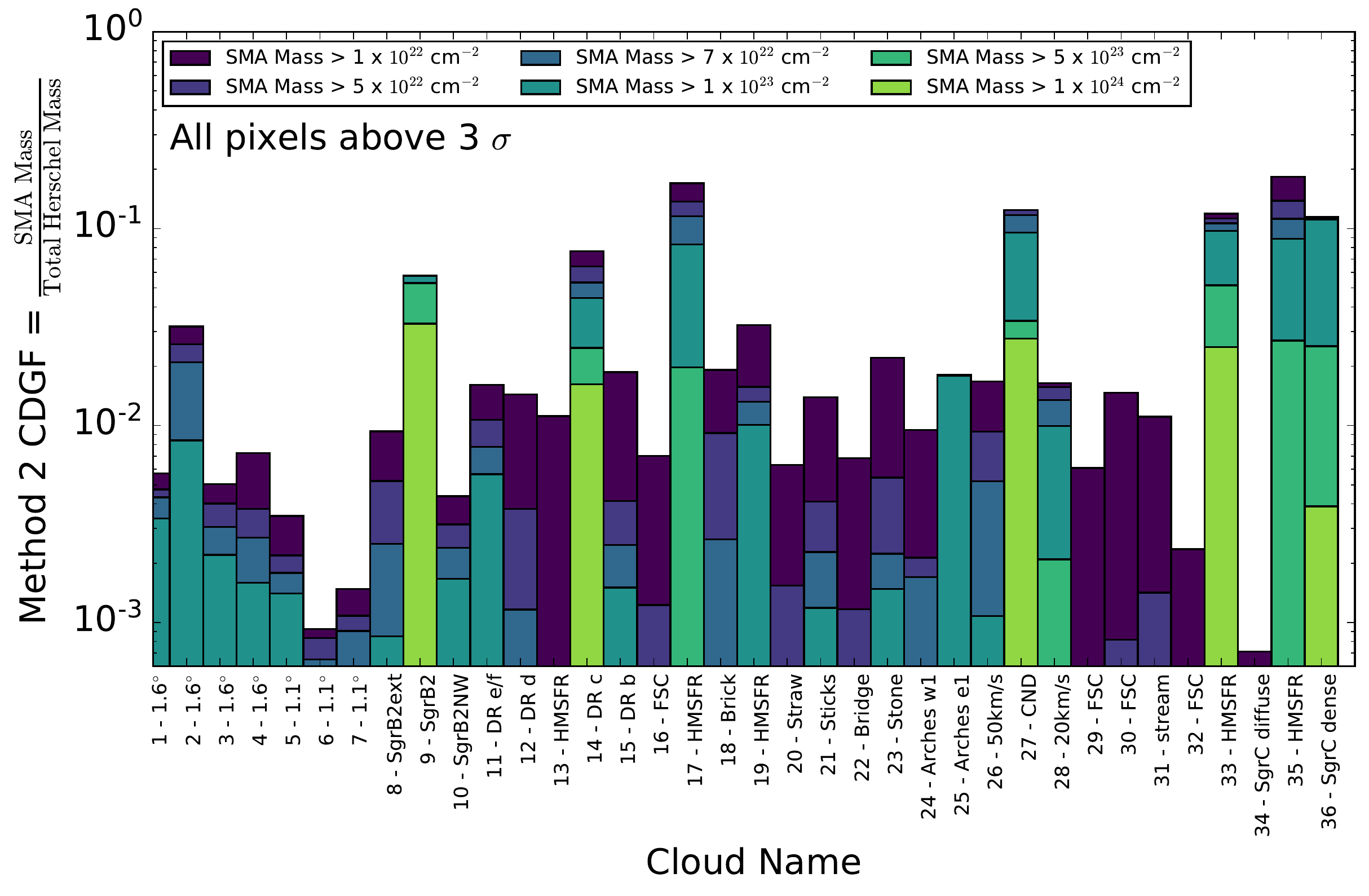}
\singlespace\caption{The method 2 compact dense gas fraction (CDGF), as measured in a comparison between the SMA and Herschel mass in each cloud. The names on the x-axis consist of a prefix, which is the mask number, and a shorthand name for each cloud, of which more details are provided in Table \ref{table-sources}. They are ordered in decreasing central Galactic longitude from left to right. Each bar shows the mass above 3 $\sigma$ in the corresponding \textit{CMZoom} SMA-mapped cloud, above various column density thresholds as shown in the legend, divided by the total mass of the cloud (at all column densities) as seen in Herschel. 
As also seen in Figure \ref{fig:dgf_sma_bgps}, the overall CDGF is less than 10\% for most of the clouds, but this figure demonstrates the varying levels of substructure in each cloud. Note that the apparently high CDGF seen toward the CND located near $\ell$ = 0\deg, is likely due to contamination from the strong synchrotron source Sgr A* and resulting imaging artifacts (see Section \ref{sec:dgf_results} for more details). The CND is removed from the remaining CDGF plots.}
\label{fig:dgf_sma_herschel}
\end{figure*}

Firstly, we compare the overall flux measured with the SMA interferometer in each region with the total flux measured with the BGPS single-dish data in Figure \ref{fig:dgf_sma_bgps}, hereafter referred to as `method 1.' The SMA data are masked on a per-region basis, as shown in Table \ref{table-sources} to study the CDGF in each cloud. We first re-project the BGPS data onto the same grid and pixel scale as the \textit{CMZoom} data using the {\sc reproject exact} function in {\sc astropy.reproject}. Next, we correct for the difference in central frequencies between the SMA dust continuum data (224.9 GHz, this work) and the BGPS dust continuum data \cite[271.1 GHz][]{agu11}, assuming a spectral index $\beta$ of 1.75 \citep[as found in ][]{bat11}, by multiplying the BGPS data by a factor of 0.5, as we did in Section \ref{sec:SD} for single dish combination. Next we convert the BGPS data from Jy beam$^{-1}$ to Jy pixel$^{-1}$ and the SMA data from Jy/sr to Jy pixel$^{-1}$, for a direct pixel-to-pixel comparison. We then sum the total SMA flux in each cloud and divide by the total BGPS flux in each cloud to derive the CDGF, as shown in Figures \ref{fig:dgf_sma_bgps} and \ref{fig:dgf_diskcomp} and Table \ref{table-DGF}. However, we only sum pixels above the 3 $\sigma$ threshold for both the SMA and BGPS data to significantly reduce noise contamination. The SMA noise is determined per pixel as described in Section \ref{sec:rms} and the BGPS noise is taken to be about 50 mJy beam$^{-1}$ in the Galactic Center, based on Figure 1 from \citet{gin13}. 

The second method for quantifying the CDGF, referred to as `method 2' for the remainder of the text, is to compare the mass of each cloud as recovered by the SMA (sensitive to the compact substructures) with the mass recovered by Herschel (sensitive to the overall mass of the cloud), using the same the Herschel column density map as the source selection \citep[][and Battersby et al., in prep.]{mol10, bat11}. We similarly reproject the Herschel column density map to the same grid and pixel scale as the \textit{CMZoom} data using {\sc reproject exact}. We then convert the Herschel column density map to a mass per pixel by assuming a Galactic Center distance of 8.15 kpc and a mean molecular weight of the gas of $\mu = 2.8$ from \citet{kau08}. The SMA flux is converted into a mass per pixel assuming the same distance and mean molecular weight, and by assuming the flux is from a single-temperature modified blackbody, where the temperature is pulled from the Herschel modified blackbody fits (the same ones that produced the Herschel column density used) at the relevant pixel. The modified blackbody calculation also assumes a gas to dust ratio of 100 and an opacity of 0.97 cm$^2$/g \citep[using the power-law interpolation from][as done in Battersby et al. 2011]{oss94}. All of the modified blackbody assumptions, including temperature, opacity, distance, and gas to dust ratio, are identical between the SMA and Herschel data (scaled to the appropriate frequencies). However, we note that these assumptions may not apply equally to the more extended emission traced by Herschel and the compact structures highlighted in the SMA data. We then sum up the mass in each cloud with the SMA (above 3 $\sigma$) and divide it by the mass in Herschel and report it as the CDGF in Figures \ref{fig:dgf_sma_herschel} and \ref{fig:dgf_diskcomp} and Table \ref{table-DGF}. We take the sum of SMA mass above a variety of column density thresholds, as shown in Figure \ref{fig:dgf_sma_herschel}, to show not only the overall CDGF, but the varying levels of substructure in each cloud. We consider Herschel emission below N(H$_2$) = 10$^{22}$ cm$^{-2}$ to be within the noise of the foreground/background (see Section \ref{sec:uncertainties}) and do not include it in the total mass calculation.

\subsection{Potential Foreground Sources}
\label{sec:foreground}

In our analysis of the CDGF of \textit{CMZoom} regions, we identified a number of potential foreground sources. The main culprits are fields that were selected as isolated regions of high-mass star formation, noted as `isolated HMSF candidate' in Table \ref{table-sources} and abbreviated as `HMSFR' in Figure \ref{fig:dgf_sma_herschel} and Table \ref{table-DGF}. The clouds in this category are G0.393$-$0.034, G0.316$-$0.201, G0.212$-$0.001, G359.615$-$0.243, and G359.137$+$0.031. We summarize the available information about whether or not these clouds are foreground below, but first describe the selection of these clouds and reason for their inclusion in the survey. To the list of potential foreground sources we add G1.670$-$0.130 due to its small and relatively isolated nature.

To select candidate high-mass protostars for the \textit{CMZoom} survey we used the methanol multibeam maser catalog of the Galactic Center region \citep{cas10} which searched for 6.7 GHz Class II methanol masers, known to be unambiguous signposts of high-mass star formation \citep{men91, urq15}. All maser sources were associated with a dense compact clump detected in submillimeter dust emission made with Herschel Hi-Gal \citep{mol10}. Those maser sources located outside the Sgr B2 star-forming region, however, were mostly isolated from or were located at the edge of large dense, molecular clouds. Since parallax-based distance measurements from the BeSSeL survey were unavailable at the time of source selection and due to the general complexity of the velocity structure that renders kinematic distances unreliable in the Galactic Center, we assumed that these sources could be located within the CMZ. We included them in the \textit{CMZoom} survey in order to understand whether these sources show physical and chemical properties distinct from that of other CMZ star-forming regions. However, since the completion of \textit{CMZoom} observations, some of these sources have been identified as being in the foreground of the CMZ. The remainder we still consider could still potentially be foreground sources due to their isolated and compact nature and relatively low column densities.

\begin{figure*}
\begin{center}
\includegraphics[width=0.6\textwidth]{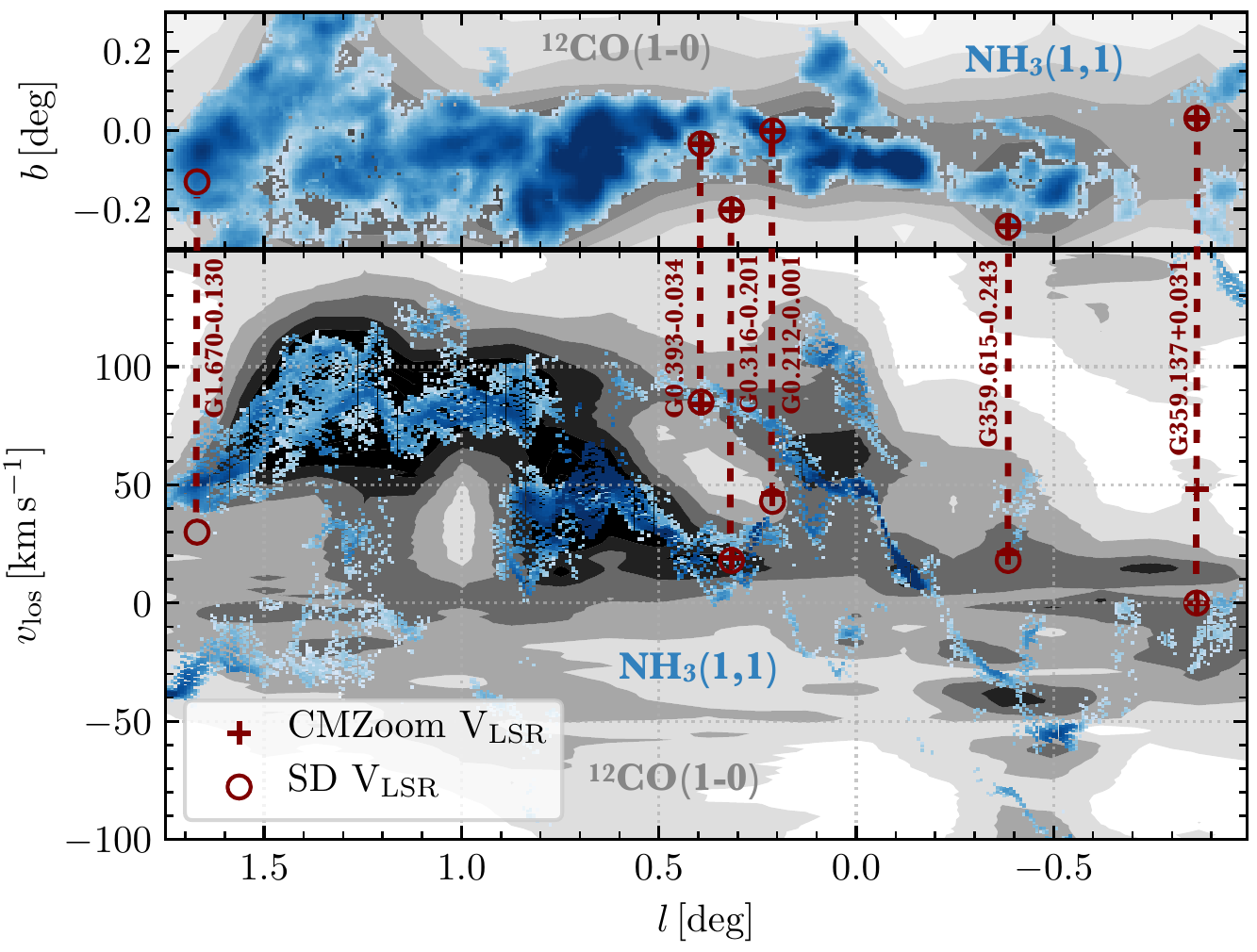}
\caption{The location of the potential foreground sources, G$1.670-0.130$, G$0.393-0.034$, G$0.316-0.201$, G$0.212-0.001$, G$359.615-0.243$, and G$359.137+0.031$, on a longitude-latitude and longitude-velocity plot of the CMZ. The gray data are CO from \citet{dam01} and the color points are the 3-D locations of Gaussian fit centroids to NH$_3$\,(1,1) data from The H$_2$O Southern Galactic Plane Survey (HOPS; \citealp{wal11,pur2012,lon17}), fit using {\sc SCOUSE} from \citet{hen16}. The circles indicate the measured peak single-dish velocity of the region and the plus sign indicates the \textit{CMZoom} preliminary peak velocity of the region as described in detail in Section \ref{sec:foreground} and summarized in Table \ref{table-foreground}. 
}
\label{fig:foreground}
\end{center}
\end{figure*}

The available information on the localization of potential foreground sources is summarized in Table \ref{table-foreground} and Figure \ref{fig:foreground}. The parallax-based distance measurements from the BeSSeL survey \citep{rei19} are of critical importance in the kinematically confused CMZ, and were used to establish that two sources, G$0.316-0.201$ and G$359.615-0.243$, are definitively not in the CMZ. \citet{rei19} suggest that source G$359.137+0.031$ may be in the Galactic Center. However, with a parallax of 0.165 $\pm$ 0.031 mas, its distance to within 2 $\sigma$ is constrained between about 4 and 10 kpc. While this is consistent with being in the Galactic Center, its velocity and linewidth do not clearly indicate a position in the CMZ and we decided that there is not definitive evidence one way or another and have left it as uncertain. The remaining three sources are not included in the \citet{rei19} survey.

In addition to the parallax information from \citet{rei19} we also investigate the velocity structure of each cloud in a relatively high-density gas tracer using single-dish and preliminary \textit{CMZoom} interferometric spectra. For the single-dish data, we use APEX H$_2$CO data from \citet{gin16} and MOPRA HCN data from \citet{jon12}. If available, we also check against the \citet{rei19} reported maser V$_{\rm{LSR}}$ centroid velocity. Based on simple spectral fits using {\sc glue}\footnote{\href{http://glueviz.org/}{http://glueviz.org/}}, we report the approximate V$_{\rm{LSR}}$ and $\sigma_v$ line-width for each source in Table \ref{table-foreground}. We also investigate preliminary \textit{CMZoom} H$_2$CO 3$_{0,3}$--2$_{0,2}$ spectra at 218.222192 GHz for each source (Callanan et al., in prep.). H$_2$CO is clearly detected toward all sources except G$1.670-0.130$ which has an imaging artifact that precludes a deep search at present. We report the centroid velocity and line-width for each source in Table \ref{table-foreground}. Source G$359.137+0.031$ has two components detected in \textit{CMZoom}, one at 0 and one at 48 \kms, both with line widths of about 2 \kms. 

In Figure \ref{fig:foreground} we compare the positions and centroid velocities for the dense gas in each source with an l-b and l-v diagram of gas in the Galactic Center. Due the complex kinematic nature of the CMZ, this investigation does not rule out that these sources could be in the foreground of the CMZ. The line width of the dense gas is another common method to discriminate between foreground and CMZ sources. The relatively small linewidths of the definitive foreground sources G$0.316-0.201$ and G$359.615-0.243$ (4 and 2 \kms~respectively) suggest that this may be a good metric and that therefore perhaps G$0.393-0.034$ and G$0.212-0.001$ (with linewidths of 8 and 7 \kms~respectively) are in the CMZ while G$1.670-0.130$ and G$359.137+0.031$ may be foreground (with a linewidth of 4 \kms~in single dish and 1.2 and 2 \kms~respectively with \textit{CMZoom}). However, we do not consider this evidence to be conclusive and therefore, we simply highlight these regions as potential foreground sources in the remainder of our analysis.

\begin{table*}
\begin{center}
{
\begin{tabular}{|l||c|c|c|c|c|l|l|}
\hline
Region Name & SD V$_{\rm{LSR}}$ & SD $\sigma_v$ & SD reference &CMZoom V$_{\rm{LSR}}$ & CMZoom $\sigma_v$ & Reid et al. (2019) & CMZ? \\
            & [km s$^{-1}$] & [km s$^{-1}$] & & [km s$^{-1}$] & [km s$^{-1}$] &  & (Y/N/?)\\
\hline
G1.670-0.130 & 30 & 4 & 1 & -- & -- & -- & ? \\ 
G0.393-0.034 & 85 & 8 & 1 & 84 & 8  & -- & ? \\
G0.316-0.201 & 18 & 4 & 2,3 & 19 & 1.2 & Not GC & N \\
G0.212-0.001 & 43 & 7 & 1 & 46 & 1.2 & -- & ? \\
G359.615-0.243 & 20 & 2 & 1,3 & 21 & 3 & Not GC & N \\
G359.137+0.031 & -1  & -- & 3 & 0 or 48 & 2 & maybe GC & ? \\
\hline
\end{tabular}}

{\singlespace\caption{This table gives the basic properties of the \textit{CMZoom} regions which were observed as part of the survey, but are now considered potential foreground sources (see Section \ref{sec:foreground} for details). The SD V$_{\rm{LSR}}$ and $\sigma_{v}$ refer to single dish measurements of the approximate central velocity and line width (within about 10\%) as measured from either APEX H$_2$CO data \citep[marked by a 1 in the reference column]{gin16} or from MOPRA HCN data \citep[marked by a 2 in the reference column]{jon12}. We also include the \citet{rei19} determinations of V$_{\rm{LSR}}$ for comparison (marked by a 3 in the reference column), however, that work does not give the line width. The \textit{CMZoom} V$_{\rm{LSR}}$ and $\sigma_v$ are the central velocity and line width as measured with preliminary \textit{CMZoom} H$_2$CO maps  (within about 10\%). Source G1.670-0.130 is not detected in H$_2$CO in our preliminary data. The \citet{rei19} column gives the determination from that work of whether or not that region is likely in the Galactic Center. The final column is our determination of whether or not each region is in the CMZ. }}
\end{center}
\label{table-foreground}
\end{table*}

\subsection{Non-CMZ data points for comparison}
\label{sec:dgf_noncmz}

In order to understand the measured CDGFs of the CMZ in the context of star formation in the Galaxy as a whole, we sought comparison points for our Galaxy's disk outside of the CMZ. For each of the available datasets, we worked with the data to make as fair a comparison as possible. However, as discussed more in Section \ref{sec:uncertainties}, there are a number of systematic effects that make inter-comparison between different datasets difficult. \citet{bat10} performed a systematic analysis of errors in dust-continuum-derived masses using similar datasets of clouds in the disk and found a typical systematic error of about a factor of two. While the uncertainties in temperature, opacity and gas-to-dust ratio may be larger for gas in the CMZ, the distances are better constrained for CMZ clouds. We therefore estimate that these datasets are uncertain by at least a factor of two in both column density and CDGF, and therefore cannot be inter-compared with a greater confidence than that. 

\begin{figure*}
\begin{center}
\subfigure{
\includegraphics[width=0.8\textwidth]{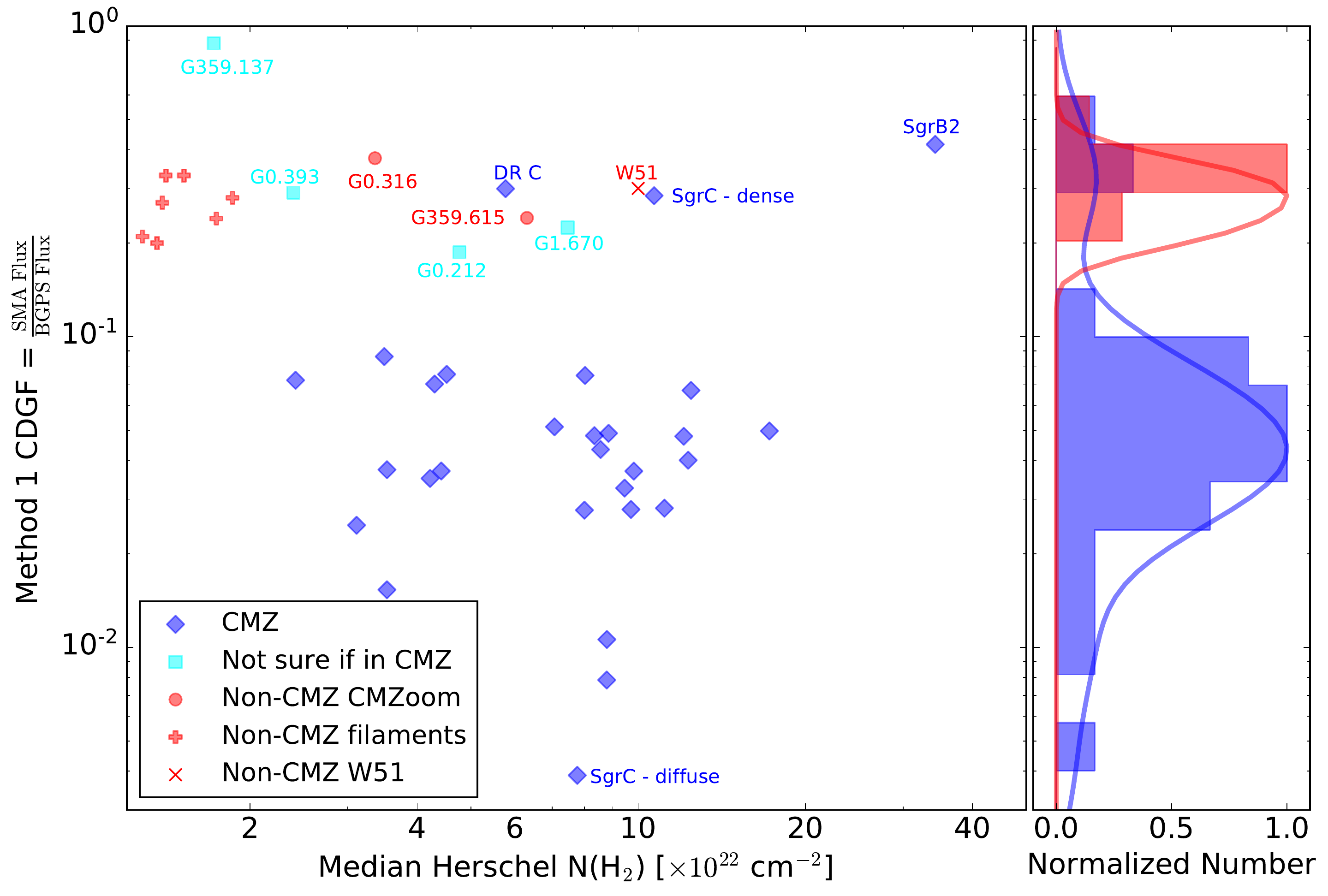}}
\subfigure{
\includegraphics[width=0.8\textwidth]{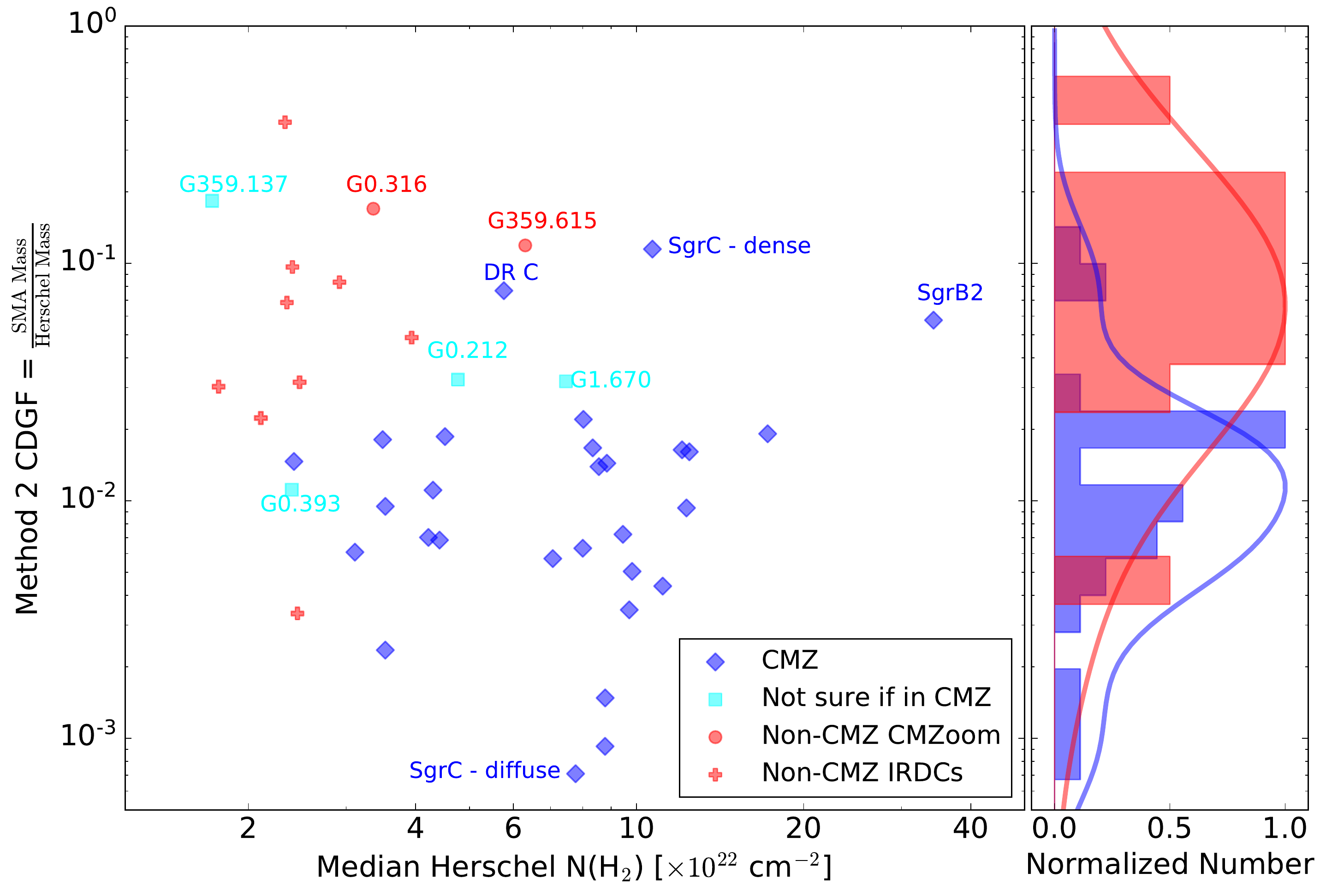}}
\singlespace\caption{Clouds in the CMZ overall have a much lower compact dense gas fraction (CDGF), with the exception of highly star-forming Dust Ridge Cloud C, SgrC - dense, and SgrB2, than clouds in the Galactic disk, despite their high average column densities. This figure shows the method 1 CDGF (SMA flux divided by BGPS flux) vs. the median Herschel column density for each cloud on the \textit{top} with the method 2 CDGF (SMA mass divided by Herschel mass) in the \textit{bottom} panel. The right panel of each plot shows the histogram of CMZ (blue) and non-CMZ (red) data points with a Kernel Density Estimator fit. Clouds marked with a dark blue diamond are located within the CMZ, those marked with cyan squares are not definitively inside or outside of the CMZ (see Section \ref{sec:foreground}), and those marked as red circles, plus signs, or an x are not in the CMZ. The Non-CMZ data points are discussed in \ref{sec:dgf_noncmz}. While we tried to measure the CDGF and median column density for the non-\textit{CMZoom} points in the same way, we advise a factor of two uncertainty in the interpretation of the relative data points (see Section \ref{sec:dgf_noncmz}).} 
\label{fig:dgf_diskcomp}
\end{center}
\end{figure*}

The first dataset we compared with in the Galactic disk is the actively star-forming W51 complex. W51  was observed with ALMA in \citet{gin17} and in BGPS single dish in \citet{gin12}. The ALMA continuum data had a spatial resolution of about 0.005 pc at a central frequency of about 226.6 GHz. The BGPS data were similarly scaled for frequency as in this work. \citet{gin17} report an ALMA to BGPS single-dish flux ratio of 30\%, which is equivalent to our method 1 CDGF and the BGPS data in \citet{gin12} give an approximate column density of 10$^{23}$ cm$^{-2}$. We include W51 as a point for comparison in Figure \ref{fig:dgf_diskcomp} and in Table \ref{table-DGF}.

The second dataset in the Galactic disk for comparison was a sample of Infrared Dark Clouds (IRDCs) observed with ALMA by Barnes et al. (in prep.) and compared with Herschel data from Hi-GAL \citep{mol10}. The ALMA campaign was to observe 10 filamentary IRDCs in the Galactic disk at 90 GHz, with about 0.05 pc spatial resolution, compared with the SMA \textit{CMZoom} observations at 230 GHz with about 0.1 pc spatial resolution. Because of the difference in frequency range, we only compared the method 2 version of measuring the CDGF with these data. We worked with the original data to ensure fair inter-comparison with the \textit{CMZoom} data as much as possible. This dataset is included in Figure \ref{fig:dgf_diskcomp} in the method 2 comparison, marked in the legend as ``Non-CMZ IRDCs." The values from Barnes et al. (in prep.) are also reported in Table \ref{table-DGF}.

The third and final dataset in the Galactic disk for comparison was a sample of filamentary high-mass star forming molecular clouds from \citet{Lu2018}. This observing program observed eight filamentary clouds with the SMA at 1.3 mm in both compact and subcompact configurations. This is exactly the same setup as the \textit{CMZoom} survey. Their beam size is about 0.07 pc, compared with a spatial resolution of about 0.1 pc for \textit{CMZoom}. We worked with the original data files to reproduce method 1 that was used for the \textit{CMZoom} data points as closely as possible, including the primary beam correction, assumed beta for MAMBO flux conversion (from 1.2 mm to 1.3 mm), and the noise thresholds. The SMA flux was then compared with single-dish data from MAMBO to derive the values reported in Figure \ref{fig:dgf_diskcomp} (in method 1, the top figure, labeled ``Non-CMZ filaments'') and Table \ref{table-DGF}.

Finally, the best comparison data between the CMZ and the Galactic disk are the foreground sources, G$0.316-0.201$ and G$359.615-0.243$, observed as part of the \textit{CMZoom} survey and processed in the same way. While we expect that the other three datasets have some systematic uncertainty in their comparison with the \textit{CMZoom} data, despite our best efforts to compare in as similar a way as possible, it is encouraging that they generally mimic the trend observed with the \textit{CMZoom} foreground sources which were analyzed and processed in an identical method to the rest of the \textit{CMZoom} survey.

\begin{table*}
\begin{center}
\begin{tabular}{|c|l|l|c|c|c|c|c|c|c|}
\hline
Mask & Cloud Names & Abbrev. & Med. N(H$_2$) & SMA Flux & BGPS Flux & SMA Mass & Herschel Mass & CDGF & CDGF \\
\#  &              &  &  10$^{22}$ cm$^{-2}$ & Jy & Jy & M$_\odot$ & M$_\odot$ & method1 & method2 \\
 \hline \hline
1 & G1.683-0.089 & 1.6$^{\circ}$ & 7.1 & 0.12 & 2.4 & 2.3e+02 & 4e+04 & 0.051 & 0.0057 \\
2 & G1.670-0.130 & 1.6$^{\circ}$ & 7.5 & 0.59 & 2.6 & 1.1e+03 & 3.4e+04 & 0.22 & 0.032 \\
3 & G1.651-0.050 & 1.6$^{\circ}$ & 9.8 & 0.29 & 7.9 & 6.1e+02 & 1.2e+05 & 0.037 & 0.005 \\
4 & G1.602+0.018 & 1.6$^{\circ}$ & 9.5 & 0.4 & 12 & 8e+02 & 1.1e+05 & 0.033 & 0.0072 \\
5 & G1.085-0.027 & 1.1$^{\circ}$ & 9.7 & 0.34 & 12 & 5.7e+02 & 1.6e+05 & 0.028 & 0.0035 \\
6 & G1.038-0.074 & 1.1$^{\circ}$ & 8.8 & 0.1 & 13 & 1.7e+02 & 1.8e+05 & 0.0079 & 0.00093 \\
7 & G0.891-0.048 & 1.1$^{\circ}$ & 8.8 & 0.33 & 31 & 4.9e+02 & 3.3e+05 & 0.011 & 0.0015 \\
8 & G0.714-0.100 & SgrB2ext & 12 & 3.1 & 78 & 4.4e+03 & 4.7e+05 & 0.04 & 0.0093 \\
9 & G0.699-0.028 & SgrB2 & 34 & 1.5e+02 & 3.7e+02 & 6.7e+04 & 1.2e+06 & 0.42 & 0.058 \\
10 & G0.619+0.012 & SgrB2NW & 1.1e+23 & 2.6 & 92 & 3e+03 & 6.9e+05 & 0.028 & 0.0044 \\
11 & G0.489+0.010 & DR e/f & 12 & 2.6 & 38 & 3.9e+03 & 2.4e+05 & 0.067 & 0.016 \\
12 & G0.412+0.052 & DR d & 8.8 & 0.71 & 15 & 1e+03 & 7.1e+04 & 0.049 & 0.014 \\
13 & G0.393-0.034 & HMSFR & 2.4 & 0.1 & 0.35 & 1e+02 & 9.2e+03 & 0.29 & 0.011 \\
14 & G0.380+0.050 & DR c & 5.8 & 2.3 & 7.7 & 2.7e+03 & 3.5e+04 & 0.3 & 0.077 \\
15 & G0.340+0.055 & DR b & 4.5 & 0.42 & 5.5 & 5.1e+02 & 2.7e+04 & 0.076 & 0.019 \\
16 & G0.326-0.085 & FSC & 4.2 & 0.33 & 9.4 & 3.6e+02 & 5.1e+04 & 0.035 & 0.007 \\
17 & G0.316-0.201 & HMSFR & 3.4 & 2 & 5.4 & 2.3e+03 & 1.3e+04 & 0.38 & 0.17 \\
18 & G0.253+0.016 & Brick & 17 & 1.2 & 24 & 1.6e+03 & 8.6e+04 & 0.05 & 0.019 \\
19 & G0.212-0.001 & HMSFR & 4.8 & 0.57 & 3 & 5.9e+02 & 1.8e+04 & 0.19 & 0.032 \\
20 & G0.145-0.086 & Straw & 8 & 0.2 & 7.3 & 2.4e+02 & 3.9e+04 & 0.028 & 0.0063 \\
21 & G0.106-0.082 & Sticks & 8.6 & 0.37 & 8.5 & 4.8e+02 & 3.4e+04 & 0.043 & 0.014 \\
22 & G0.070-0.035 & Bridge & 4.4 & 0.51 & 14 & 5.8e+02 & 8.6e+04 & 0.037 & 0.0068 \\
23 & G0.068-0.075 & Stone & 8 & 0.83 & 11 & 1.1e+03 & 4.8e+04 & 0.075 & 0.022 \\
24 & G0.054+0.027 & Arches w1 & 3.5 & 0.14 & 3.7 & 1.2e+02 & 1.2e+04 & 0.037 & 0.0095 \\
25 & G0.014+0.021 & Arches e1 & 3.5 & 0.12 & 1.3 & 1e+02 & 5.6e+03 & 0.086 & 0.018 \\
26 & G0.001-0.058 & 50km/s & 8.3 & 2 & 42 & 2.2e+03 & 1.3e+05 & 0.048 & 0.017 \\
27 & G359.948-0.052 & CND & 3.8 & 12 & 29 & 8.8e+03 & 7.1e+04 & 0.4 & 0.12 \\
28 & G359.889-0.093 & 20km/s & 12e & 4.1 & 86 & 6e+03 & 3.7e+05 & 0.048 & 0.016 \\
29 & G359.865+0.022 & FSC & 3.1 & 0.1 & 4.2 & 97 & 1.6e+04 & 0.025 & 0.0061 \\
30 & G359.734+0.002 & FSC & 2.4 & 0.18 & 2.5 & 1.8e+02 & 1.2e+04 & 0.072 & 0.015 \\
31 & G359.648-0.133 & stream & 4.3 & 0.4 & 5.7 & 4.8e+02 & 4.3e+04 & 0.07 & 0.011 \\
32 & G359.611+0.018 & FSC & 3.5 & 0.048 & 3.1 & 53 & 2.3e+04 & 0.015 & 0.0024 \\
33 & G359.615-0.243 & HMSFR & 6.3 & 2.5 & 10 & 3.2e+03 & 2.7e+04 & 0.24 & 0.12 \\
34 & G359.484-0.132 & SgrC diffuse & 7.8 & 0.061 & 16 & 73 & 1e+05 & 0.0039 & 0.00071 \\
35 & G359.137+0.031 & HMSFR & 1.7 & 1.2 & 1.4 & 1.3e+03 & 6.9e+03 & 0.88 & 0.18 \\
36 & G359.484-0.132 & SgrC dense & 11 & 2.4 & 8.3 & 2.8e+03 & 2.4e+04 & 0.28 & 0.11 \\
\hline
 & &   & & & & & & & \\
Non- & CMZoom & Sources   & & & & & & & \\
\hline
 & W51 & Ginsburg et al.  & 10  & --  & --  & --      & --      & 0.30 & -- \\
 & Disk filament & Lu et al. & 1.5 & -- & -- & --      & --     & 0.33  & -- \\
 & Disk filament & Lu et al. & 1.3 & -- & -- & --      & --     & 0.21 & -- \\
 & Disk filament & Lu et al. & 1.9 & -- & -- & --      & --     & 0.28 & -- \\
 & Disk filament & Lu et al. & 1.7 & -- & -- & --      & --     & 0.24 & -- \\
 & Disk filament & Lu et al. & 1.4 & -- & -- & --      & --     & 0.20 & -- \\
 & Disk filament & Lu et al. & 1.4 & -- & -- & --      & --     & 0.27 & -- \\
 & Disk filament & Lu et al. & 1.5 & -- & -- & --      & --     & 0.33 & -- \\
 & Disk IRDC & Barnes et al. & 2.5 & -- & -- & --      & --     & -- & 0.03 \\
 & Disk IRDC & Barnes et al. & 2.4 & -- & -- & --      & --     & -- & 0.10 \\
 & Disk IRDC & Barnes et al. & 3.9 & -- & -- & --      & --     & -- & 0.05 \\
 & Disk IRDC & Barnes et al. & 2.9 & -- & -- & --      & --     & -- & 0.08 \\
 & Disk IRDC & Barnes et al. & 2.4 & -- & -- & --      & --     & -- & 0.00 \\
 & Disk IRDC & Barnes et al. & 2.3 & -- & -- & --      & --     & -- & 0.39 \\
 & Disk IRDC & Barnes et al. & 2.1 & -- & -- & --      & --     & -- & 0.02 \\
 & Disk IRDC & Barnes et al. & 2.3 & -- & -- & --      & --     & -- & 0.07 \\
 & Disk IRDC & Barnes et al. & 1.8 & -- & -- & --      & --     & -- & 0.03 \\

\hline
\end{tabular}

{\singlespace\caption{This table gives the compact dense gas fraction (CDGF) for each region, calculated using methods 1 and 2. We remind the reader that the term CDGF is not well-defined in the literature and its measurement will vary widely depending on the exact definition used. See more details in Section \ref{sec:dgf_methods}. We discuss sources of uncertainty in Section \ref{sec:uncertainties} and estimate that the mass measurements are uncertain by about a factor of two. This table also includes the mask \#, cloud name and abbreviation, median Herschel column density, total SMA flux, total BGPS flux, total SMA mass, and total Herschel mass for each region. The Non-CMZoom sources are from \citet{gin17}, Barnes et al. in prep, and \citet{Lu2018} as discussed in Section \ref{sec:dgf_methods}.}}
\end{center}
\label{table-DGF}
\end{table*}

\subsection{Compact Dense Gas Fractions (CDGF)}
\label{sec:dgf_results}
We find that most regions in the \textit{CMZoom} survey have CDGFs of less than 10\%, using both methods at the 3$\sigma$ level. The method 1 (SMA flux divided by BGPS flux) results are shown in Figure \ref{fig:dgf_sma_bgps}, while the method 2 (SMA mass divided by Herschel mass) results are shown in Figure \ref{fig:dgf_sma_herschel}. The quantities for both are tabulated in Table \ref{table-DGF}. Both methods are shown in Figure \ref{fig:dgf_diskcomp} and compared with regions from the Galactic disk as discussed in Section \ref{sec:dgf_noncmz}. 

The method 1 CDGFs are uniformly higher than method 2. This could be due to the fact that some of the BGPS large-scale extended structure is removed by the atmospheric corrections \citep[see][for details]{gin13},  or to other systematic effects. The methods also differ in that method 1 is a simple flux comparison, whereas method 2 attempts to correct for temperature variations by comparing the masses. While there is a systematic difference between the two methods, the overall trends are the same. The majority of clouds in the CMZ have low levels of compact substructure as measured by their CDGFs, with a few notable exceptions discussed in more detail below.  

Figure \ref{fig:dgf_sma_herschel} provides more information on the dense gas substructures detected with the SMA in each region. Each region (a bar on the x-axis) is split up into how much of the recovered SMA mass originates at each of the labelled SMA column densities. Figure \ref{fig:dgf_sma_herschel} shows that two regions can have very similar overall CDGFs, but with gas that is substructured in very different ways. For example, G1.670$-$0.130 (cloud 2) and G0.699$-$0.028 (Sgr B2, cloud 9) have similar overall CDGFs (3\% and 6\%, respectively), yet the highest column density gas in G1.670$-$0.130 is 10$^{23}$ cm$^{-2}$ while much of the gas in Sgr B2 is closer to 10$^{24}$ cm$^{-2}$. 

We compile the CDGFs from both methods and plot them against the median column density of each cloud (as measured by Herschel) in Figure \ref{fig:dgf_diskcomp}. In this figure we highlight a number of sources as potential foreground sources (cyan) as well as several that have been identified as not being in the CMZ (red). All of the sources in Figure \ref{fig:dgf_diskcomp} were observed as part of the \textit{CMZoom} survey and therefore processed and reduced to a CDGF in the same way, except for the W51, IRDC, and filaments comparison points discussed in more detail in Section \ref{sec:dgf_noncmz}. We discuss the potential foreground sources in more detail in Section \ref{sec:foreground}. 

All of the clouds that stand out in Figures \ref{fig:dgf_sma_bgps}, \ref{fig:dgf_sma_herschel}, and \ref{fig:dgf_diskcomp} for their unusually high CDGFs, can be placed into one of three categories: 1) not in the CMZ (W51, G$0.316$, G$359.615$, non-CMZ IRDCs, and non-CMZ filaments), 2) isolated high-mass star-forming regions that are potentially foreground sources (G$359.137$, G$0.393$, G$0.212$, G$1.670$), and 3) the most active sites of star-formation in the CMZ (Dust Ridge cloud C, Sgr B2, and SgrC dense). 

There is one additional high CDGF data point that was removed from Figure \ref{fig:dgf_diskcomp}, the CND. The cloud imaged as part of \textit{CMZoom} and labelled as the CND contains the supermassive black hole at the center of our Galaxy, SgrA*, whose flux at these wavelengths is known to be variable and largely due to synchrotron emission \citep{Zhao03,Marrone06}, which is $\approx$ 3 $\sigma$ time variable at 1mm \citep{Serabyn97}. Since no corrections for time or contaminating synchrotron (as opposed to dust continuum) emission were made as part of the \textit{CMZoom} survey, we do not consider the recovered SMA flux or cloud mass to be reliable and exclude it from the remainder of our discussion about the CDGF and caution interpretations of the flux in this region. Finally, we also caution the reader that the dust continuum data for clouds G0.054$+$0.027 and G0.014$+$0.021 (Arches w1 and e1) are amongst the noisiest in the entire survey and their relatively high CDGFs may not be reliable. G0.014$+$0.021 in particular had only one pointing, whereas other regions had overlapping data from other pointings, so these data are of overall poor quality.

\subsection{Uncertainties and Systematic Effects}
\label{sec:uncertainties}
As discussed already, a major uncertainty in measuring a CDGF is simply the extent to which the use of the term `dense gas fraction' varies throughout the literature. The term has been used to mean many different things in the literature and we emphasize again here that our focus is on the fraction of gas that it is contained in compact substructures (small on the sky) compared with the total amount of gas in the same region that is seen on large scales. We urge care in the interpretation of our reported CDGFs and caution in inter-comparison with other datasets. We release the script used to produce our CDGFs on the \textit{CMZoom} GitHub page$^3$.

Another large uncertainty for consideration in all of the measurements derived from the \textit{CMZoom} dust continuum, including a CDGF, are the dust properties. While we have assumed the same dust temperatures, spectral index, dust to gas ratio, and dust opacity laws for the computation of SMA and Herschel masses, these properties may vary from the small size scales probed with the SMA to the large size scales seen with Herschel. Additionally, the spectral index, dust to gas ratio, and dust opacity laws may vary from cloud to cloud and therefore affect the absolute masses derived in each cloud. 

Another major uncertainty is the assumption that most, if not all, of the continuum emission at 1.3 mm is due to thermal dust continuum. In some parts of the CMZ, free-free or synchrotron emission may be significant at these wavelengths. We estimate the contamination from free-free or synchrotron emission using VLA C-band (5.56~GHz) continuum data that cover a total of 26 clouds between SgrB2ext and SgrC in Table~\ref{table-DGF} \citep{lu19b}. For each cloud, we extrapolate the 5.56~GHz flux to that at 230~GHz with an assumed spectral index, and use the ratio between it and the SMA flux reported in Table~\ref{table-DGF} to estimated the fraction of potential contamination. In the case of synchrotron emission with a spectral index of $-0.7$, the mean contamination ratio is found to be 1.2\%, which is negligible. For optically thin free-free emission with a spectral index of $-0.1$, the mean contamination ratio is 10.7\%, which is smaller than the absolute flux calibration error and therefore does not affect our result either.

Other uncertainties which affect our measurements of mass from the dust continuum, and therefore the method 2 CDGFs, include the absolute flux uncertainty (about 10-20\% and more in very noisy regions) and the distance uncertainty (a few \%). The measured uncertainty in the dust temperature is generally about 10-20\% \citep{bat11}, however this does not account for varying temperatures along the line-of-sight, within a beam, or from the small scales of the SMA to the larger scales probed with Herschel. Herschel column density uncertainties can be about 40-50\% \citep{bat11} and depend upon the subtraction of foreground/background cirrus emission. Typical uncertainties in the dust opacities, spectral index, and dust to gas ratios are not well-understood a priori and the reasonable range for these is relatively large, giving an overall uncertainty to these parameters of about a factor of two. \citet{bat10} performed a systematic analysis of errors in dust-continuum-derived masses using similar datasets and found a typical systematic error of about a factor of two. We adopt this value as our figure of merit for the dust-continuum-derived masses in this work, but acknowledge that it is simply an estimate which is dependent upon a number of `unknown unknowns.' The measurements of the CDGF have all of these uncertainties, but across multiple datasets. However, the CDGF is also computed by directly comparing datasets where the same assumptions have been made, so some of these uncertainties would cancel out. To be cautious, we suggest a factor of two should be considered a minimum for the uncertainty of the CDGF. 

Finally, for this work, we chose one cloud, SgrC which is well-known to be star-forming \citep[e.g.][]{ken13} and we separated out the dense, star-forming region (Sgr C - dense) from the rest of the cloud (Sgr C - diffuse) which was not previously studied, but most of which was still above the 10$^{23}$ cm$^{-2}$ threshold. Previously the SgrC overall data point was of moderate CDGF (method 1: 0.10, method 2: 0.02), we can see that by separating it, the well-known star-forming portion, SgrC - dense, has a very high CDGF (method 1: 0.28, method 2: 0.11), while the rest of the cloud, SgrC - diffuse, has the lowest CDGF in the sample (method 1: 0.0039, method 2: 0.00071). This case study demonstrates that it matters over which area the CDGF is calculated. One thing that is unique about the \textit{CMZoom} survey is the large-area coverage of relatively unknown regions. However, we do not believe that this is responsible for the overall low CDGFs of gas in the CMZ observed. While the Galactic disk data points for the IRDCs and filaments were mostly of smaller areas, every single one had a lower average column density than the SgrC - diffuse data point. Therefore, there is something unique about all clouds in the CMZ, including SgrC - diffuse, in that they have an incredibly small amount of compact substructure (CDGF) given their high column densities.

\subsection{Discussion}
\label{sec:discussion}

\textit{CMZoom} is the first blind survey of all of the highest column density gas in the CMZ able to probe the compact substructure of the gas. Our observations reveal that the majority of this high column density CMZ gas is either partially or completely devoid of compact substructure on 0.1 - 2 pc scales. The inefficiency in forming compact structures as seen in this work with \textit{CMZoom} may be responsible for the lower overall star formation rate of the CMZ compared with expectations \citep[e.g.][]{lon13a, imm12, kru14, bar17, kau17a, kau17b}.  

A number of well-studied clouds in the CMZ, such as Sgr B2 \citep[e.g.][]{gin18}, Sgr C \citep[e.g.][]{ken13}, and the Dust Ridge clouds \citep[e.g.][]{wal18} show higher levels of compact substructure. By contrast, some very large areas of high column density gas in the CMZ, such as G0.891$-$0.048 or G1.038$-$0.074, show almost \textbf{no} substructure on 0.1-2 pc scales. There are several potential explanations in the literature that have been put forward to explain these differences, from a combination of orbital motions and the compressive gravitational potential \citep[e.g.][]{lon13b,kdl15,hen16,kru19,dal19} to inflow of gas onto the CMZ \citep{sor18,sor20}.

Clouds at high Galactic longitude, G$1.602+0.018$ and G$1.651-00.05$, show almost no substructure except for a few small cores. This lack of substructure could be due to high density gas that is relatively smooth without high contrast clumps to be picked up by an interferometer or clouds which are highly extended along the line-of-sight, leading to high column densities, but low volume densities and lack of clumpy substructures. Due to the widespread detections of molecules with high critical densities in the CMZ in general \citep[e.g.][]{jon12, mil13, gin16, lon17}, we interpret the lack of substructure as being explained by the former, though the latter interpretation may be relevant for some individual regions (perhaps the 1.1\deg~clouds G1.038$-$0.074 and G0.891$-$0.048 for example).


The analysis presented in this section shows that the large unbiased \textit{CMZoom} survey makes a number of consequential conclusions. 1) The CDGF in the CMZ is low compared with the Galactic disk. 2) Where the CDGF is high (i.e., similar to typical values in star-forming regions in the Galactic disk), the regions are actively star-forming. Therefore, 3) the surprisingly low star formation efficiency of high column density clouds in the CMZ is likely a direct result of the inability of this gas to fragment into compact substructures. Where these compact substructures do form, star formation seems to proceed as expected. 


The CMZ is not just an anomalous region of interest, it is the closest laboratory for studying the process of star formation in the type of environment (high densities and turbulent energies) common in the early universe \citep{kru13}. The \textit{CMZoom} survey has demonstrated that despite the high overall density of CMZ gas, much of that gas is not substructured on small scales. Where the gas is highly substructured, it is actively forming stars. Therefore, it is likely that low CDGFs are responsible for the paucity of star formation in the CMZ given its amount of dense gas. Figure \ref{fig:dgf_diskcomp} demonstrates that for a given median column density, we see a range of about 2 orders of magnitude in the CDGF in the CMZ. Uncovering the factors responsible for a cloud's CDGF, and the role that other physical mechanisms such as magnetic fields play in contributing to the low overall star formation efficiency \citep{fed16,pil15}, is key for understanding the star formation process in extreme environments across the cosmos. These observations also call into question the idea of any sort of universal density threshold for star formation, which needs to be much higher than elsewhere in the Galactic disk, or a universal star formation efficiency per free fall time, which needs to be much lower in CMZ clouds with a low CDGF \citep[e.g.][]{hei10, kau10, lad10, lad12}.

\section{Conclusion}

We present the 1.3 mm dust continuum data from the large SMA survey of our Galactic Center, \textit{CMZoom}, and the associated data release (on the \textit{CMZoom} Dataverse, \href{https://dataverse.harvard.edu/dataverse/cmzoom}{https://dataverse.harvard.edu/dataverse/cmzoom}), to occur with publication of this work.  \textit{CMZoom} surveyed almost all of the highest column density gas (N(H$_2$) $\ge 10^{23}$ cm$^{-2}$) plus a few additional regions of interest with lower column density, in the inner 5\deg~longitude $\times$ 1\deg~latitude (about 700 $\times$ 150 pc) of the Galactic Center. The survey covered about 350 square arcminutes over 974 mosaic pointings, observed over 61 tracks. The full survey time is estimated to be about 550 hours and took place at the SMA over the course of 4 years (2013-2017).

\textit{CMZoom} observed the high column density gas in the CMZ in both compact and subcompact configurations, achieving a median angular resolution of about 3$\arcsec$~(0.1 pc), with a sensitivity to larger scales of about 45$\arcsec$~(1.8 pc). The RMS noise of the \textit{CMZoom} survey had a good deal of natural variation due to observing complexity, but about 75\% of regions have RMS noise values between 11-18 MJy Sr$^{-1}$, with a median of 13 MJy Sr$^{-1}$. This paper focuses on the 1.3 mm dust continuum observed with \textit{CMZoom} and its associated data release. However, this paper also describes the observation and calibration of vast spectral line data cubes, spanning 8 GHz at minimum to 16 GHz at maximum in the 211 - 238 GHz range over the course of the survey. The properties of the reduced spectral line data cubes and their public release will be announced in a forthcoming publication (Callanan et al., in prep.) and the \textit{CMZoom} source catalog is presented in a companion publication (Hatchfield et al., in prep.). Toward many regions, the \textit{CMZoom} survey reveals rich and complex substructure, including dense cores and filaments. Some of these complexes have been detected previously, but many are newly discovered with \textit{CMZoom}. The fully-reduced data products are released publicly with this publication to maximize the scientific return of this rich dataset.

\textit{CMZoom} is the first blind, high-resolution survey of the high-column density gas in the inner Galaxy at wavelengths suitable for identifying the next generation of high-mass stars. A key result of this work is the overall deficit of compact substructure in clouds on 0.1 - 2 pc scales, measured by low compact dense gas fractions (CDGFs). We quantitatively measure this CDGF in two ways, both of which compare the compact, dense emission recovered by the SMA on 0.1 - 2 pc scales with the emission on larger scales from single dish telescopes. By both methods, the CDGFs are less than about 10\% across nearly all of the CMZ. 

By comparing with clouds in the Galactic disk observed and analyzed in similar ways, we find that the CDGF in the CMZ is substantially lower, despite having significantly higher column densities. The few locations of the CMZ that have comparably high CDGFs (i.e. similar to star-forming regions in the Galactic disk) are well-known sites of active star formation. We therefore conclude that the low star formation efficiency of the high column density clouds in the CMZ is likely a direct result of the inability of the gas to fragment into compact substructures, since where these substructures form, star formation seems to proceed as expected. 

These results have ramifications for our understanding of how star formation proceeds in an extreme environment. In the CMZ, the inability of the gas to form compact substructures is inhibiting the formation of stars. The factors controlling the CDGF, whether environmental or something else, are therefore key to understanding any possible variation in the star formation process across different environments. These results are a critical addition to the growing body of evidence showing that the CMZ rules out a universal density threshold for star formation (which needs to be much higher than elsewhere in the Galactic disk) or a universal star formation efficiency per free-fall time (which needs to be much lower in CMZ clouds with a low CDGF). We expect the continued analysis of the unique \textit{CMZoom} data to provide further insights into the physics of star formation under extreme conditions.

\acknowledgments
We acknowledge and thank the staff of the SMA for their assistance. First and foremost, we thank Ray Blundell for his ongoing support of the \textit{CMZoom} project. We thank Ken Young (Taco), Charlie Qi, and Mark Gurwell for their help working with the SMA data. We thank Glen Petitpas and Ryan Howie for scheduling these observations and Shelbi Holster and the SMA operators for making them happen. We thank Margaret Simonini for unending logistical assistance with this project and associated travel.  
The authors wish to recognize and acknowledge the very significant cultural role and reverence that the summit of Mauna Kea has always had within the indigenous Hawaiian community. We are most fortunate to have had the opportunity to conduct observations from this mountain. We thank Stephanie Santillo for her assistance and expertise composing Tables 1 and 2 as well as Jimmy Casta\~no, Liz Gehret, Mark Graham, Elizabeth Guti\'errez, Dennis Lee, and Irene Vargas-Salazar for their early investigations into these data which helped to guide its progress and improve its quality. This research made use of APLpy, an open-source plotting package for python hosted at \href{http://aplpy.github.com}{http://aplpy.github.com}, astropy, a community-developed core Python package for astronomy \citep{astropy13,astropy18}, glue-viz \citep{robitaille_glue, beaumont_glue}, CASA \citep{casa07}, MIR \href{https://www.cfa.harvard.edu/~cqi/mircook.html}{https://www.cfa.harvard.edu/~cqi/mircook.html}, and Miriad \citep{miriad95}. This research has made use of NASA's Astrophysics Data System. The National Radio Astronomy Observatory is a facility of the National Science Foundation operated under cooperative agreement by Associated Universities, Inc. This paper makes use of the following ALMA data: ADS/JAO.ALMA\#2013.1.00617.S, ALMA\#2017.1.00687.S, and ALMA\#2018.1.00850.S. ALMA is a partnership of ESO (representing its member states), NSF (USA) and NINS (Japan), together with NRC (Canada), MOST and ASIAA (Taiwan), and KASI (Republic of Korea), in cooperation with the Republic of Chile. The Joint ALMA Observatory is operated by ESO, AUI/NRAO and NAOJ.

CB gratefully acknowledges support from the National Science Foundation under Award Nos. 1602583 and 1816715. ATB would like to acknowledge the funding provided from the European Union's Horizon 2020 research and innovation programme (grant agreement No 726384). HPH thanks the LSSTC Data Science Fellowship Program, which is funded by LSSTC, NSF Cybertraining Grant \#1829740, the Brinson Foundation, and the Moore Foundation; his participation in the program has benefited this work. JMDK gratefully acknowledges funding from the Deutsche Forschungsgemeinschaft (DFG, German Research Foundation) through an Emmy Noether Research Group (grant number KR4801/1-1) and the DFG Sachbeihilfe (grant number KR4801/2-1), as well as from the European Research Council (ERC) under the European Union's Horizon 2020 research and innovation programme via the ERC Starting Grant MUSTANG (grant agreement number 714907). XL acknowledges support by JSPS KAKENHI grant No.\ JP18K13589. EACM gratefully acknowledges support by the National Science Foundation under grant No. AST-1813765. LH acknowledges support by the National Science Foundation of China (11721303, 11991052) and the National Key R\&D Program of China (2016YFA0400702).

\facilities{SMA, CSO, ALMA, Herschel} 
\software{Numpy, MIR IDL, CASA, MIRIAD, Astropy, Aplpy, Glue-viz}

\bibliographystyle{yahapj}
\bibliography{CMZoom_references}

\appendix
\section{SMA Configurations}
\label{appendix-configs}

During the course of the \textit{CMZoom} survey, observations were conducted in a number of configurations (see Table \ref{table-obs}). The broad categories of these configurations are `compact' and `subcompact' configurations, however, since the array was re-configured a number of times over the course of the survey, the exact placement of the antennas varied. In Table \ref{table-config} we outline the approximate antenna positions, and dates, for the various version of the `compact' configuration throughout the \textit{CMZoom} survey. While there are variations in these configurations, they are not substantive and we do not expect any effect on the overall data properties, however, for the sake of posterity we record the configurations for each observation in Table \ref{table-obs} and the antenna positions corresponding to those configurations here in Table \ref{table-config}. We note that even this list is not comprehensive as there were slight variations throughout, especially around the time of antenna moves when, for some tracks, not all antennas had been moved to their final positions.

\begin{table*}
\begin{center}
{\footnotesize
\begin{tabular}{cccccc}
\hline
Array Configuration as noted in Table \ref{table-obs} &  Antenna number & Pad number & x [m] & y [m] & z [m]\\
\hline \hline
\textbf{compact-1} & 1 & 	11 & 	-16.56370 &   	-27.02601 &   	+30.76931 \\    
May 25, 2014 to & 2 & 	1  &	+0.00000  &	+0.000000 & 	+0.00000  \\ 
July 10, 2014 & 3 &	9  &     -6.40022 &	-68.00149 &	+3.63770  \\
& 4 & 	12 & 	+11.73945 & 	-53.66415 & 	-40.72509 \\ 
& 5 & 	23 & 	-17.91678 & 	-59.55727 & 	+30.07000 \\ 
& 6 & 	7  &	+5.22813  &	-20.07633 & 	-14.84290 \\ 
& 7 & 	10 & 	-21.50842 & 	-51.01853 & 	+39.95824 \\ 
& 8 & 	8  &	+4.44256  &	+4.44256  &	-21.83625 \\
\hline
\textbf{compact-2} & 1 & 	17 & 	+26.51706 &  	-133.59605 &  	-104.17552 \\ 
April 14, 2015 to & 2 & 	1  &	+0.00000  &	+0.00000   &      +0.00000 \\ 
June 2, 2015 & 3 & 	5  &	-5.70045  &	-18.98608  &	+15.61049  \\ 
& 4 & 	4  &	-0.50844  &	-25.15213  &	+1.28090   \\ 
& 5 & 	8  &	+4.44086  &	-63.87470  &	-21.83592  \\ 
& 6 & 	9  &	-6.40032  &	-68.00154  &	+3.63828   \\ 
& 8 & 	12 & 	+11.74035 & 	-53.66340  &	-40.72437  \\  
\hline
\textbf{compact-3} & 1 & 	9  &	-6.40028  &	-68.00141  &	+3.63791   \\ 
July 10, 2015 to  & 2 & 	1  &	+0.00000  &	+0.00000   &      +0.00000 \\   
July 28, 2015& 3 & 	5  &	-5.70051  &	-18.98613  &	+15.61050  \\ 
& 4 & 	4  &	-0.50847  &	-25.15215  &	+1.28082   \\
& 5 & 	23 & 	-17.91691 & 	-59.55850  &	+30.06983  \\  
& 6 & 	7  &	+5.22837  &	-20.07633  &	-14.84313  \\  
\hline
\textbf{compact-4} & 1 & 	5  &	-5.70169  &	-18.98447  &	+15.61072  \\ 
March 16, 2016 to  & 2 & 	1  &	+0.00000  &	+0.00000   &      +0.00000 \\   
March 29, 2016 & 3 & 	12 & 	+11.74055 & 	-53.66549  &	-40.72453  \\  
& 4 & 	4  &	-0.50836  &	-25.15217  &	+1.28065   \\ 
& 5 & 	11 & 	-16.56428 & 	-27.02544  &	+30.77050  \\ 
& 6 & 	23 & 	-17.91585 & 	-59.55776  &	+30.06966  \\ 
& 7 & 	7  &	+5.22795  &	-20.07641  &	-14.84155  \\ 
& 8 & 	9  &	-6.39960  &	-68.00048  &	+3.63829   \\
\hline
\textbf{compact-5} & 1 & 	8  &	+4.44190  &	-63.87476  &	-21.83623  \\ 
April 30, 2016 to & 2 & 	1  &	+0.00000  &	+0.00000   &      +0.00000 \\  
June 17, 2016 & 3 & 	12 & 	+11.74080 & 	-53.66558  &	-40.72437  \\ 
& 4 & 	23 & 	-17.91590 & 	-59.55770  &	+30.06971  \\ 
& 5 & 	11 & 	-16.56472 & 	-27.02547  &	+30.77028  \\ 
& 6 & 	7  &	+5.22816  &	-20.07642  &	-14.84265  \\ 
& 7 & 	10 & 	-21.50851 & 	-51.01856  &	+39.95844  \\ 
& 8 & 	9  &	-6.39946  &	-68.00051  &	+3.63842   \\
\hline
\textbf{compact-6} & 1 &	8  & 	+4.44110  & 	-63.87358  & 	-21.83693 \\ 
July 15, 2017 to & 2 &	23 & 	-17.91629 & 	-59.55685  & 	+30.06933 \\ 
July 31, 2017 & 3 &	1  & 	+0.00000  & 	+0.00000   &	+0.00000  \\
& 4 &	4  & 	-0.50911  & 	-25.15099  & 	+1.28002  \\
& 5 &	5  & 	-5.70350  & 	-18.98436  & 	+15.61066 \\ 
& 6 &	9  & 	-6.40099  & 	-68.00024  & 	+3.63762  \\
& 7 &	7  & 	+5.22698  & 	-20.07498  & 	-14.84246 \\ 
& 8 &	12 &  	+11.73953 & 	-53.66158  & 	-40.72467 \\ 
\hline
\end{tabular}
}

\singlespace
\singlespace\caption{Over the course of the \textit{CMZoom} survey, the SMA underwent a number of array configuration changes, and in any given `compact' configuration, the antenna setup was not identical. We record here the approximate antenna positions throughout the survey, connected with their relevant observations in Table \ref{table-obs}, but note that there are slight variations around the times of antenna moves. The slightly different versions of compact configuration throughout the survey should not affect the overall data properties.}
\end{center}
\label{table-config}
\end{table*}

\section{Comparison with the CARMA 3mm Survey of the CMZ}
\label{appendix-carma}

\begin{figure*}
\centering
\subfigure{
\includegraphics[trim = 0mm 45mm 0mm 35mm, clip, width=0.9\textwidth]{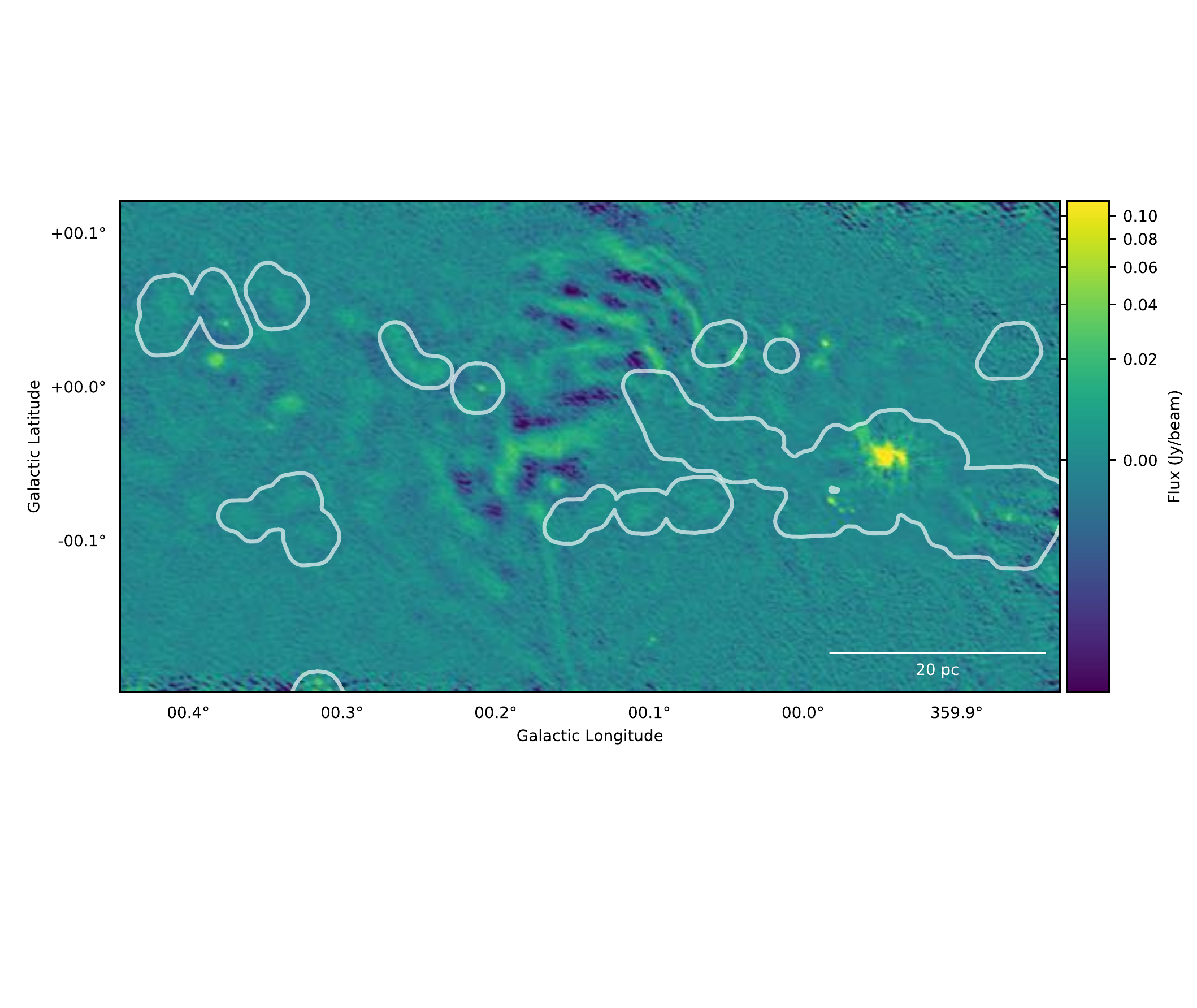}}
\subfigure{
\includegraphics[trim=31mm 22mm 10mm 0mm, clip, width=0.35\textwidth]{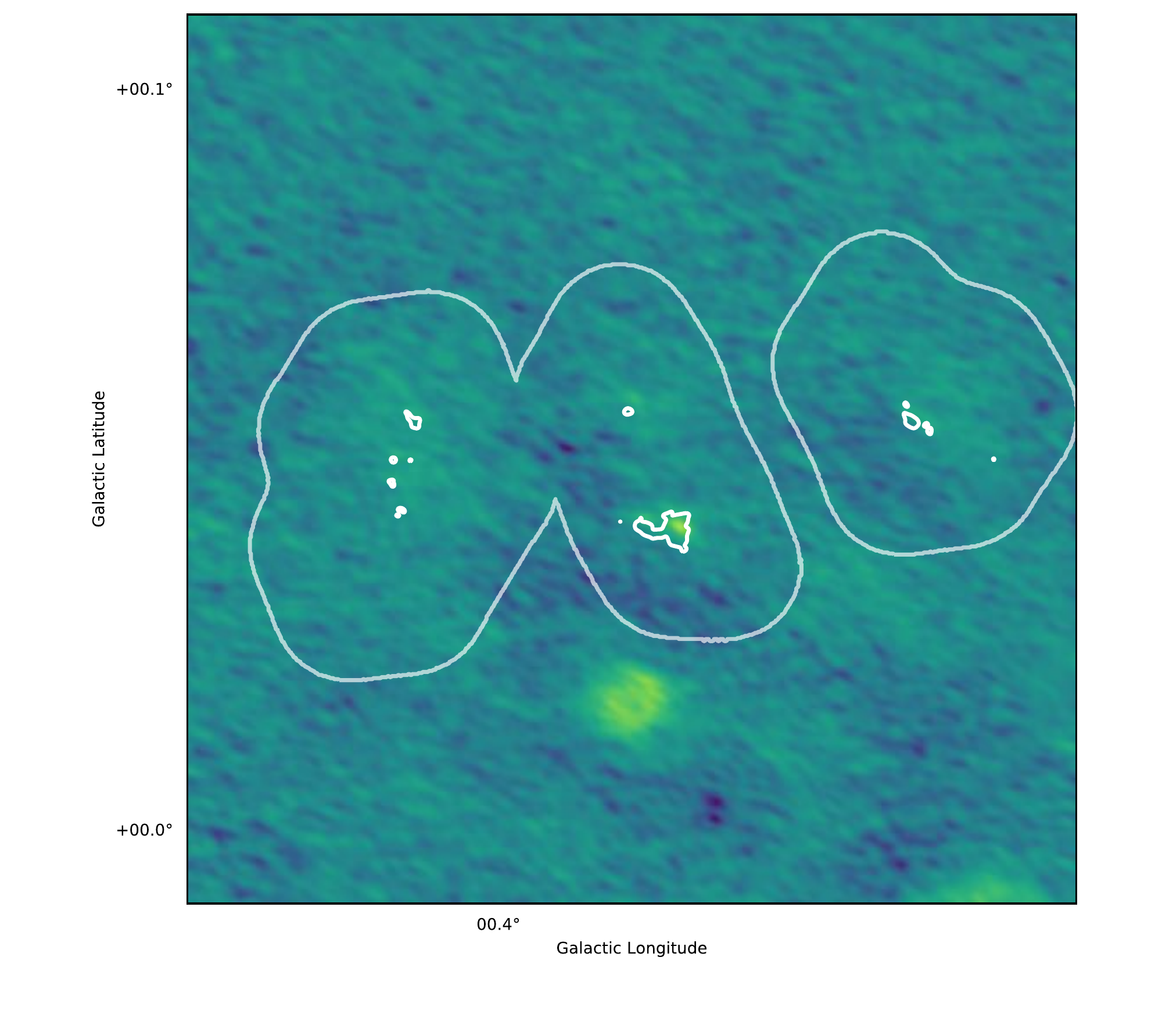}}
\subfigure{
\includegraphics[trim=31mm 22mm 10mm 0mm, clip, width=0.35\textwidth]{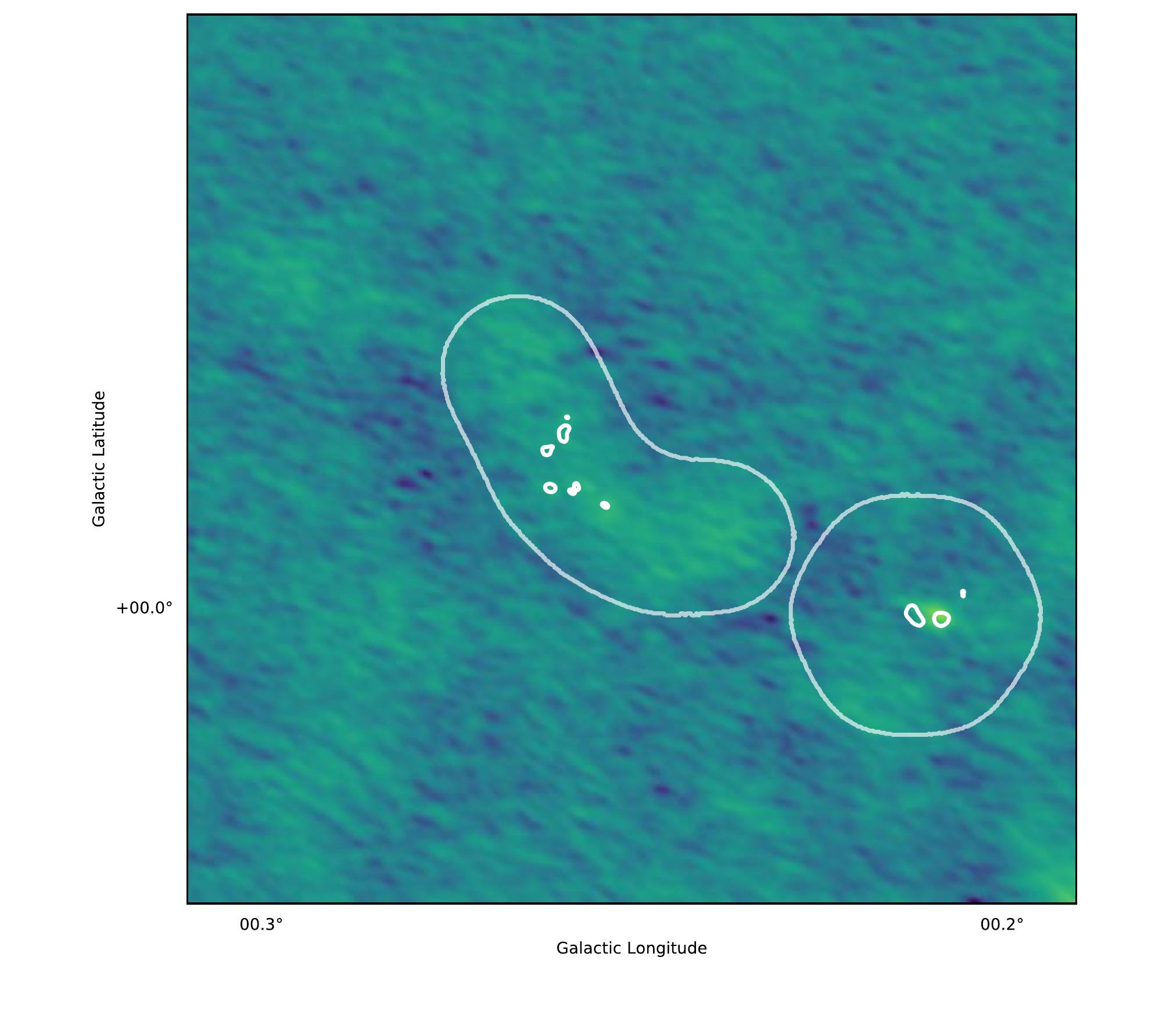}}
\subfigure{
\includegraphics[trim=31mm 22mm 10mm 0mm, clip, width=0.35\textwidth]{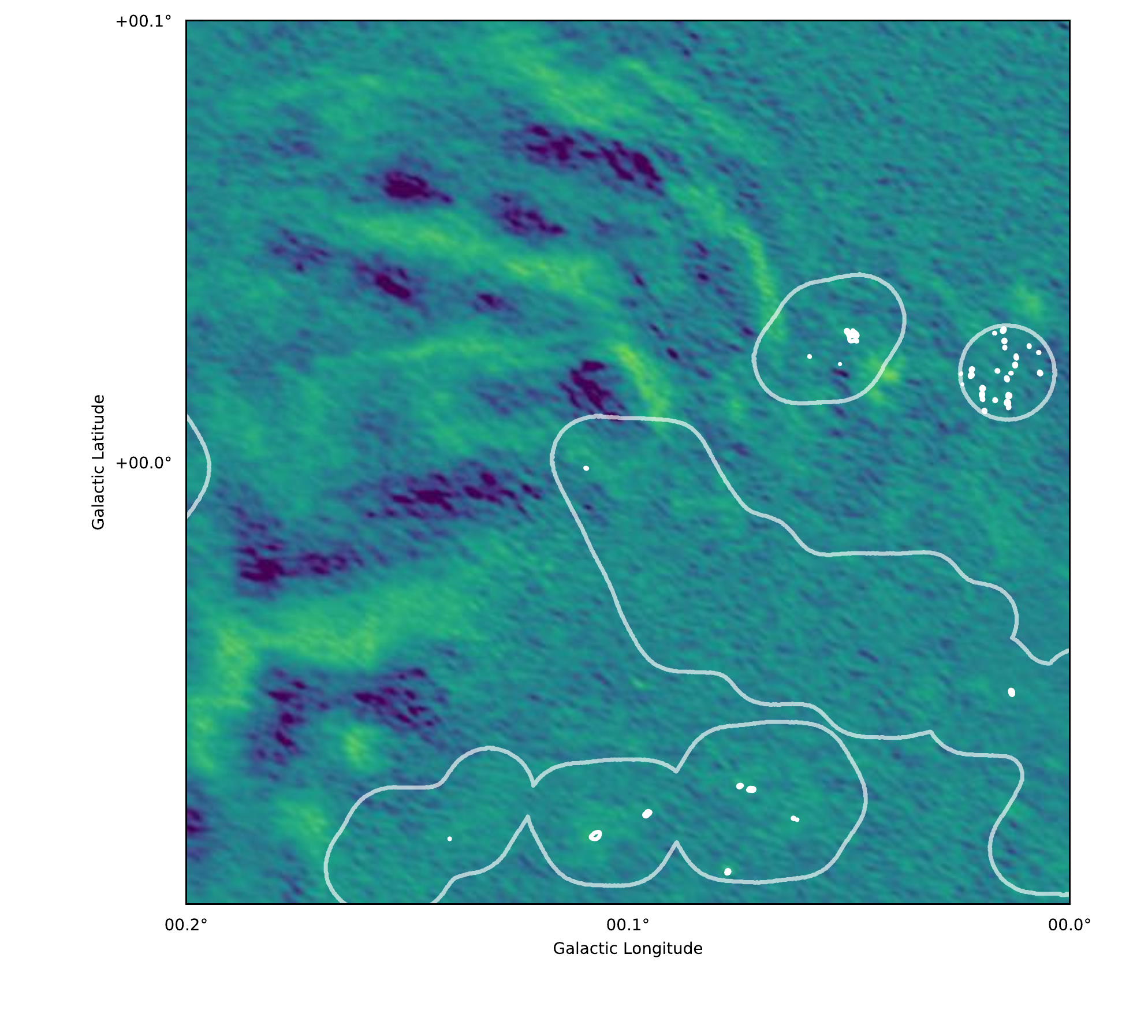}}
\subfigure{
\includegraphics[trim=31mm 22mm 10mm 0mm, clip, width=0.35\textwidth]{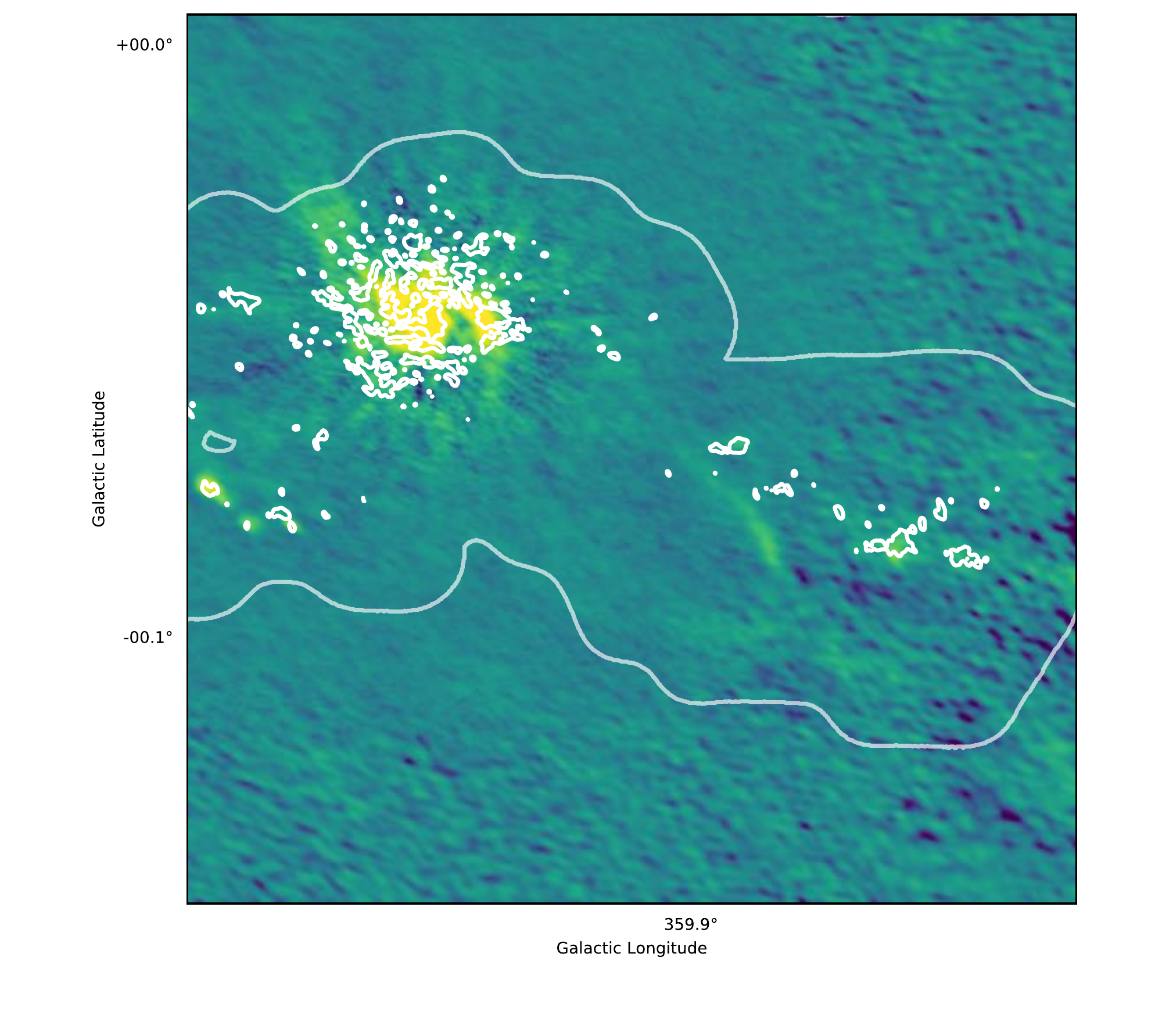}}
\singlespace\caption{Comparison of \textit{CMZoom} data with CARMA 3mm survey from \citet{pou18}. All panels show the CARMA 3mm dust continuum 15+8 \citep{pou18}, which includes CARMA-15 (the 15-element array with six 10.4 m antennas and nine 6.1 m antennas) as well as CARMA-8 (the 8-element array of 3.5 m antennas) in the same colorscale. The \textit{CMZoom} survey outline is shown in white. The bottom four figures show zoom-ins on several regions with 5-$\sigma$ \textit{CMZoom} contours (without primary beam correction) in white. }
\label{fig:carma}
\end{figure*}

The CARMA 3mm Survey of the CMZ observed the inner 0.7\deg $\times$ 0.4\deg~of the Galaxy at 3mm in continuum and a variety of spectral lines \citep{pou18}. We compare the continuum-only datasets in this section and in Figure \ref{fig:carma}, but suggest that further study and comparison of the spectral line data from CARMA (including SiO, HCO$^+$, HCN, N$_2$H$^+$, CS, etc.) with \textit{CMZoom} spectral line data in different tracers would be worthwhile. 

While \textit{CMZoom} covers a larger area of the CMZ, it only does so in select, high-column density regions. By contrast, the CARMA 3mm survey covers a complete region of the CMZ, but only the inner portion. The spatial resolution of the CARMA data is  $\sim$10\arcsec, while \textit{CMZoom} has a resolution of $\sim$3\arcsec~(0.1 pc). Additionally, the wavelength difference between CARMA (3mm) and \textit{CMZoom} (1.3mm) means that the continuum of the \textit{CMZoom} survey is sensitive to primarily cold dust emission, while the CARMA survey continuum is complicated by additional contributions from synchrotron and free-free emission, the latter of which dominates the emission in many regions.

In our brief visual comparison, we see a number of associations between \textit{CMZoom}  and CARMA emission, notably, toward Dust Ridge Cloud C (G0.380$+$0.050), and near the Brick (see Figure \ref{fig:carma}). The CARMA emission in these locations is likely due to some combination of dust continuum and free-free emission. Toward the bridge of emission (G0.070$+$0.035) Northeast from the 50\kms~cloud, we see a tiny amount of overlap in the \textit{CMZoom} coverage with the non-thermal filaments clearly detected with CARMA. There is some hint that there may be some faint \textit{CMZoom} structure following the CARMA 3mm non-thermal filaments. The continuum from the SgrA is clearly detected, with interesting substructure in both datasets. The forthcoming \textit{CMZoom} catalog paper will report CARMA 3mm fluxes within \textit{CMZoom} structures.

\section{Image Gallery}
\label{appendix-images}
We include here images of each region surveyed. Each image is on the same logarithmic color scale and contains a 3 pc scale bar as well as the beam shape and orientation for each region in the lower-left panel of the image. All these images have been primary-beam-corrected, hence the noisier map edges. We note that SgrB2 (G$0.699-0.028$) has locally high noise due to dynamic range issues and that the CND (G$359.948-0.052$) has similarly locally high noise due to dynamic range issues, and also suffers from potential contamination of synchrotron and time variable emission which is not corrected for here, but is discussed in more detail in Section \ref{sec:dgf_results}. Additionally, the two Arches regions, G$0.054+0.027$ and G$0.014+0.021$ suffer from some of the worst RMS noise in the survey due to some observing peculiarities in these very small regions and these images should be interpreted with caution.

\begin{figure*}
\begin{center}
\subfigure{
\includegraphics[width=0.48\textwidth]{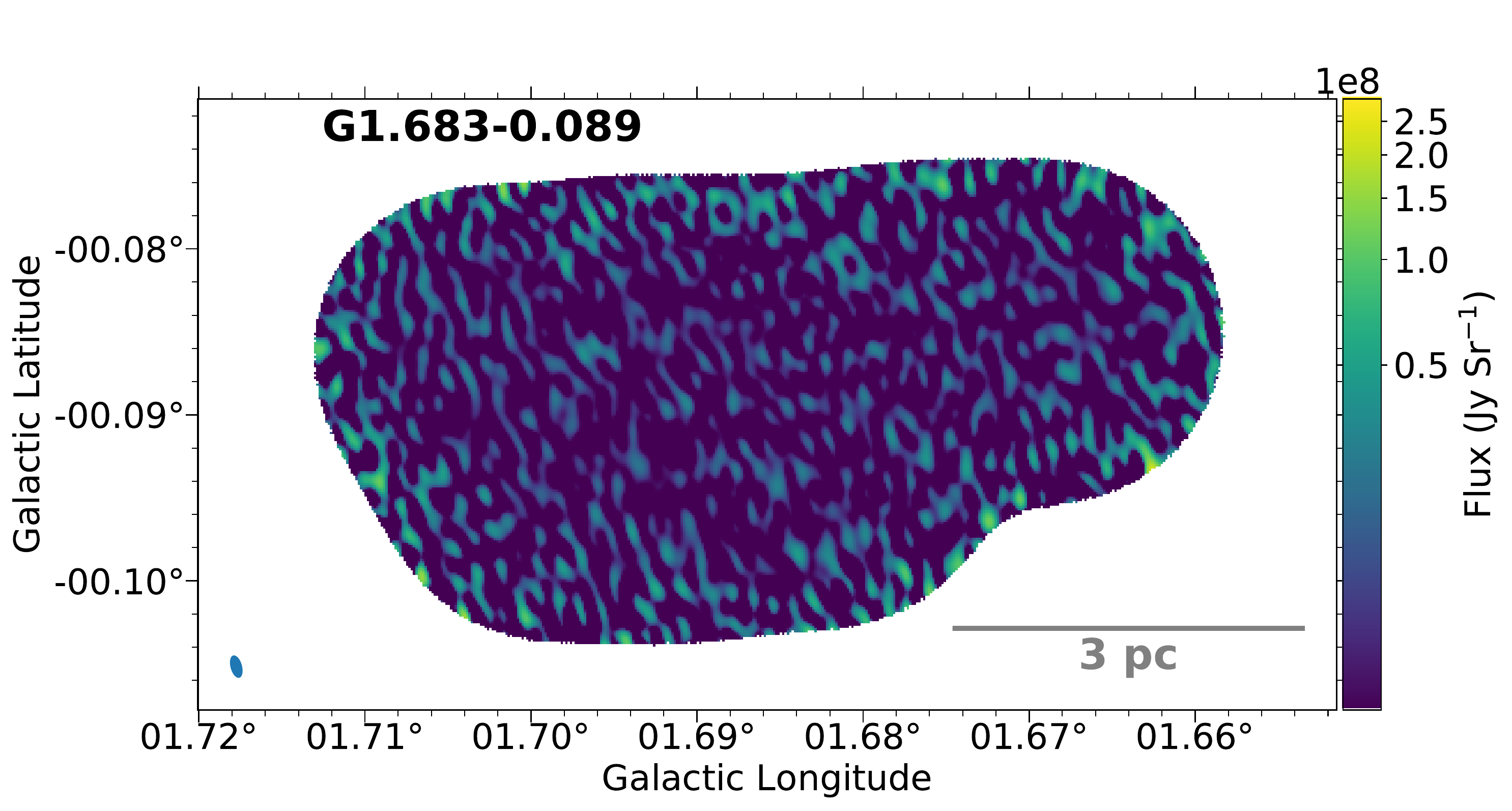}}
\subfigure{
\includegraphics[width=0.48\textwidth]{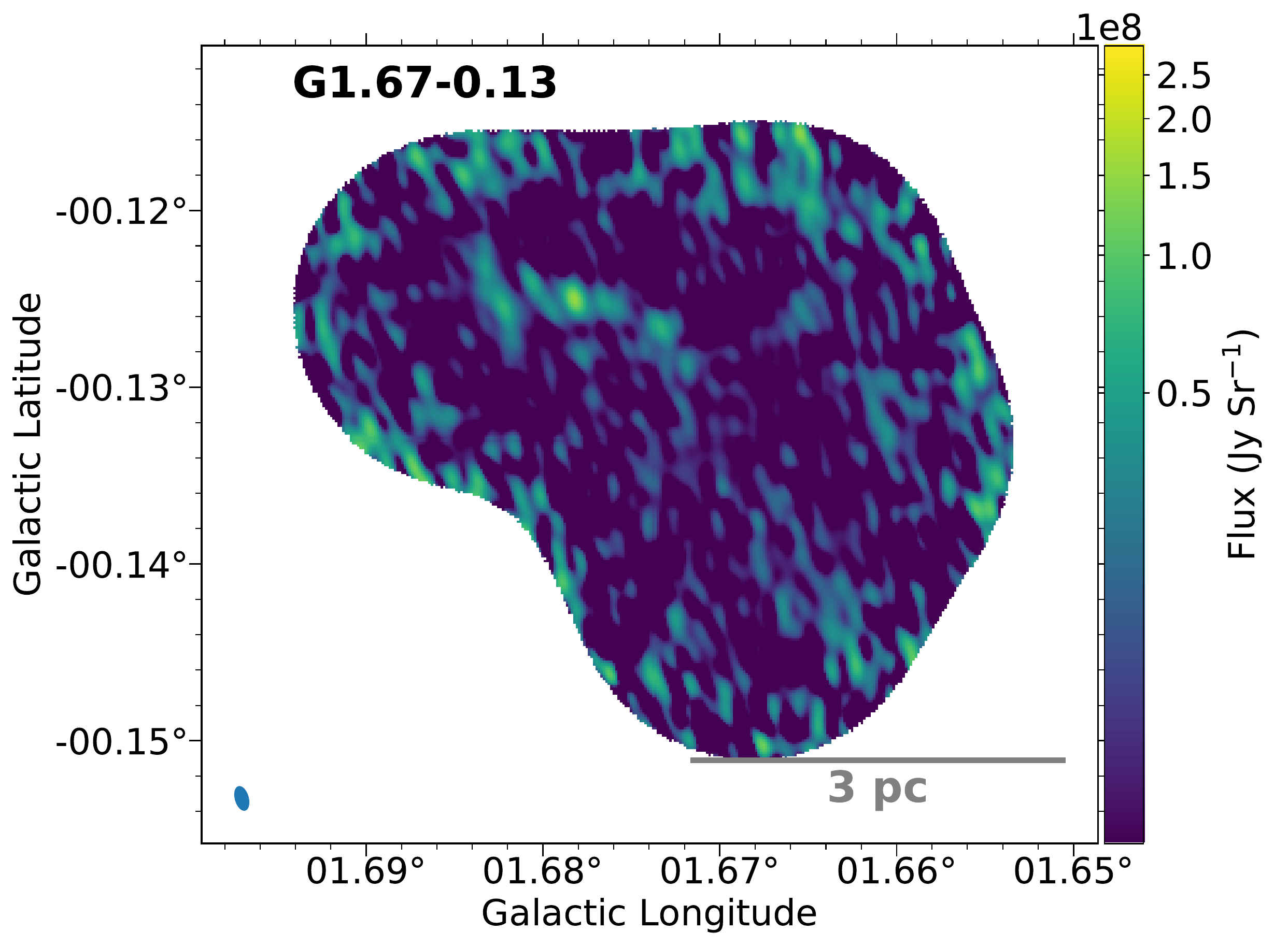}}\\
\subfigure{
\includegraphics[width=0.48\textwidth]{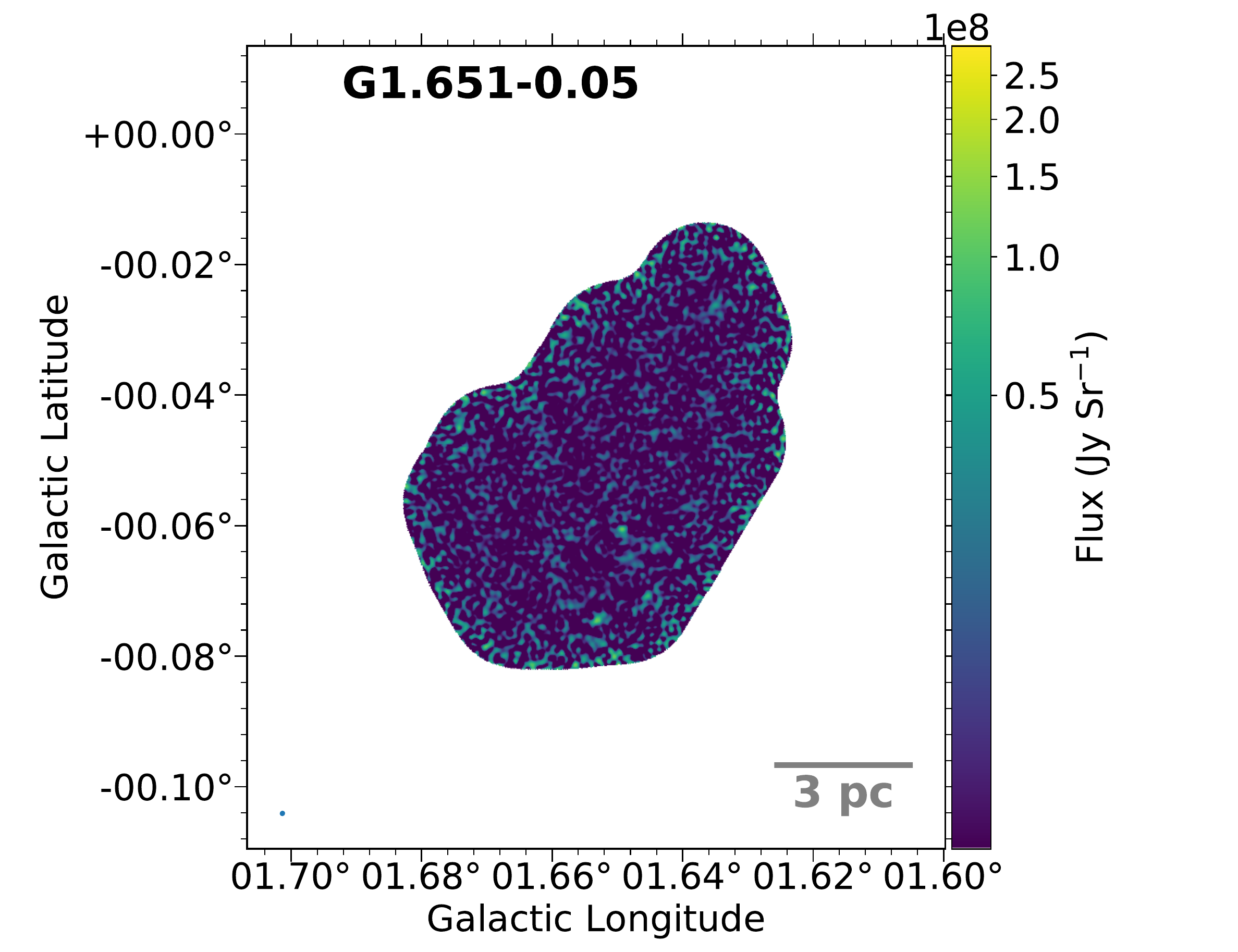}}
\subfigure{
\includegraphics[width=0.48\textwidth]{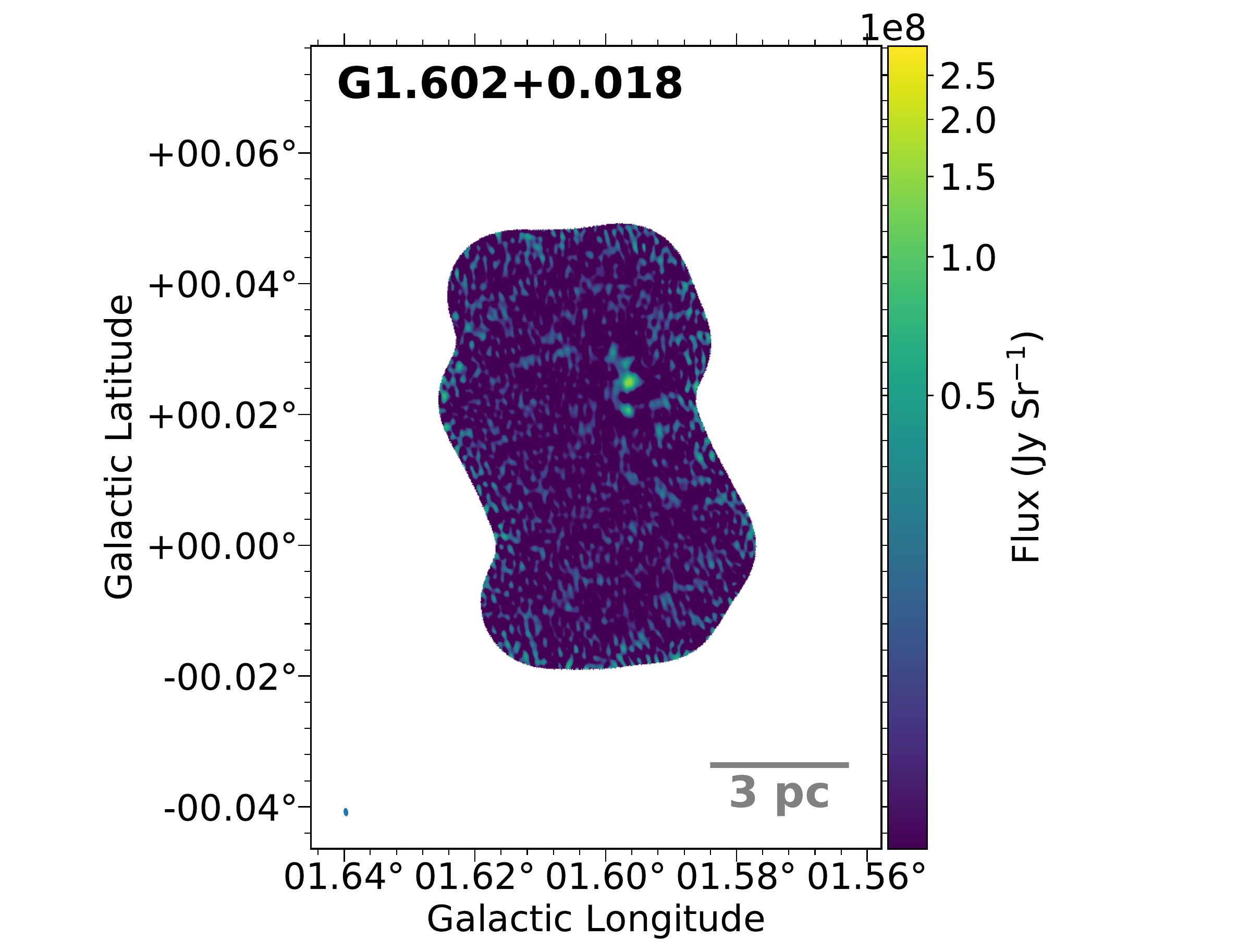}}\\
\subfigure{
\includegraphics[width=0.48\textwidth]{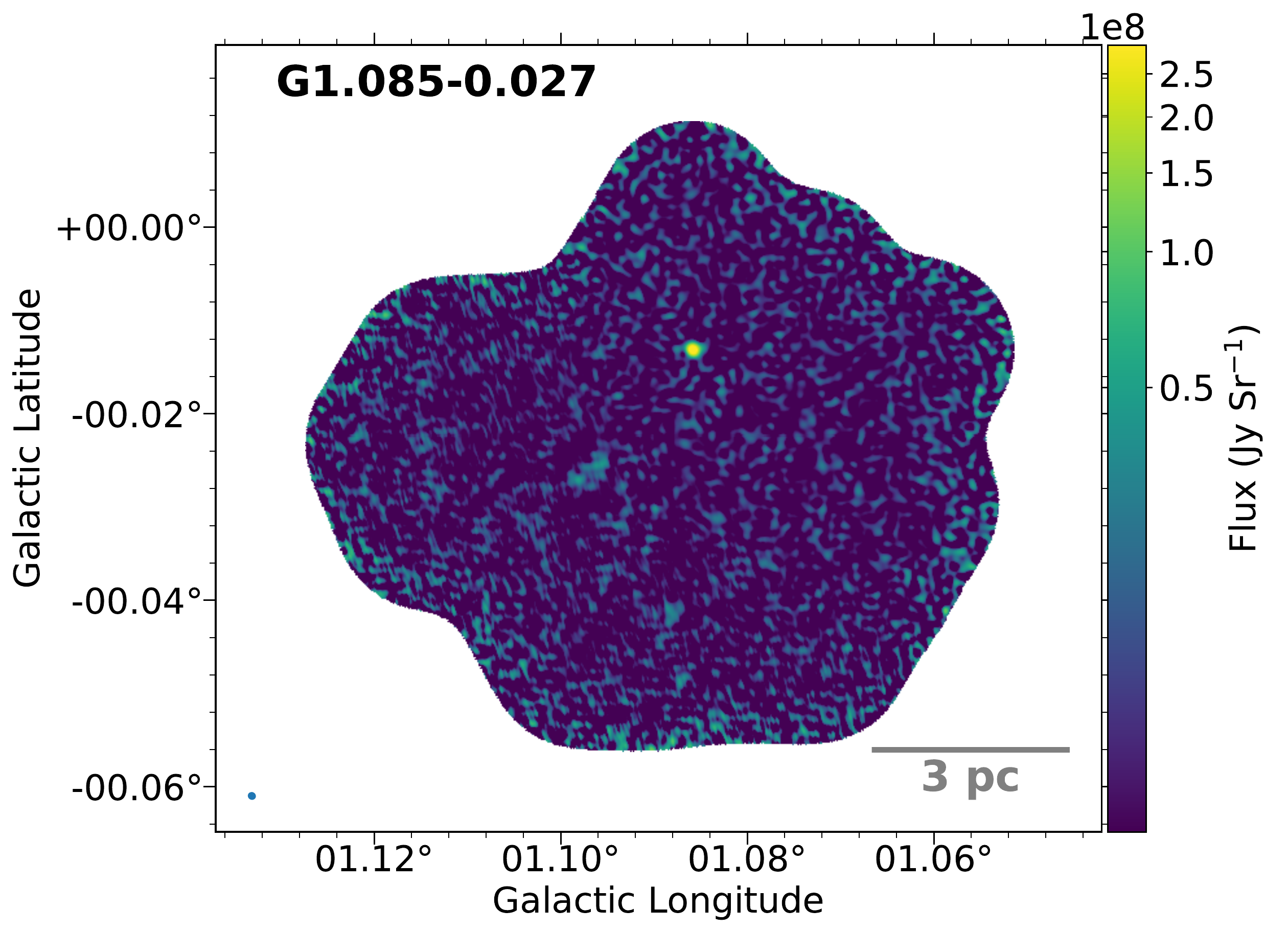}}
\subfigure{
\includegraphics[width=0.48\textwidth]{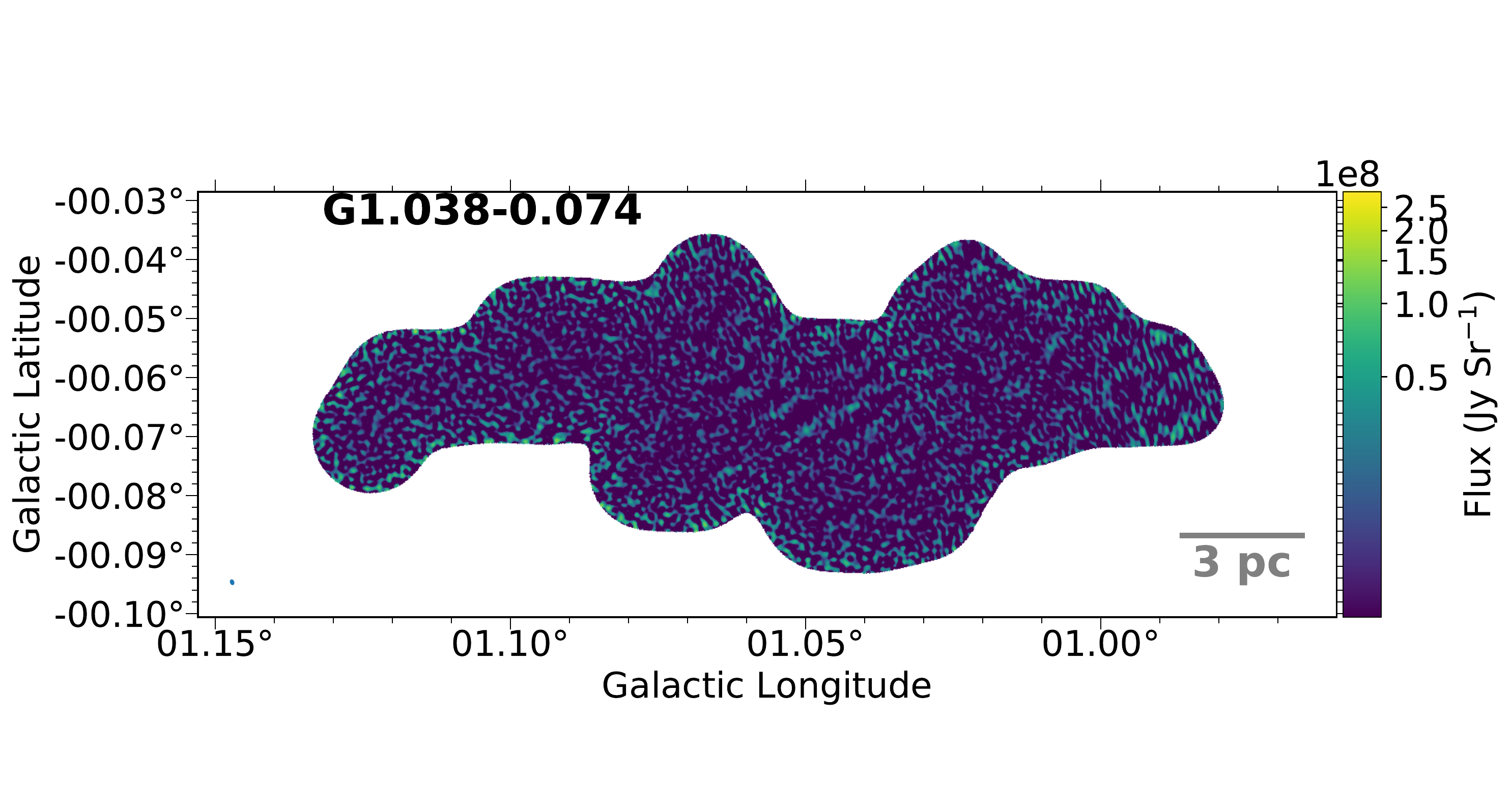}}\\
\singlespace\caption{Images show the 1.3 mm dust continuum from the \textit{CMZoom} Survey with a 3 pc scale-bar. The white contour shows the approximate 5-$\sigma$ level while the black contour shows the approximate 10-$\sigma$ level. All the images are displayed on same color scale from 2 - 290 MJy Sr$^{-1}$.} 
\label{fig:img_gallery}
\end{center}
\end{figure*}

\begin{figure*}
\begin{center}
\subfigure{
\includegraphics[width=0.48\textwidth]{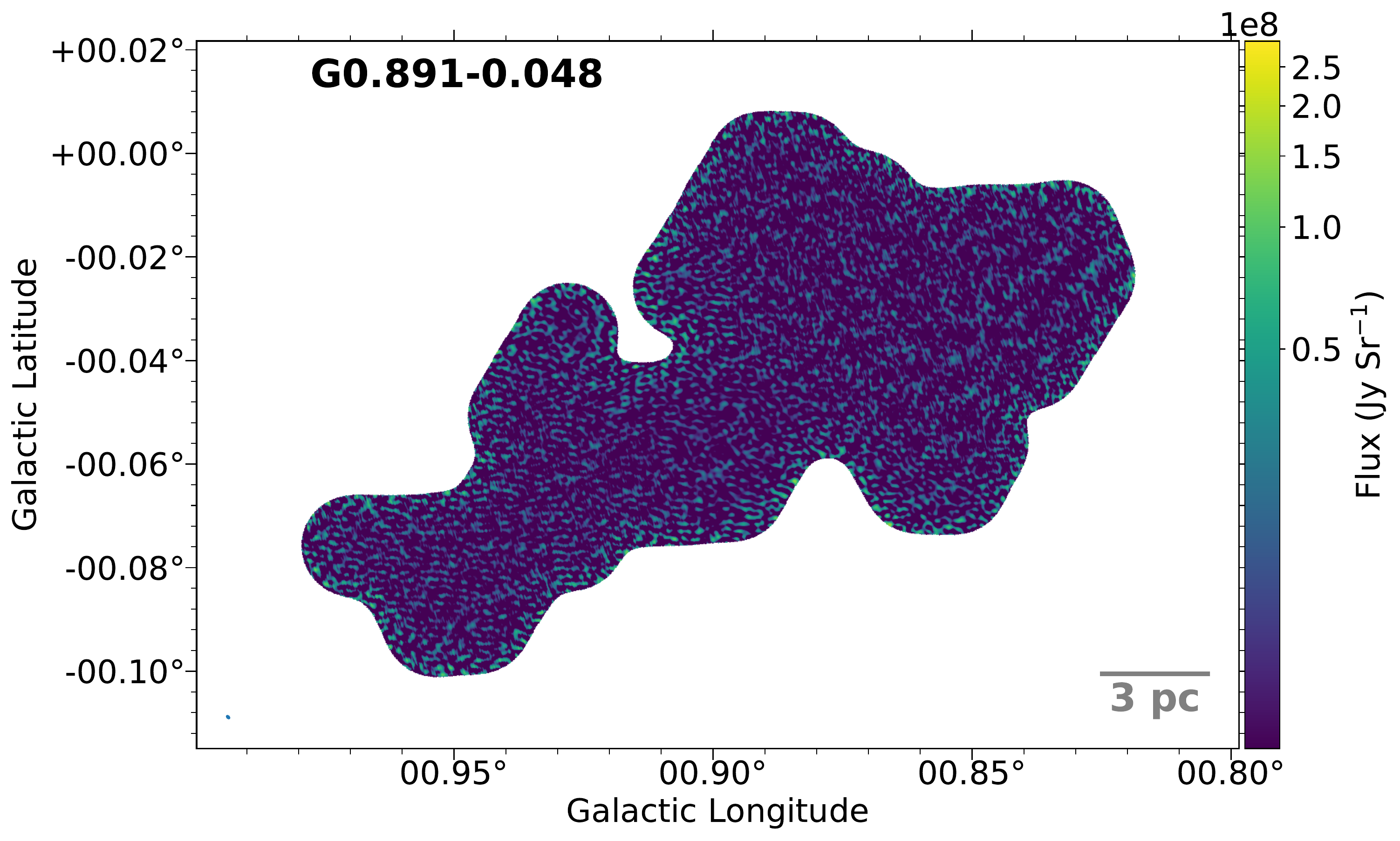}}
\subfigure{
\includegraphics[width=0.48\textwidth]{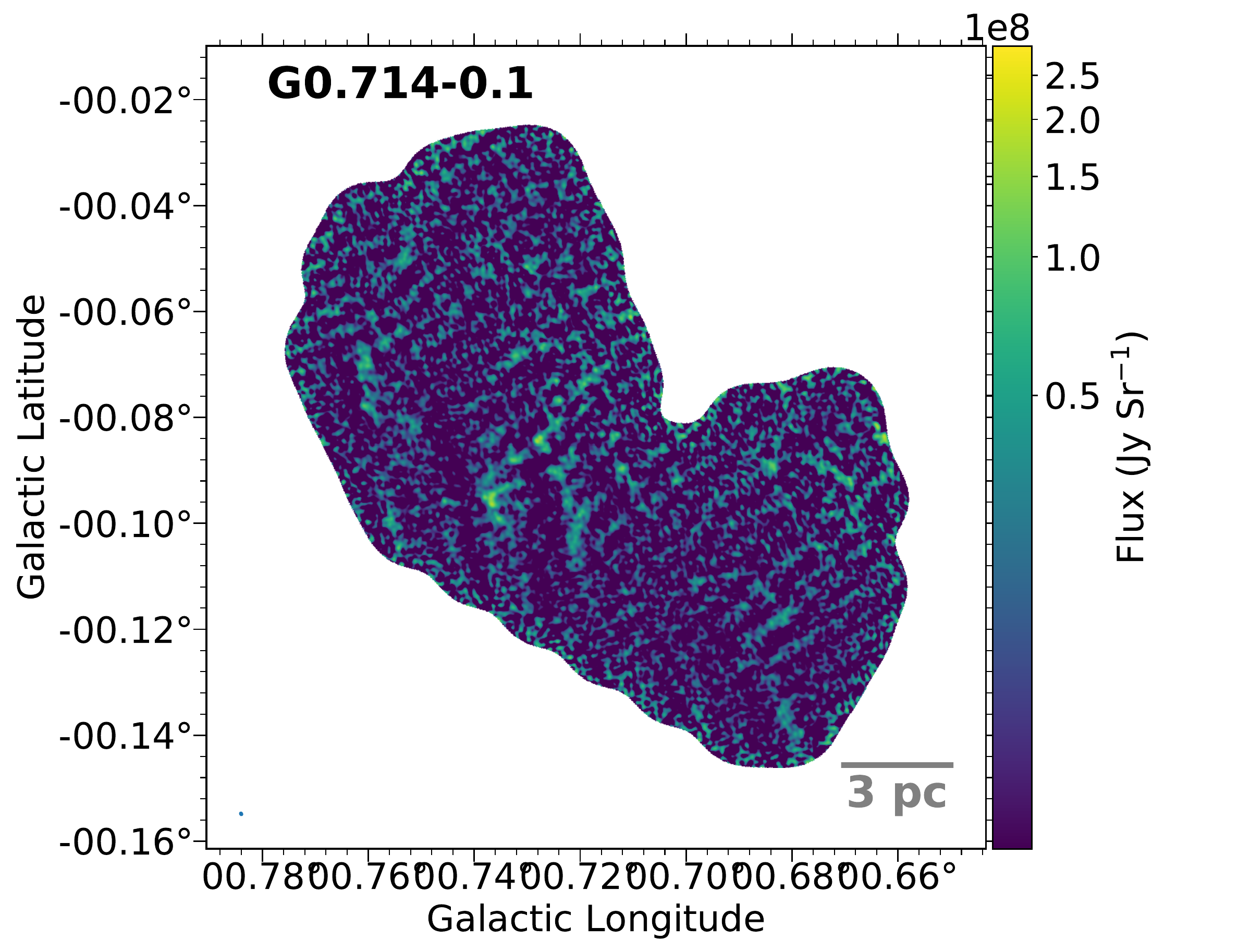}}\\
\subfigure{
\includegraphics[width=0.48\textwidth]{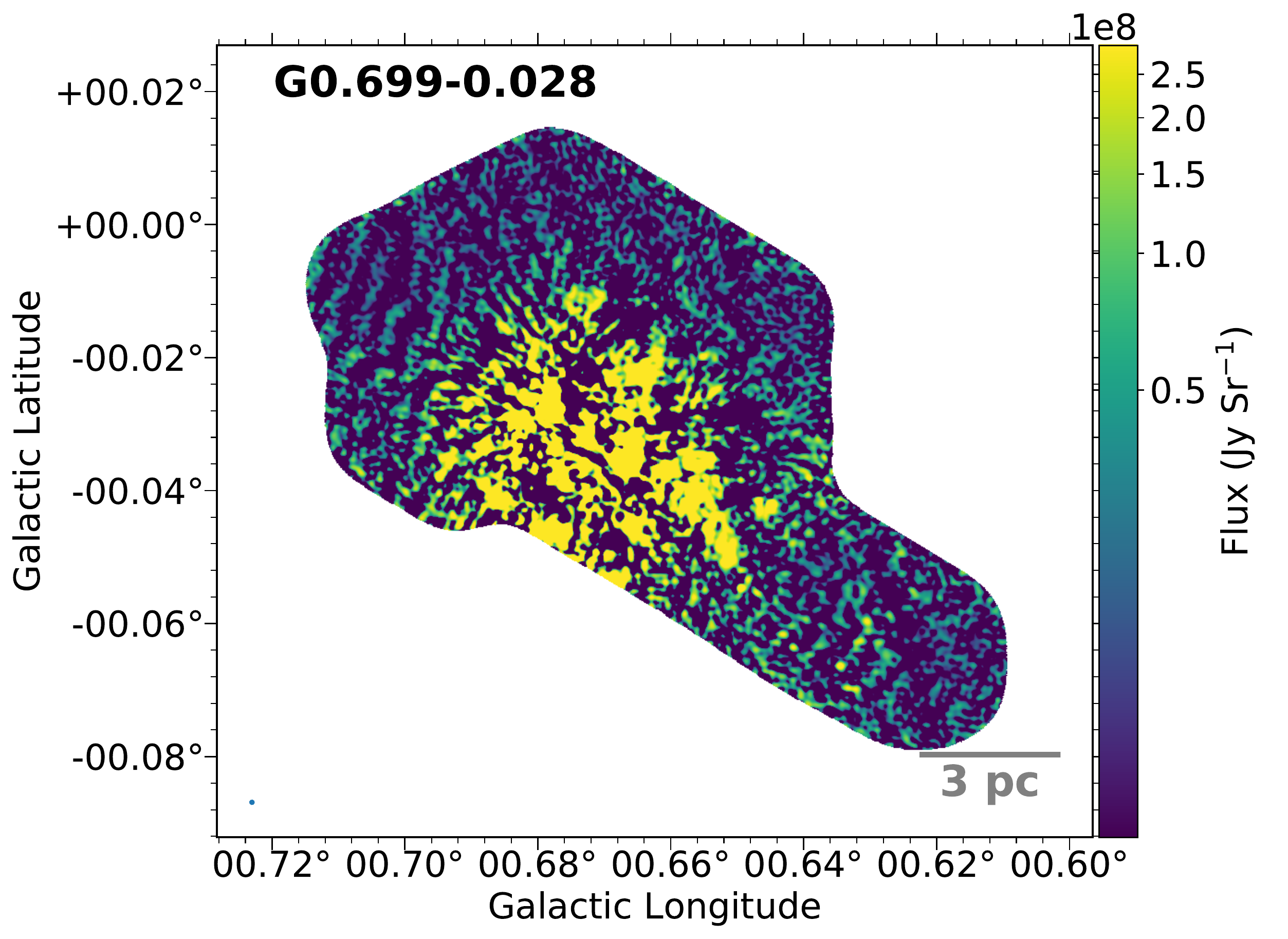}}
\subfigure{
\includegraphics[width=0.48\textwidth]{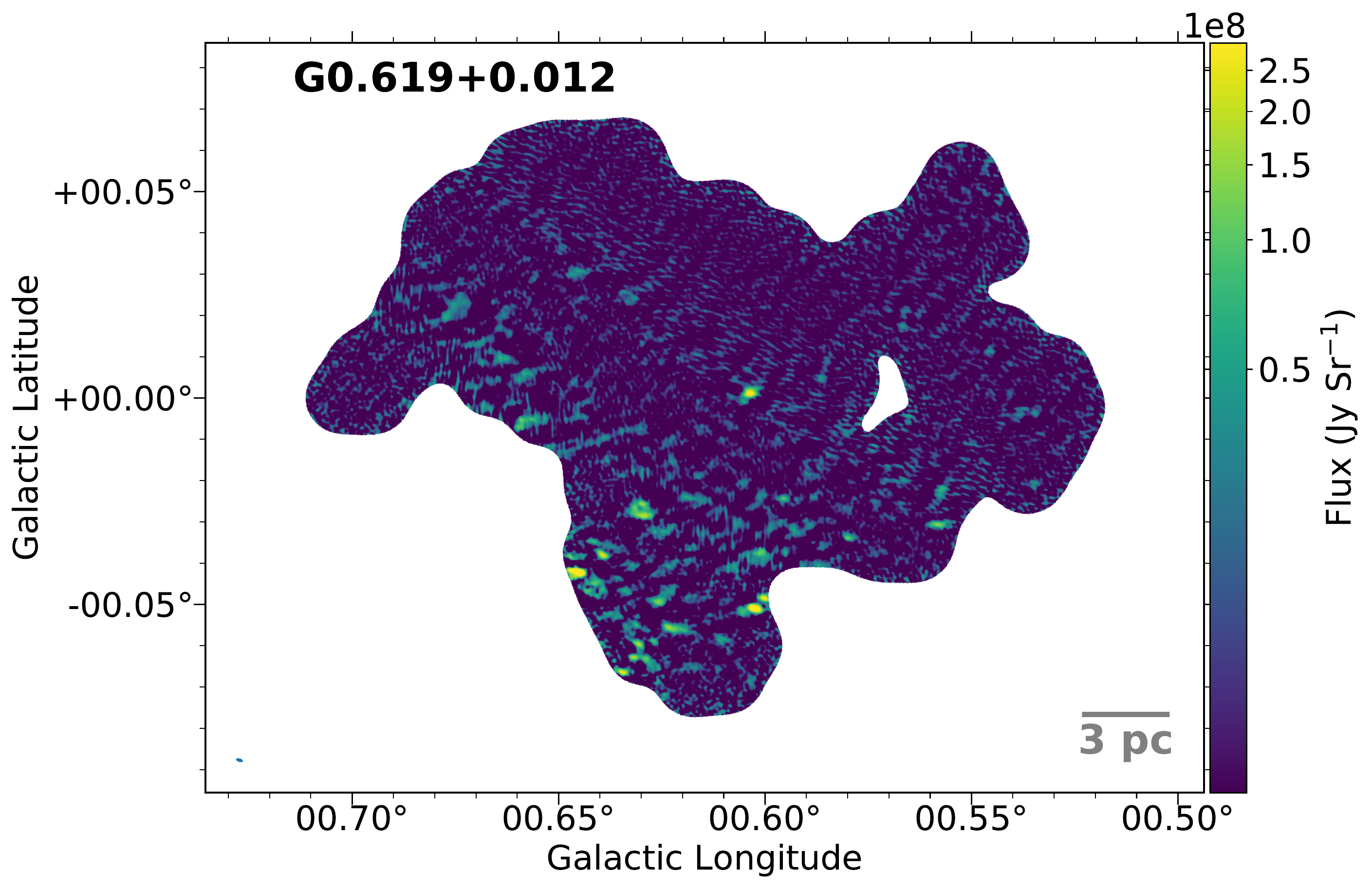}}\\
\subfigure{
\includegraphics[width=0.48\textwidth]{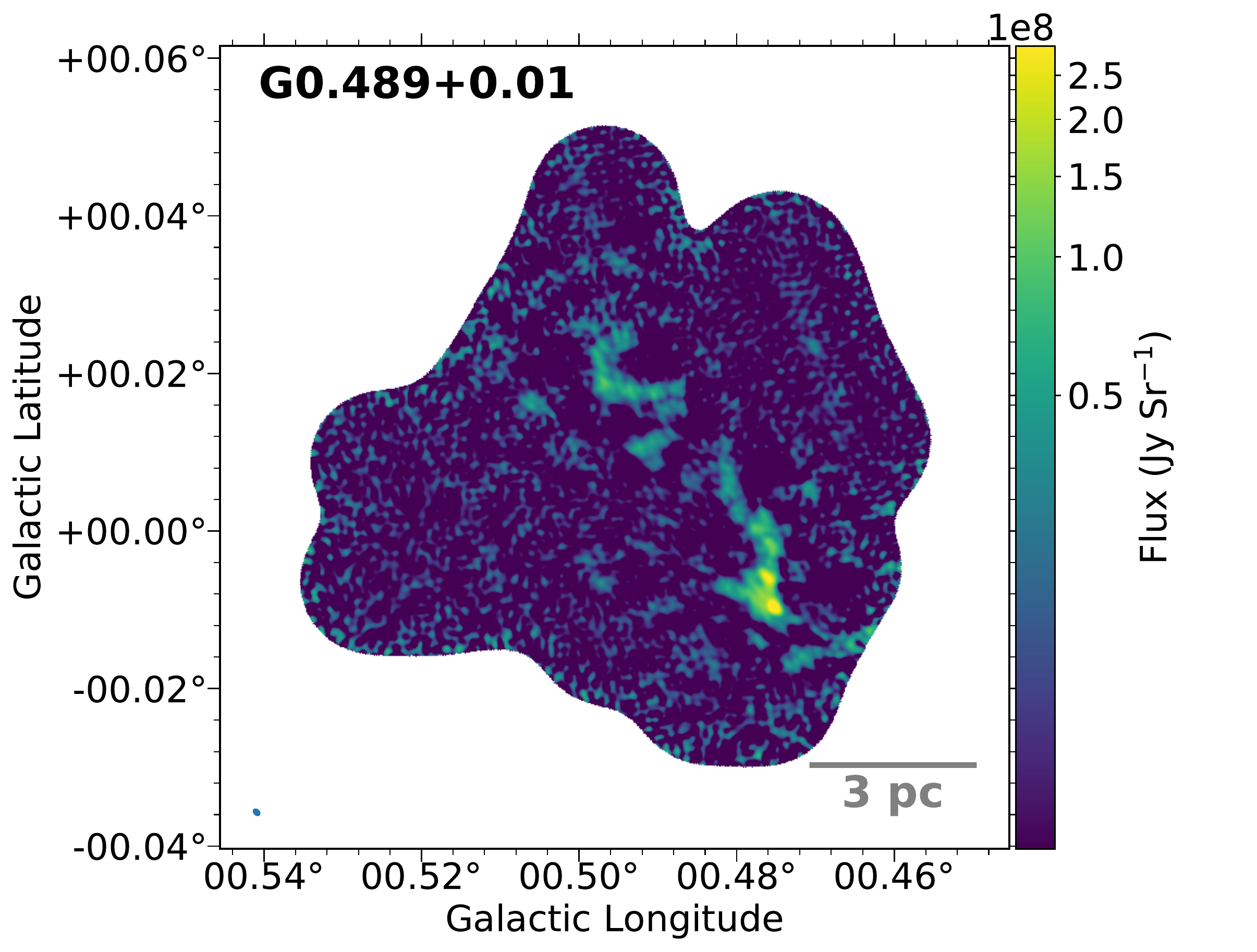}}
\subfigure{
\includegraphics[width=0.48\textwidth]{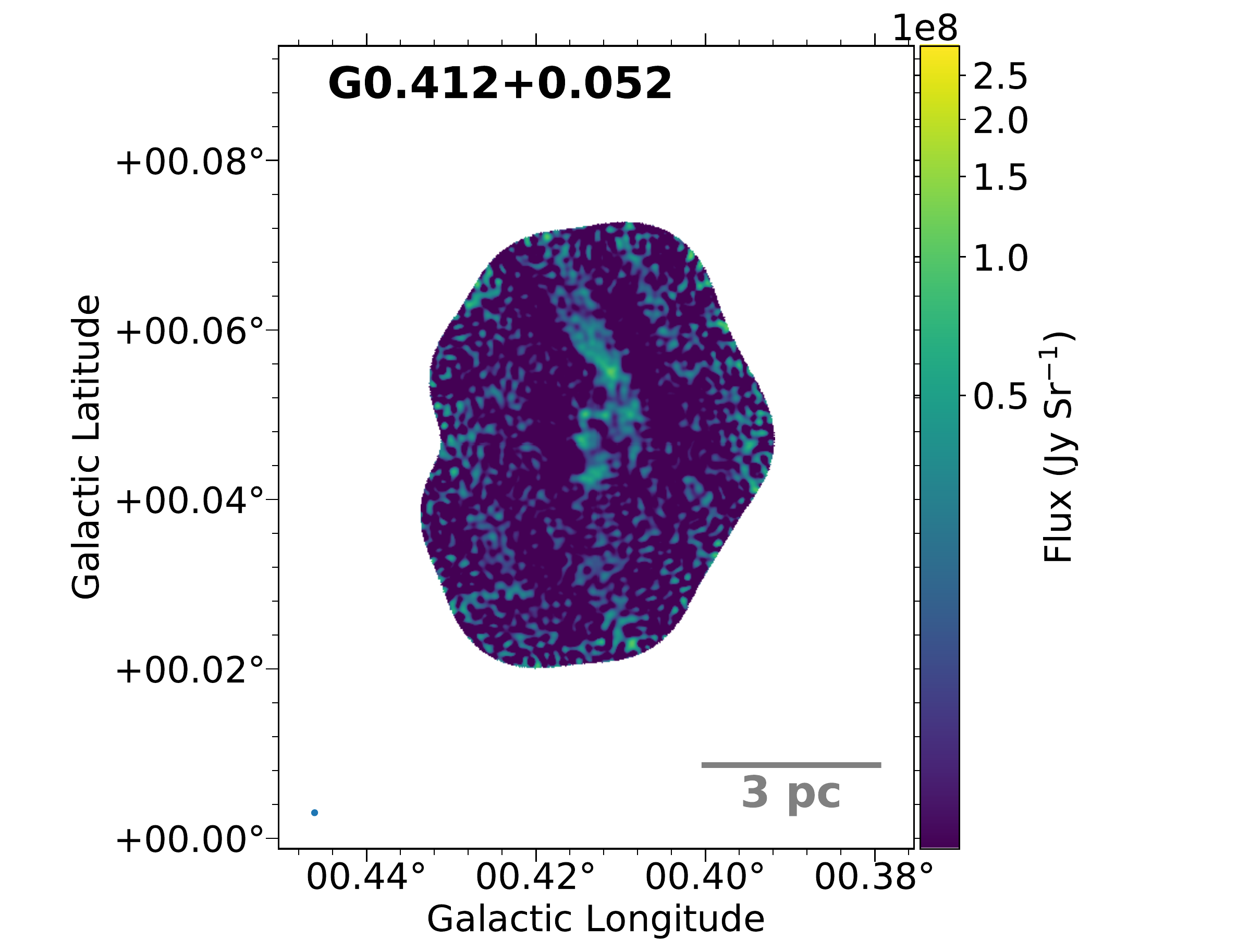}}\\
\singlespace\caption{Images show the 1.3 mm dust continuum from the \textit{CMZoom} Survey with a 3 pc scale-bar. The white contour shows the approximate 5-$\sigma$ level while the black contour shows the approximate 10-$\sigma$ level. All the images are displayed on same color scale from 2 - 290 MJy Sr$^{-1}$.} 
\label{fig:img_gallery}
\end{center}
\end{figure*}

\begin{figure*}
\begin{center}
\subfigure{
\includegraphics[width=0.48\textwidth]{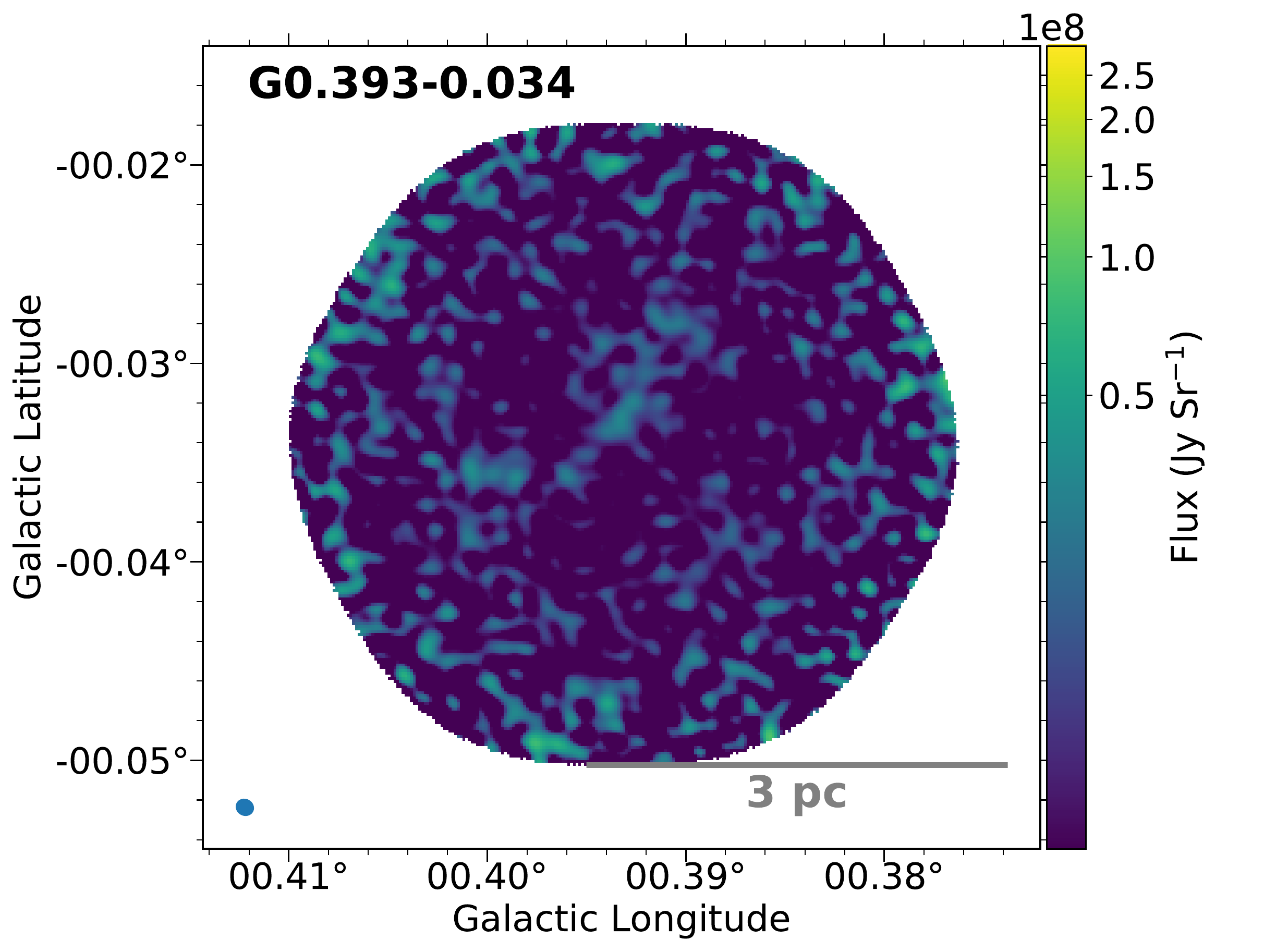}}
\subfigure{
\includegraphics[width=0.48\textwidth]{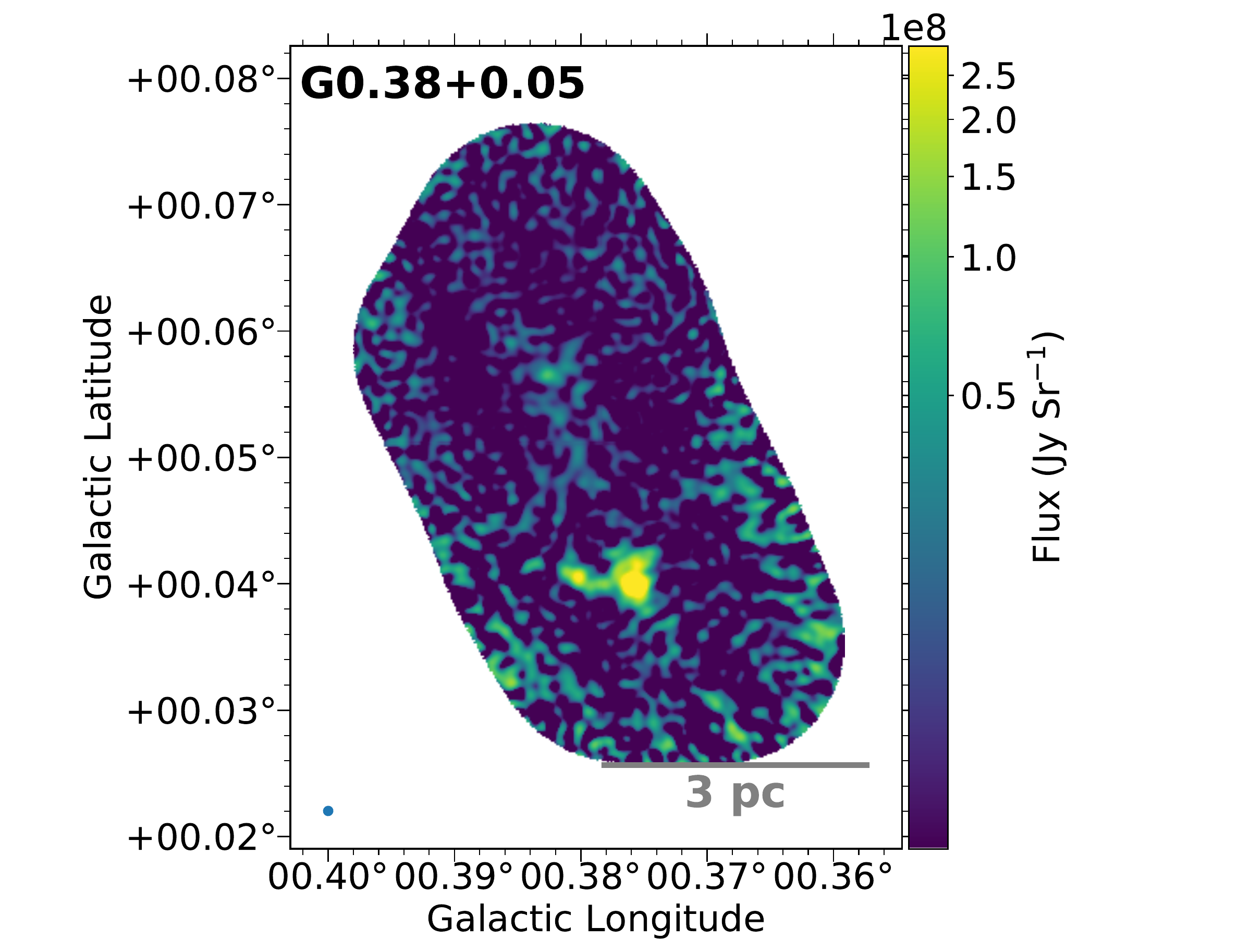}}\\
\subfigure{
\includegraphics[width=0.48\textwidth]{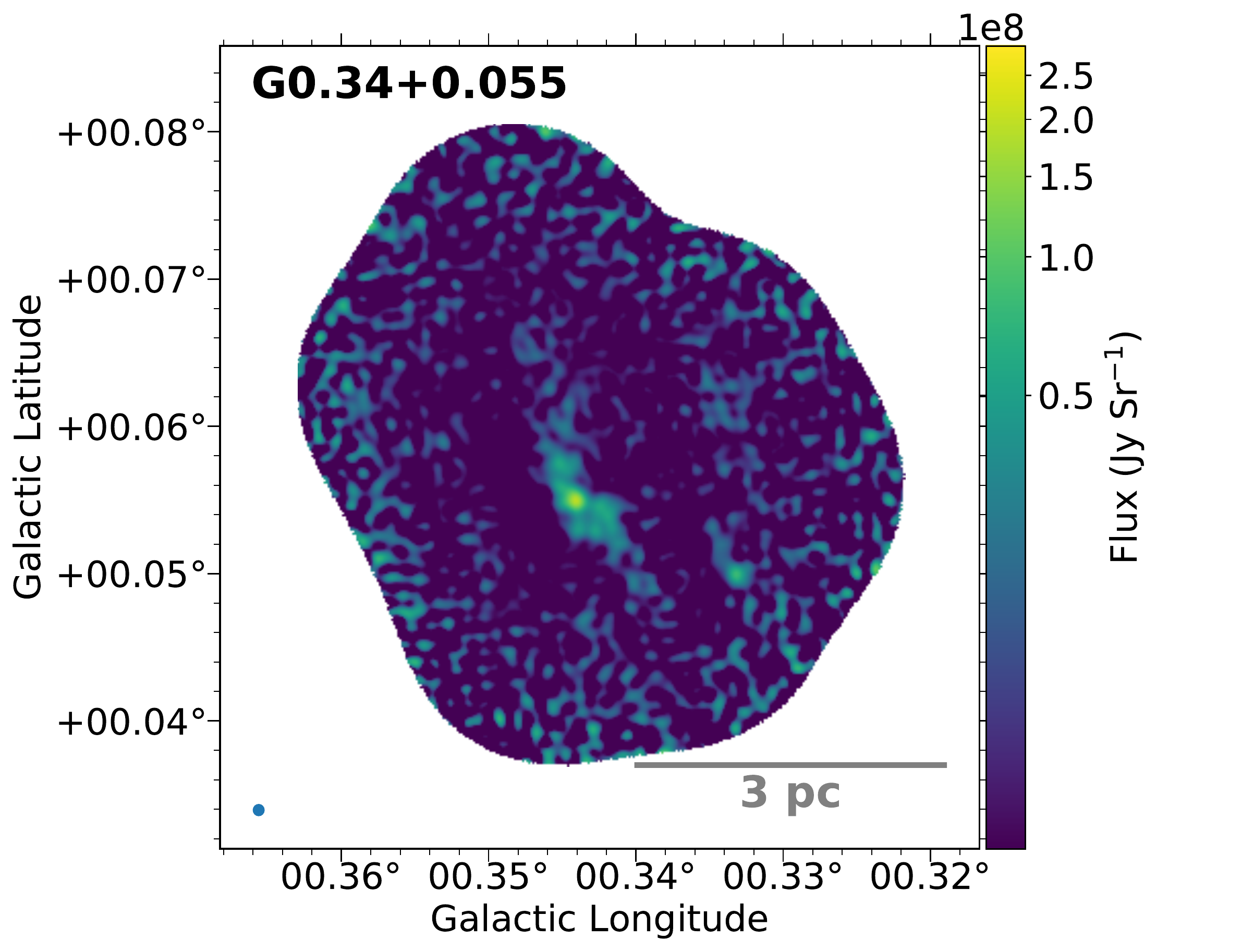}}
\subfigure{
\includegraphics[width=0.48\textwidth]{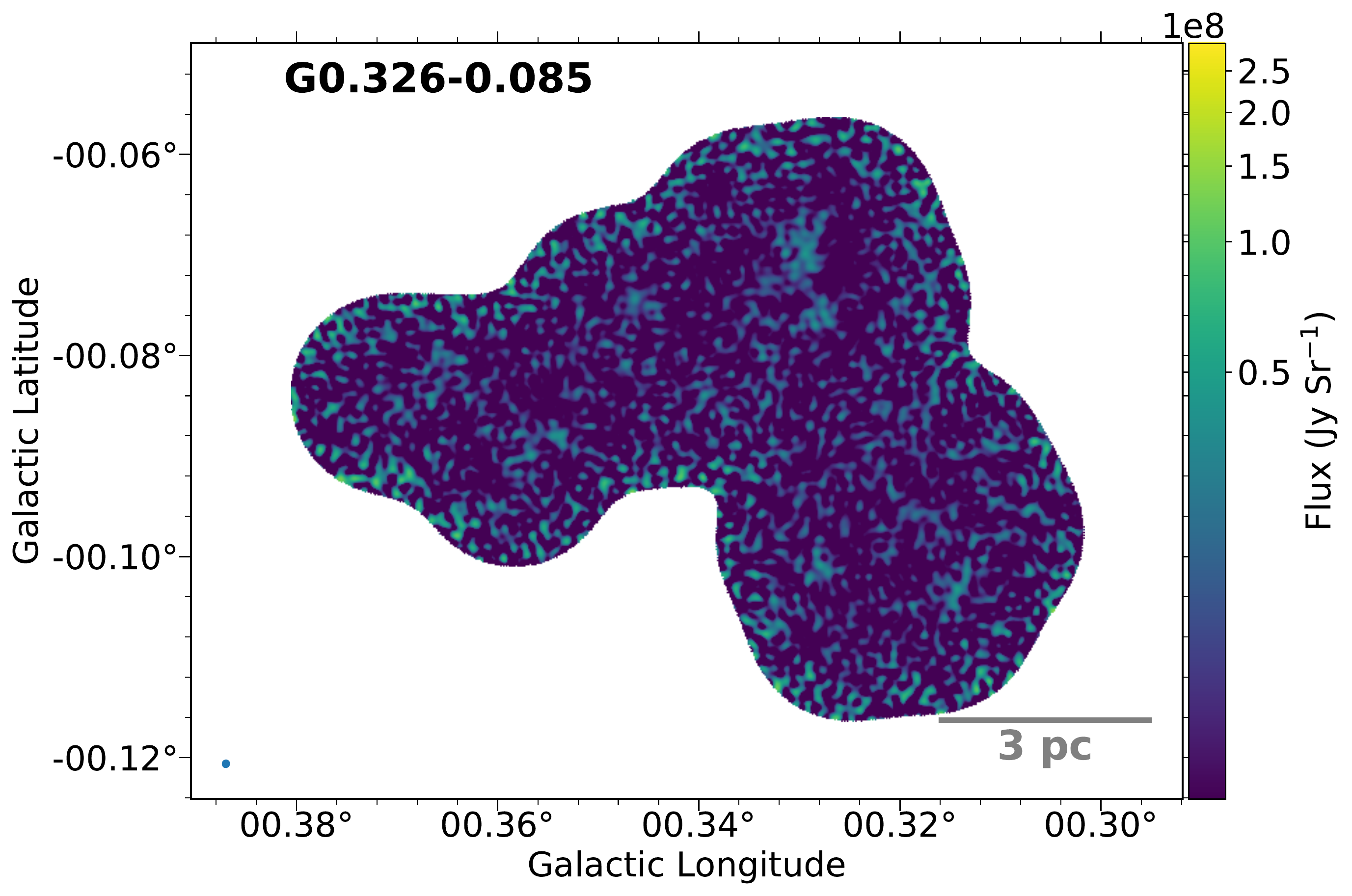}}\\
\subfigure{
\includegraphics[width=0.48\textwidth]{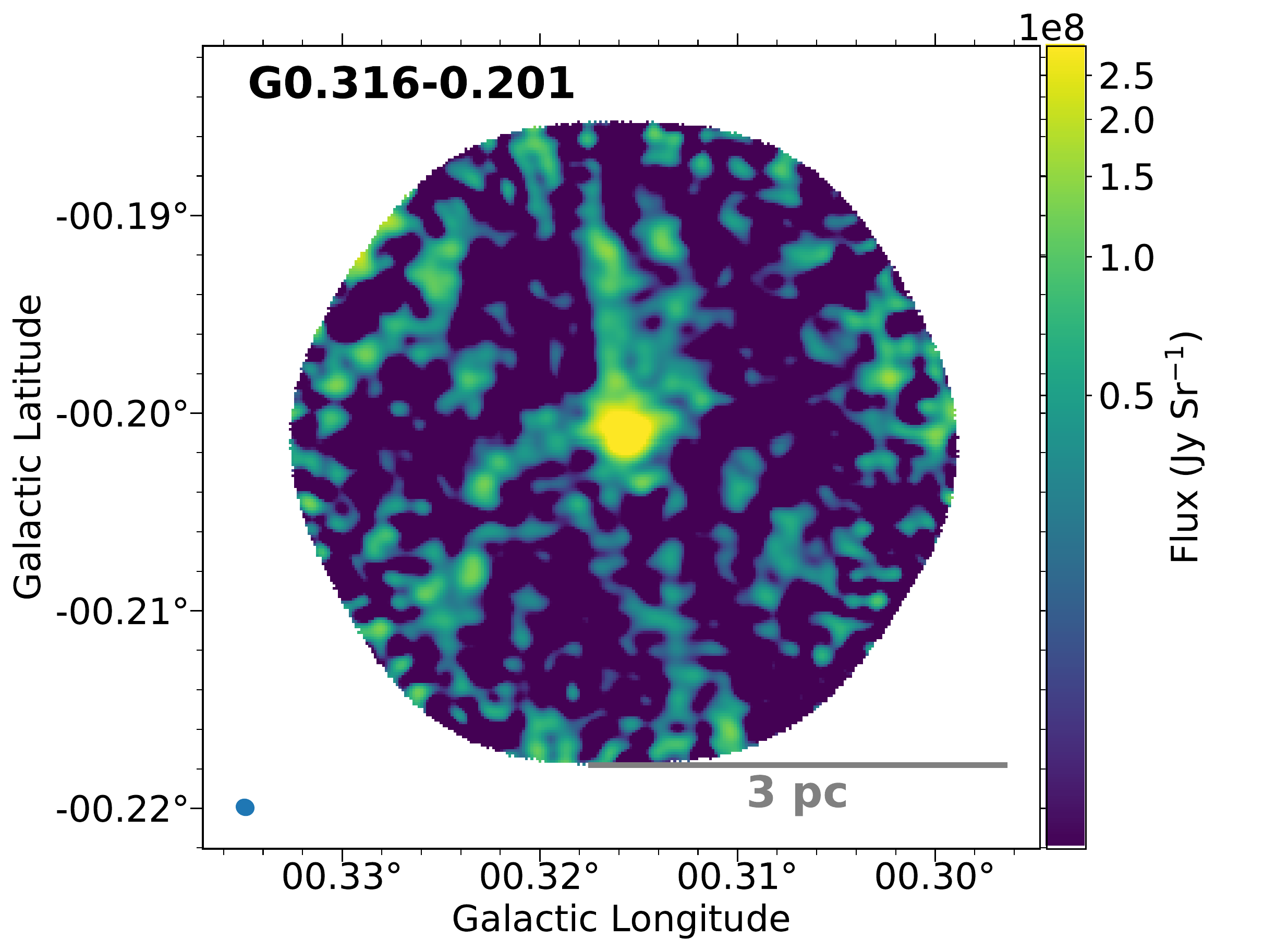}}
\subfigure{
\includegraphics[width=0.48\textwidth]{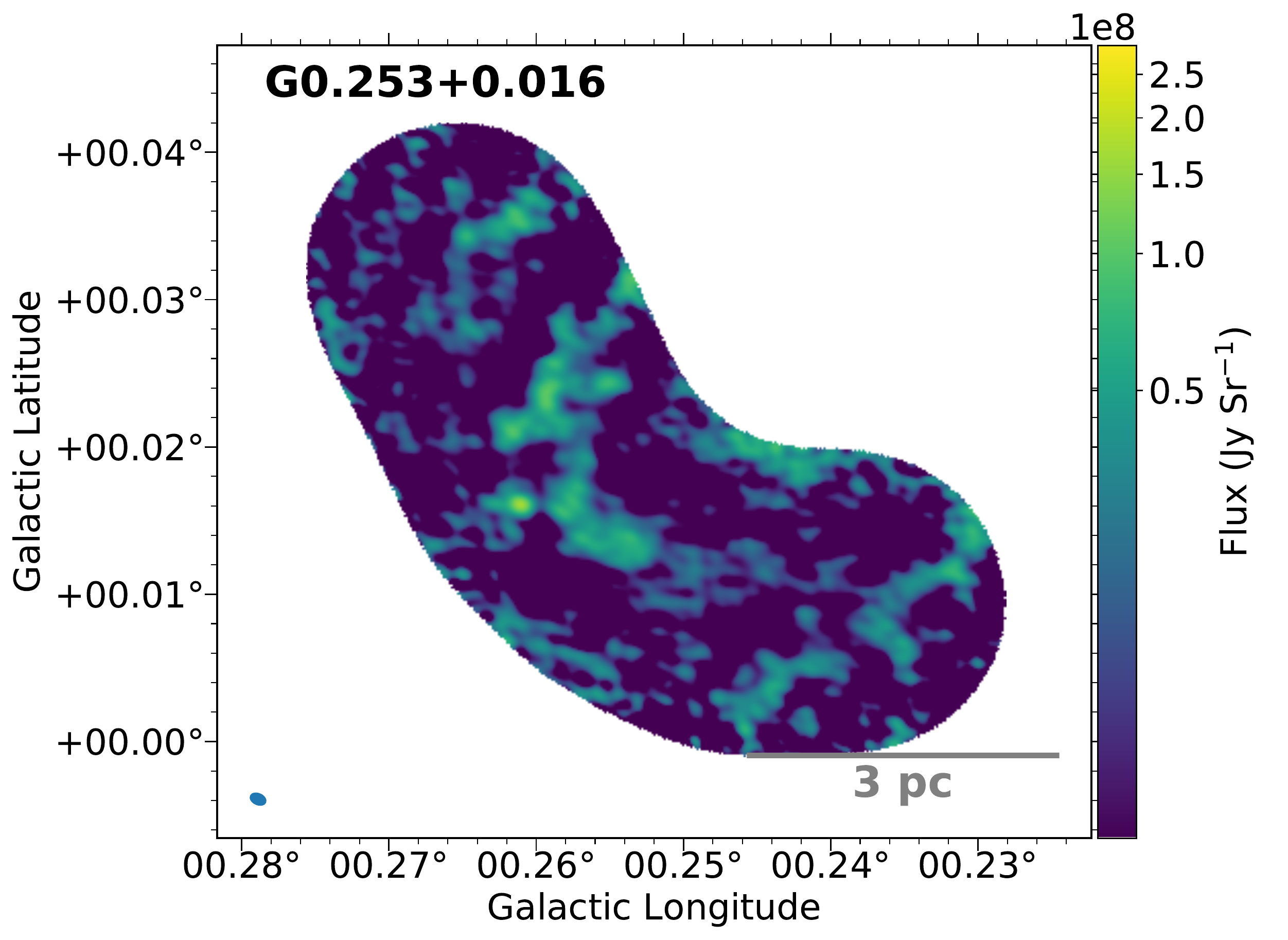}}\\
\singlespace\caption{Images show the 1.3 mm dust continuum from the \textit{CMZoom} Survey with a 3 pc scale-bar. The white contour shows the approximate 5-$\sigma$ level while the black contour shows the approximate 10-$\sigma$ level. All the images are displayed on same color scale from 2 - 290 MJy Sr$^{-1}$.} 
\label{fig:img_gallery}
\end{center}
\end{figure*}

\begin{figure*}
\begin{center}
\subfigure{
\includegraphics[width=0.48\textwidth]{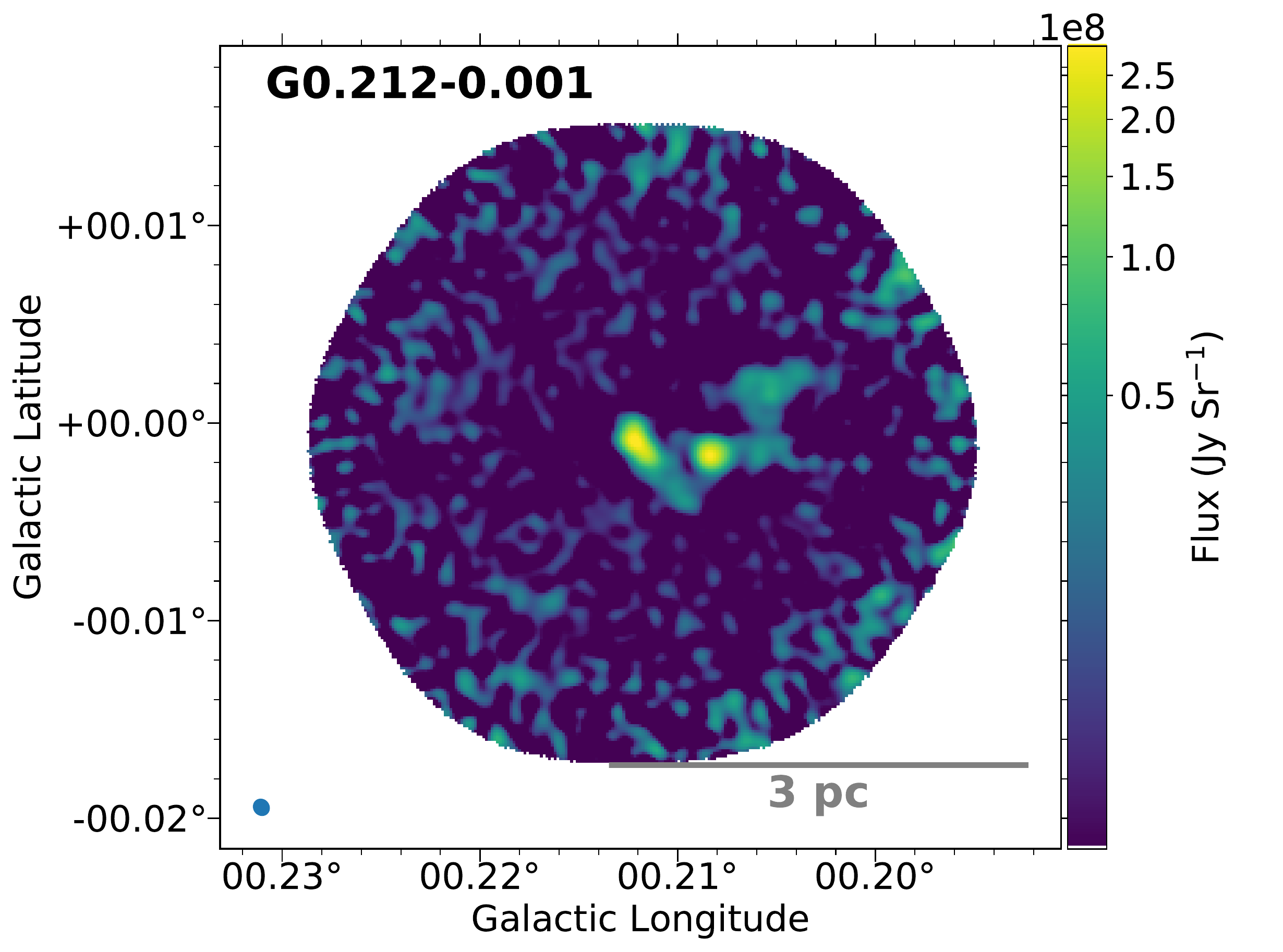}}
\subfigure{
\includegraphics[width=0.48\textwidth]{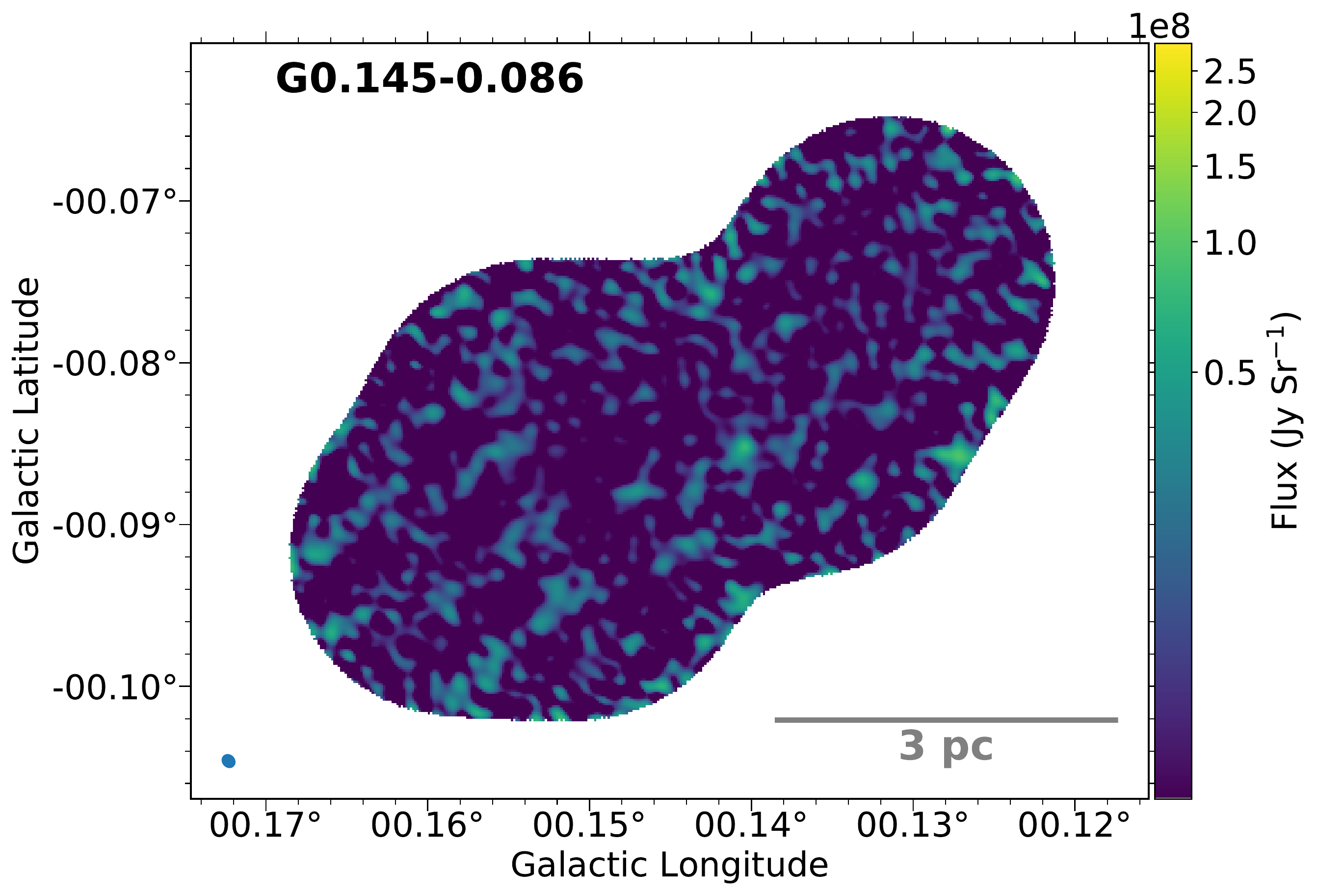}}\\
\subfigure{
\includegraphics[width=0.48\textwidth]{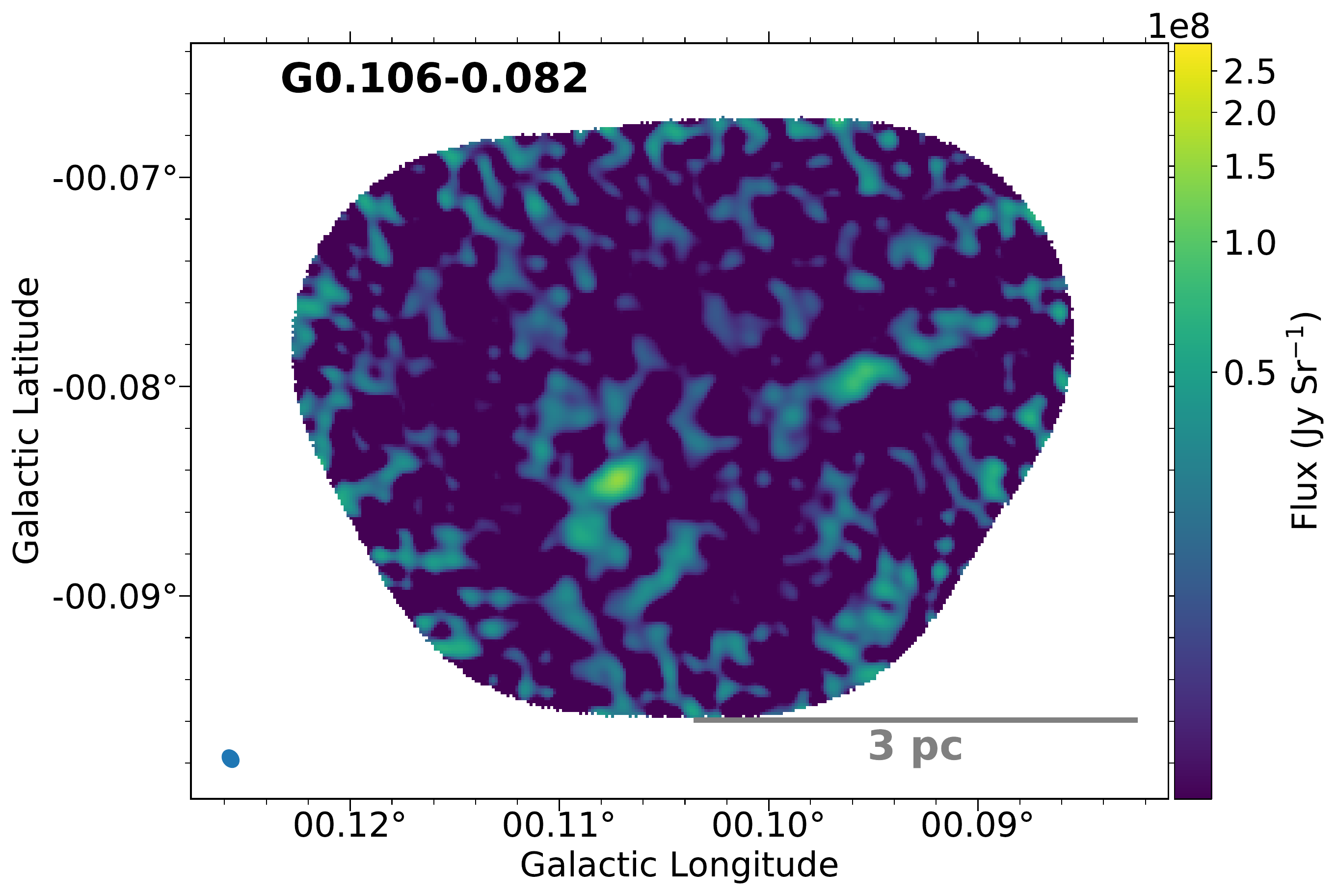}}
\subfigure{
\includegraphics[width=0.48\textwidth]{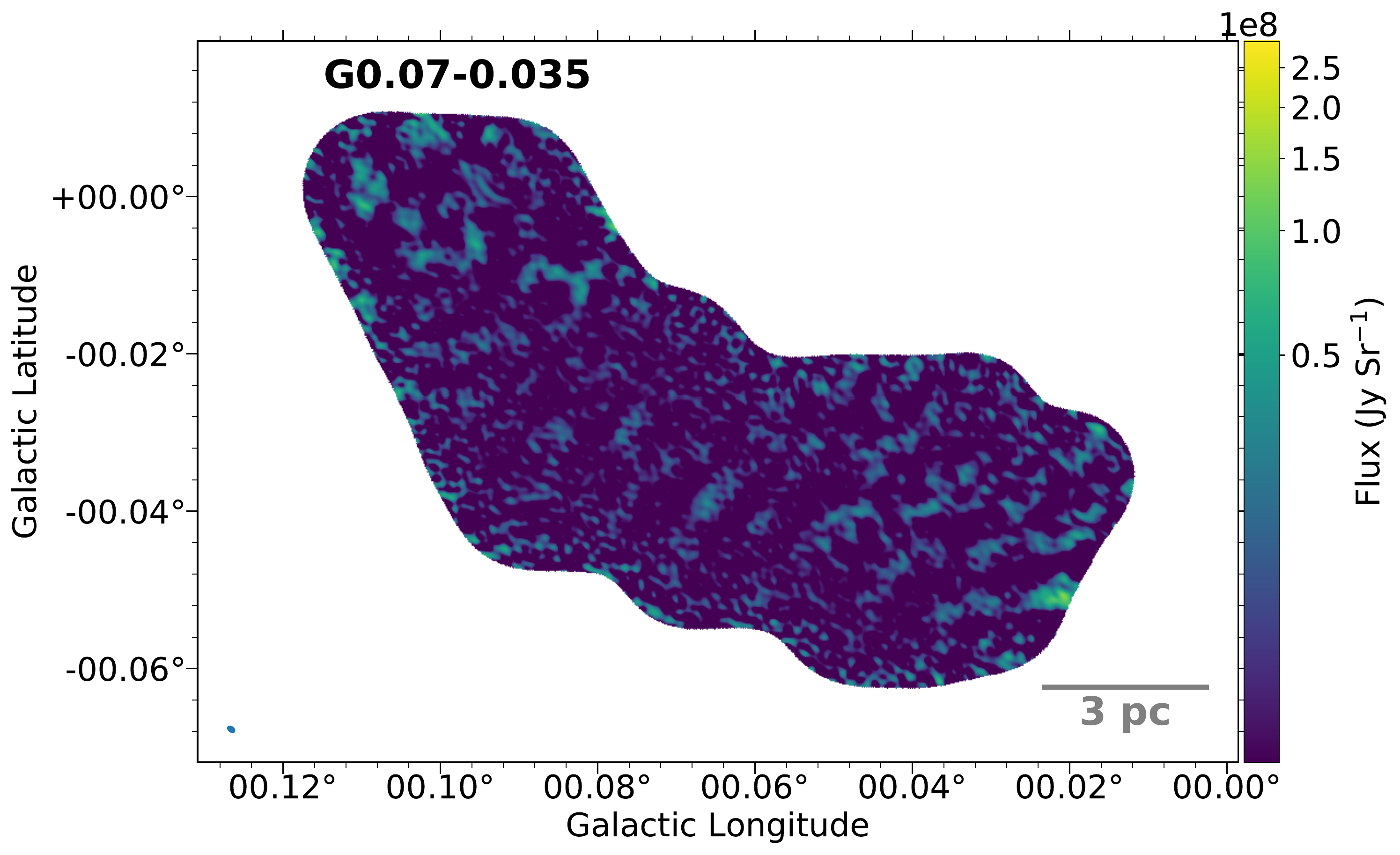}}\\
\subfigure{
\includegraphics[width=0.48\textwidth]{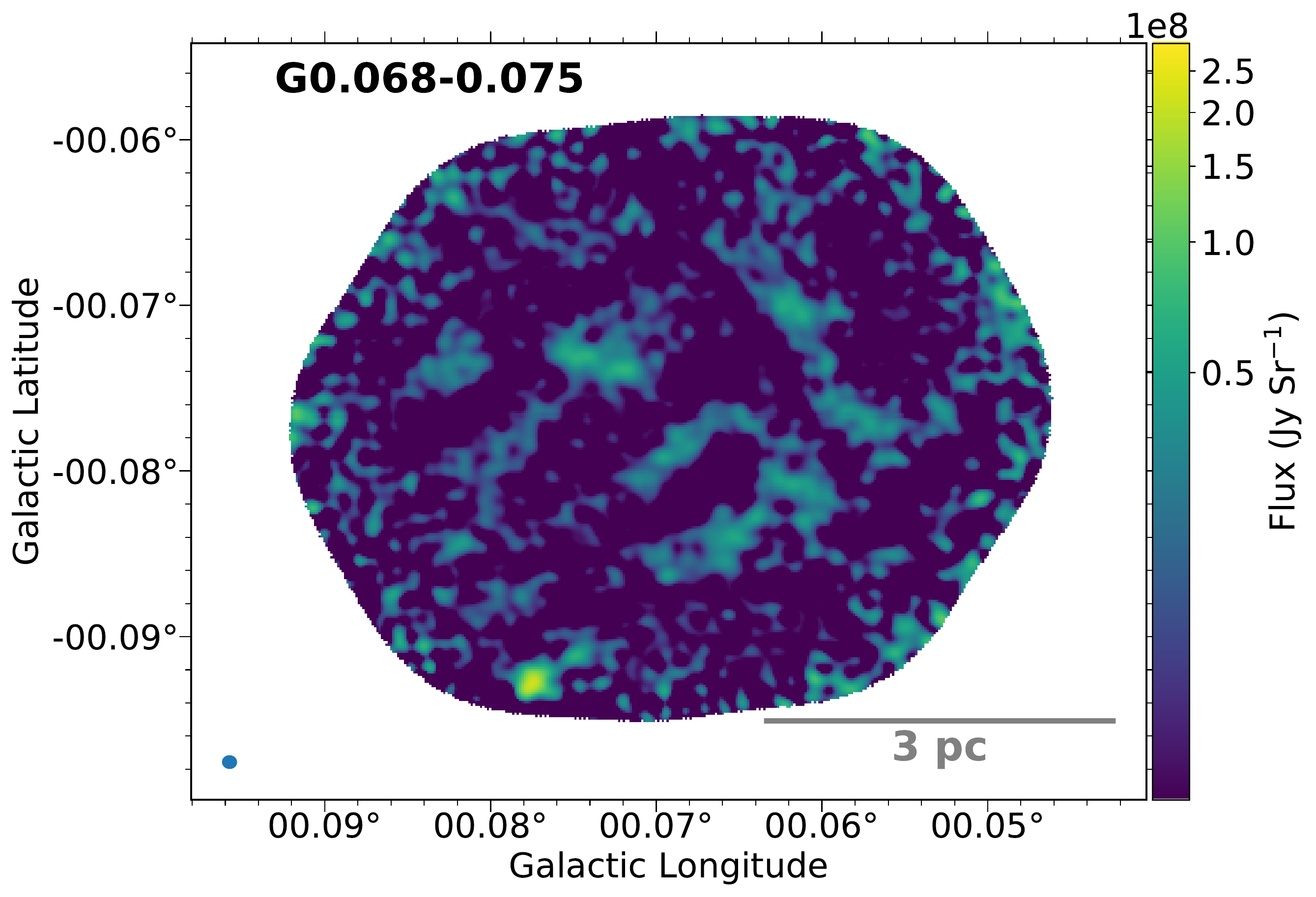}}
\subfigure{
\includegraphics[width=0.48\textwidth]{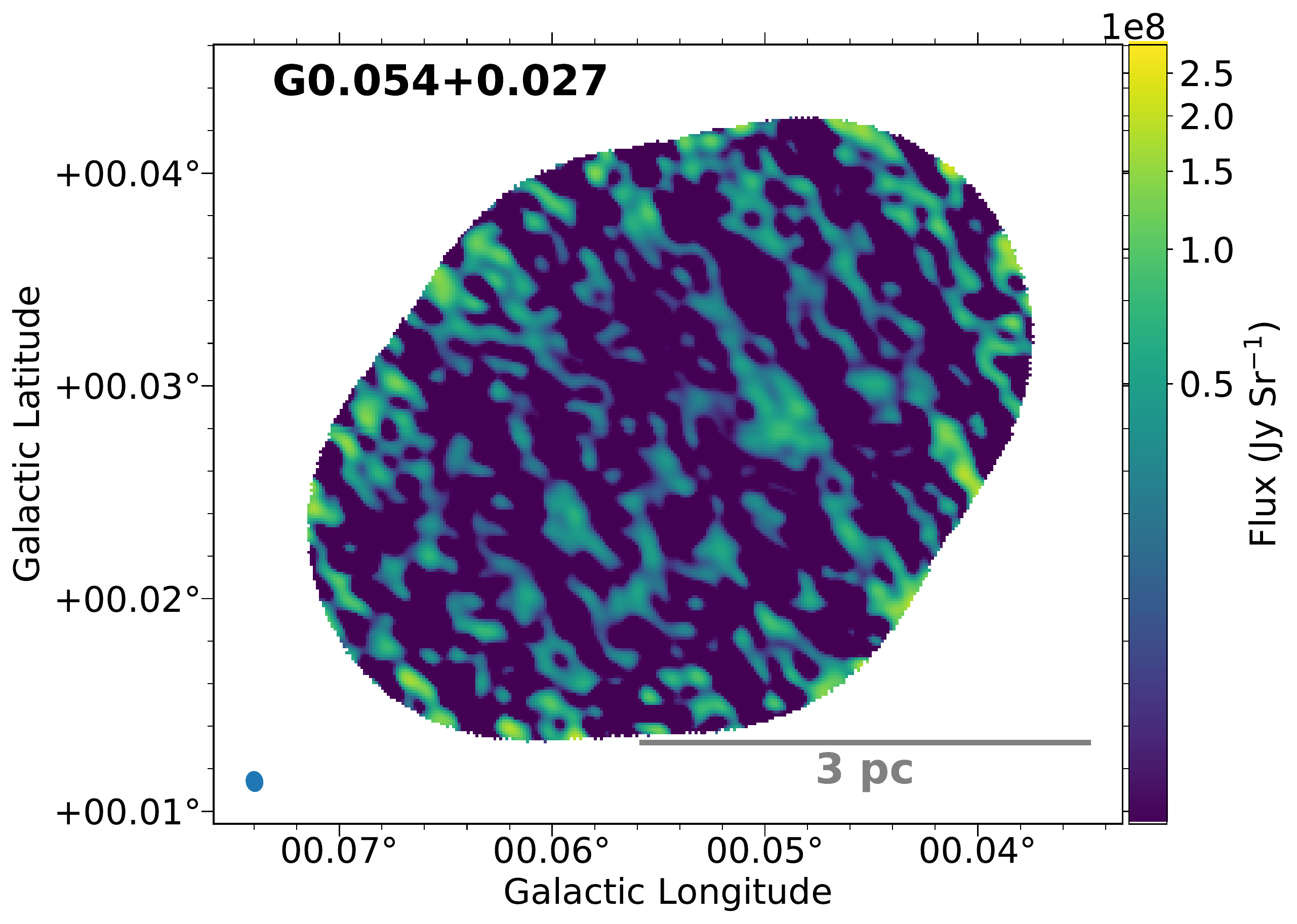}}\\
\singlespace\caption{Images show the 1.3 mm dust continuum from the \textit{CMZoom} Survey with a 3 pc scale-bar. The white contour shows the approximate 5-$\sigma$ level while the black contour shows the approximate 10-$\sigma$ level. All the images are displayed on same color scale from 2 - 290 MJy Sr$^{-1}$.} 
\label{fig:img_gallery}
\end{center}
\end{figure*}

\begin{figure*}
\begin{center}
\subfigure{
\includegraphics[width=0.48\textwidth]{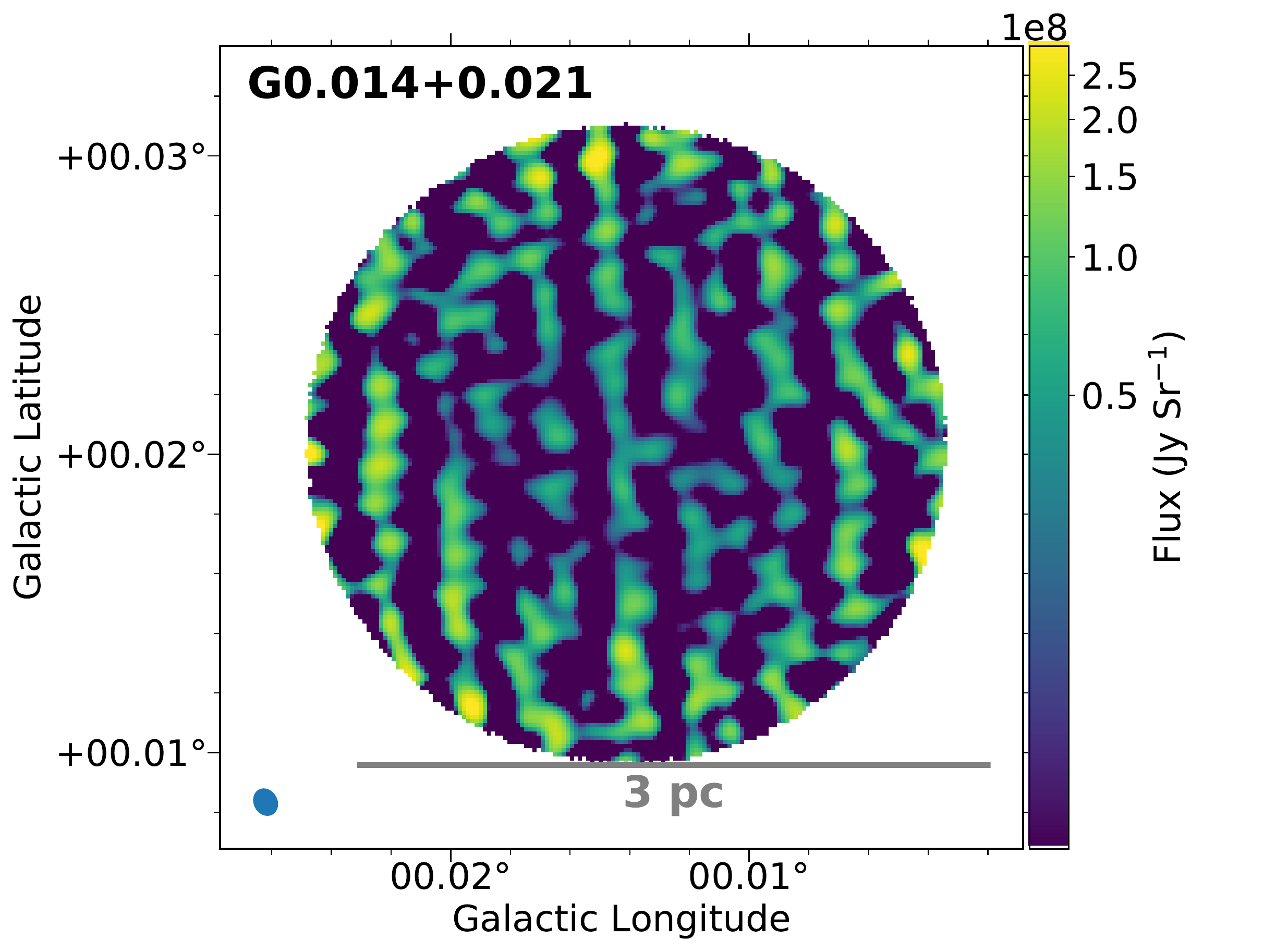}}
\subfigure{
\includegraphics[width=0.48\textwidth]{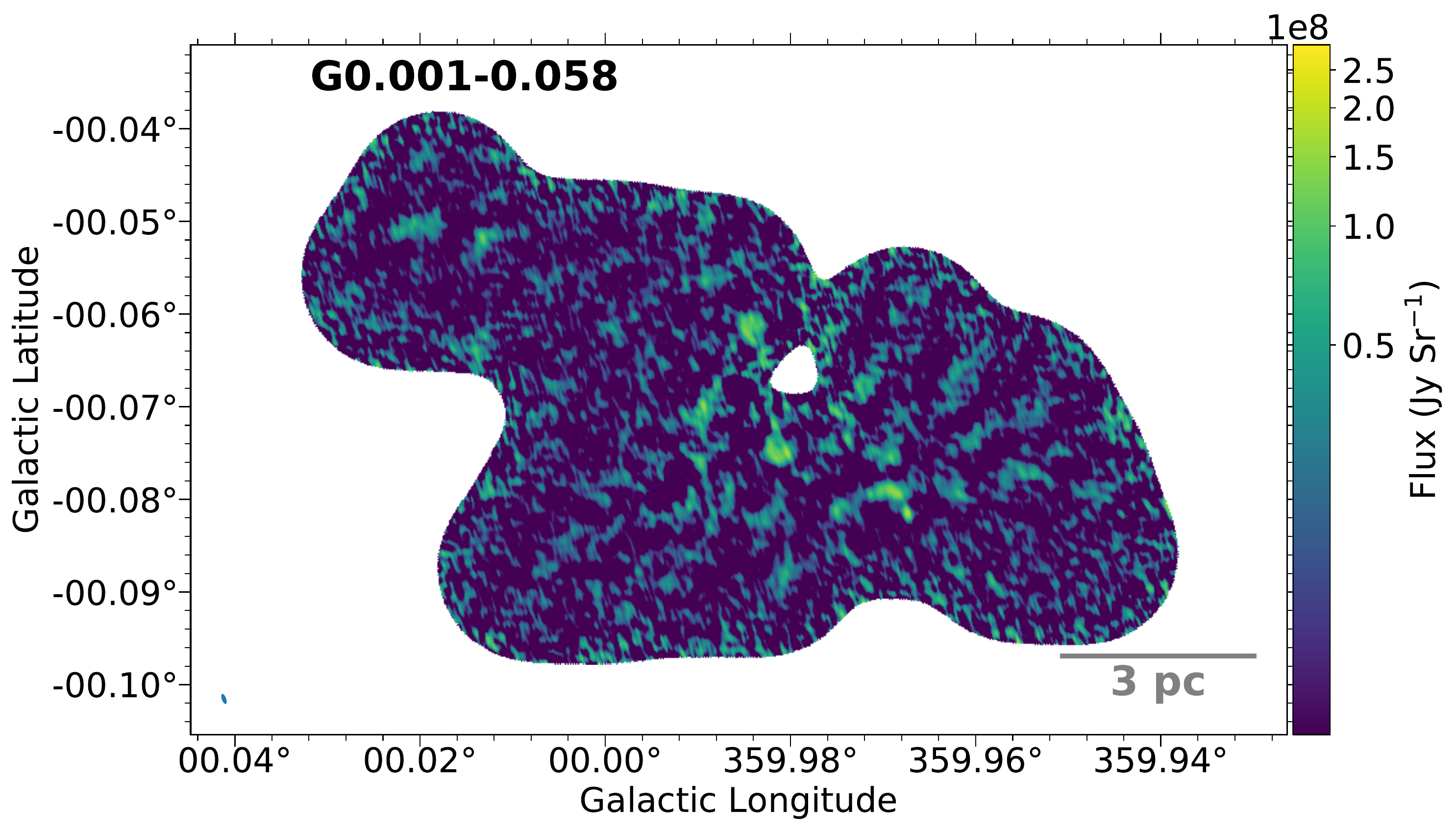}}\\
\subfigure{
\includegraphics[width=0.48\textwidth]{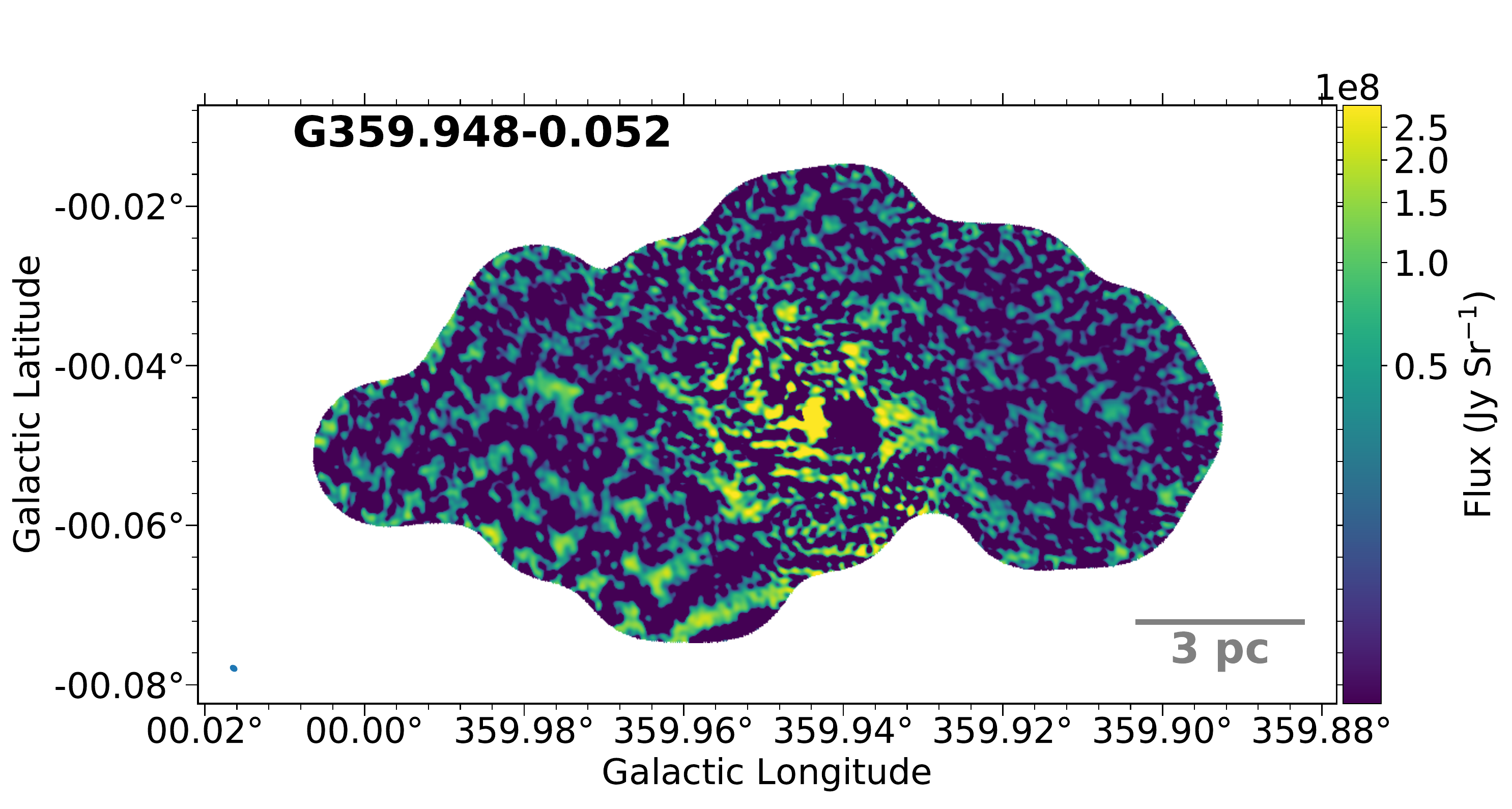}}
\subfigure{
\includegraphics[width=0.48\textwidth]{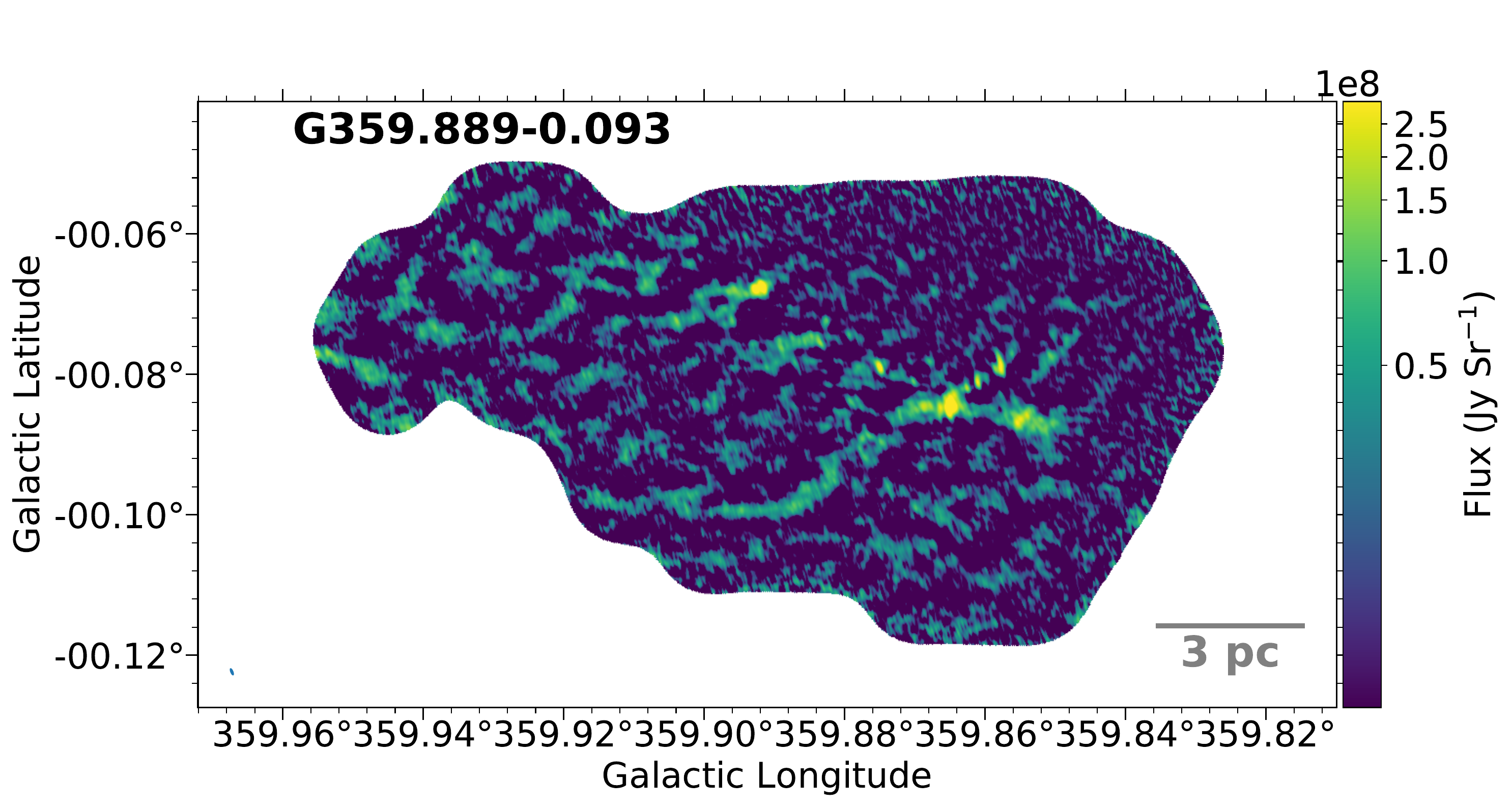}}\\
\subfigure{
\includegraphics[width=0.48\textwidth]{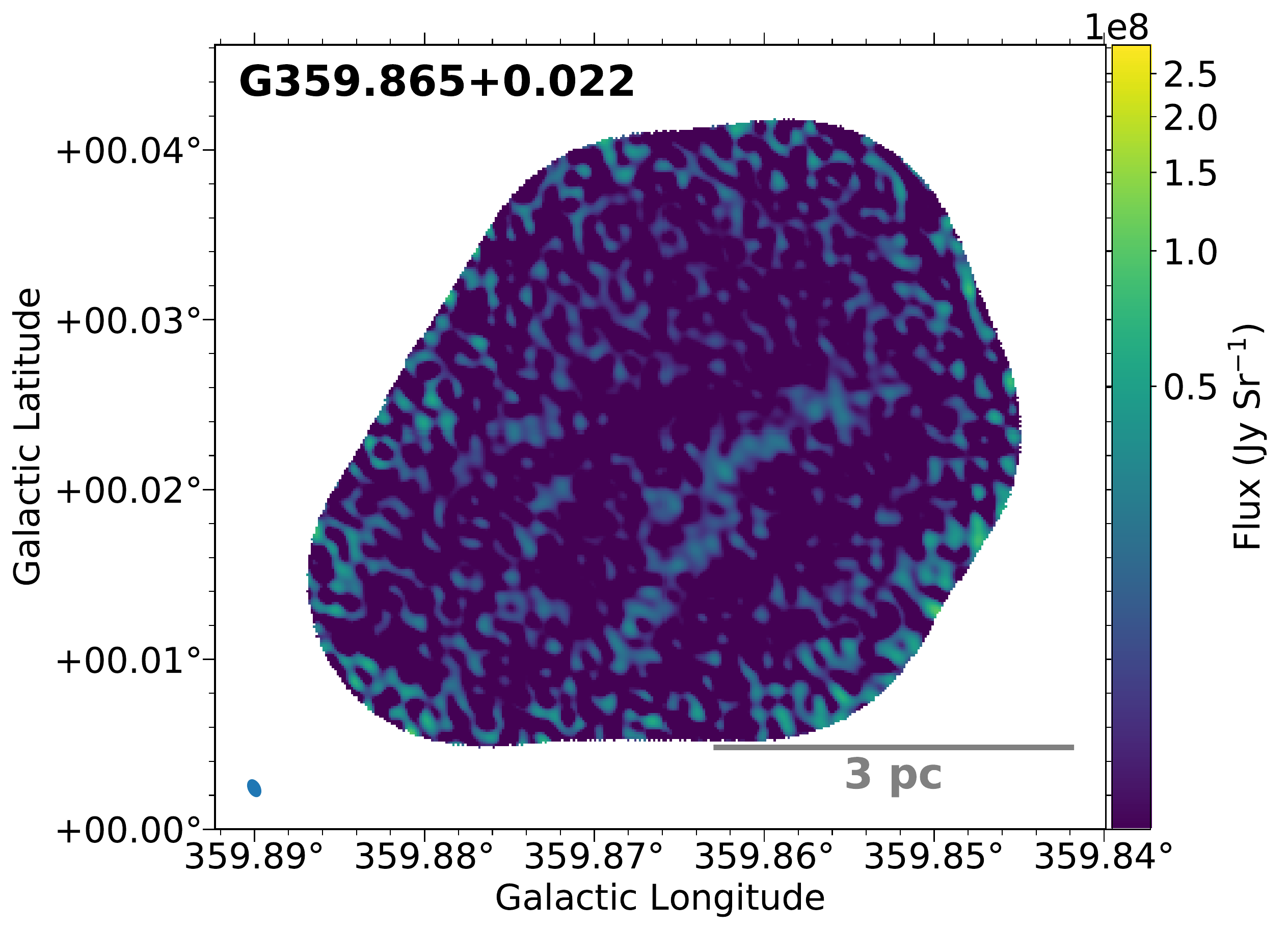}}
\subfigure{
\includegraphics[width=0.48\textwidth]{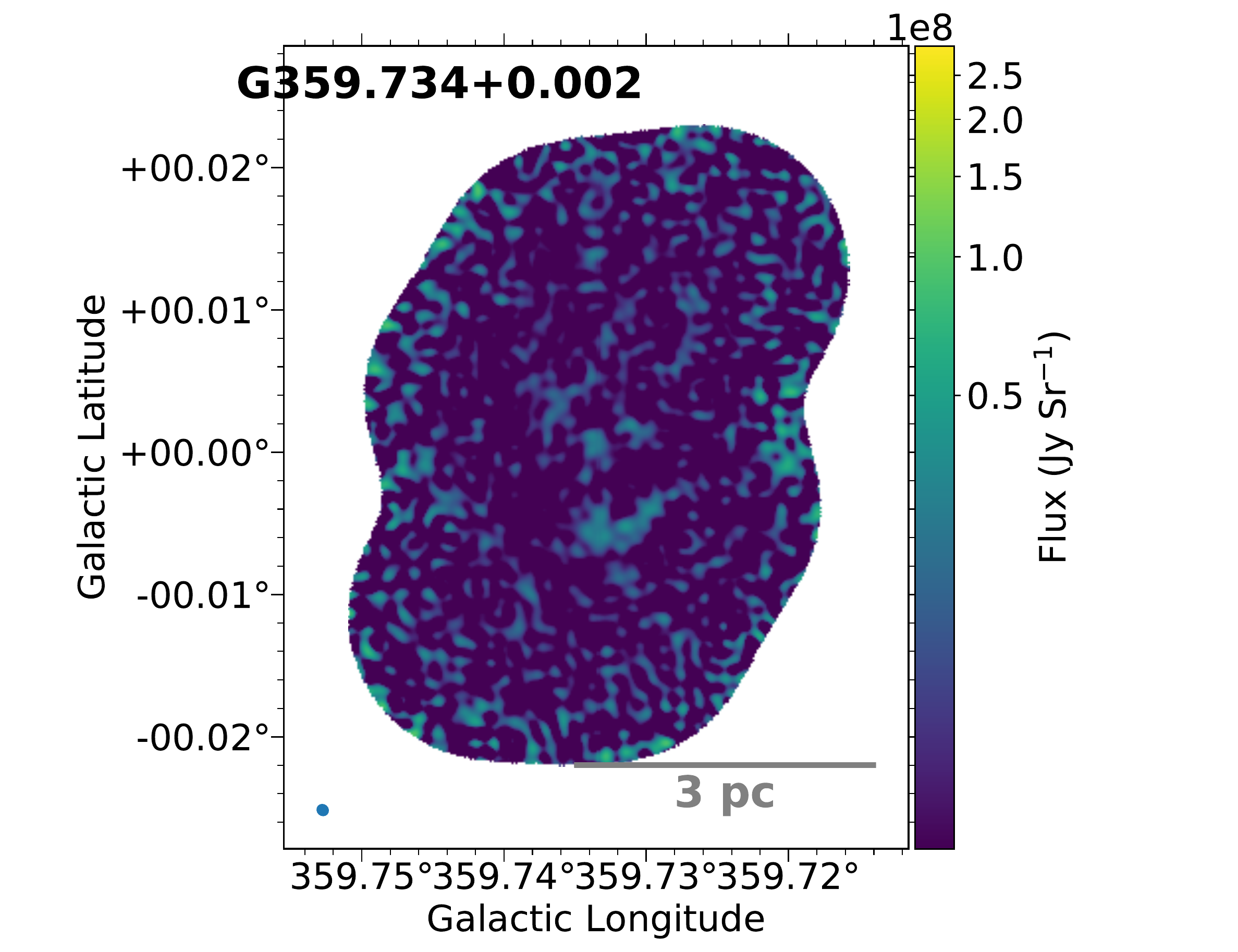}}\\
\singlespace\caption{Images show the 1.3 mm dust continuum from the \textit{CMZoom} Survey with a 3 pc scale-bar. The white contour shows the approximate 5-$\sigma$ level while the black contour shows the approximate 10-$\sigma$ level. All the images are displayed on same color scale from 2 - 290 MJy Sr$^{-1}$.} 
\label{fig:img_gallery}
\end{center}
\end{figure*}

\begin{figure*}
\begin{center}
\subfigure{
\includegraphics[width=0.48\textwidth]{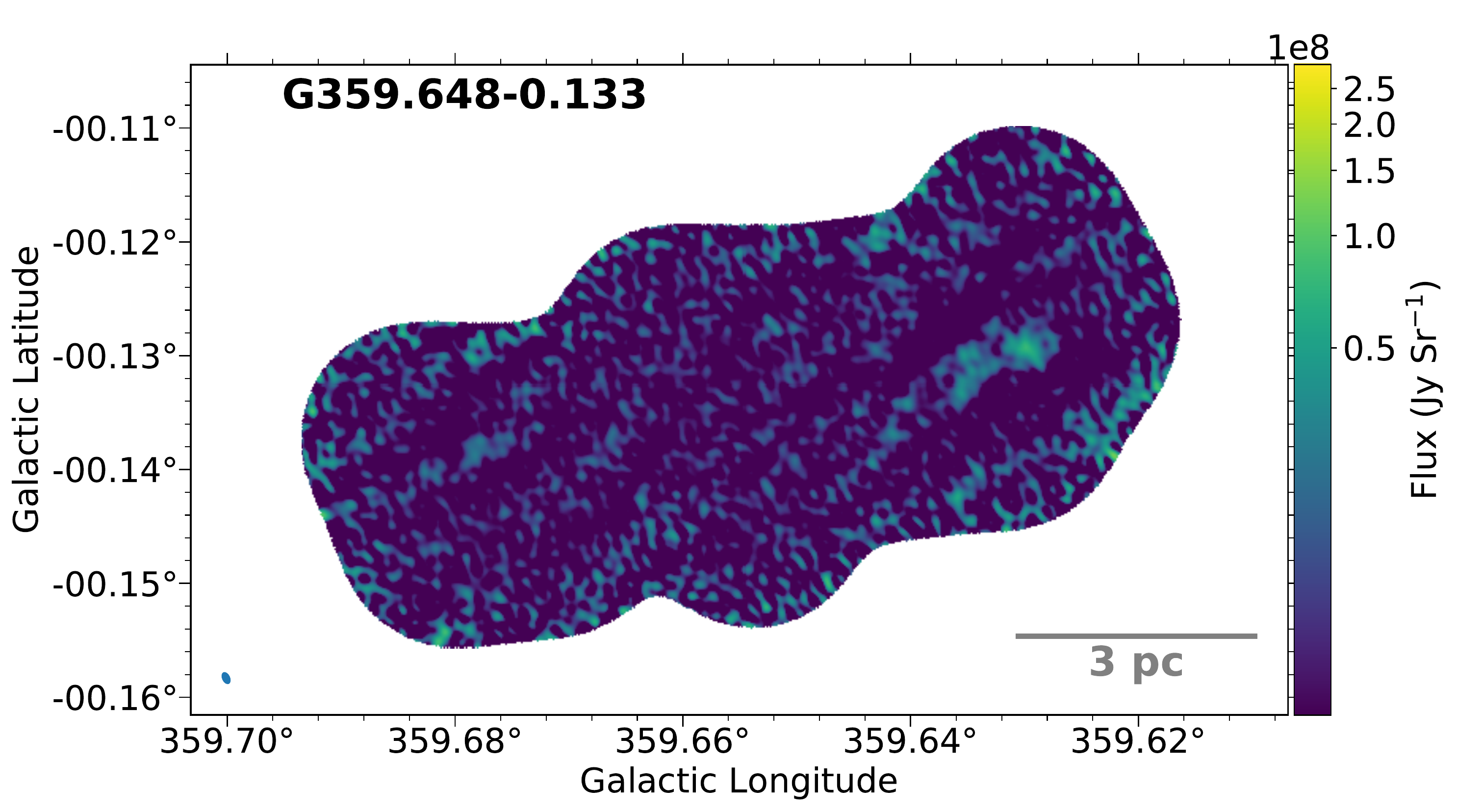}}
\subfigure{
\includegraphics[width=0.48\textwidth]{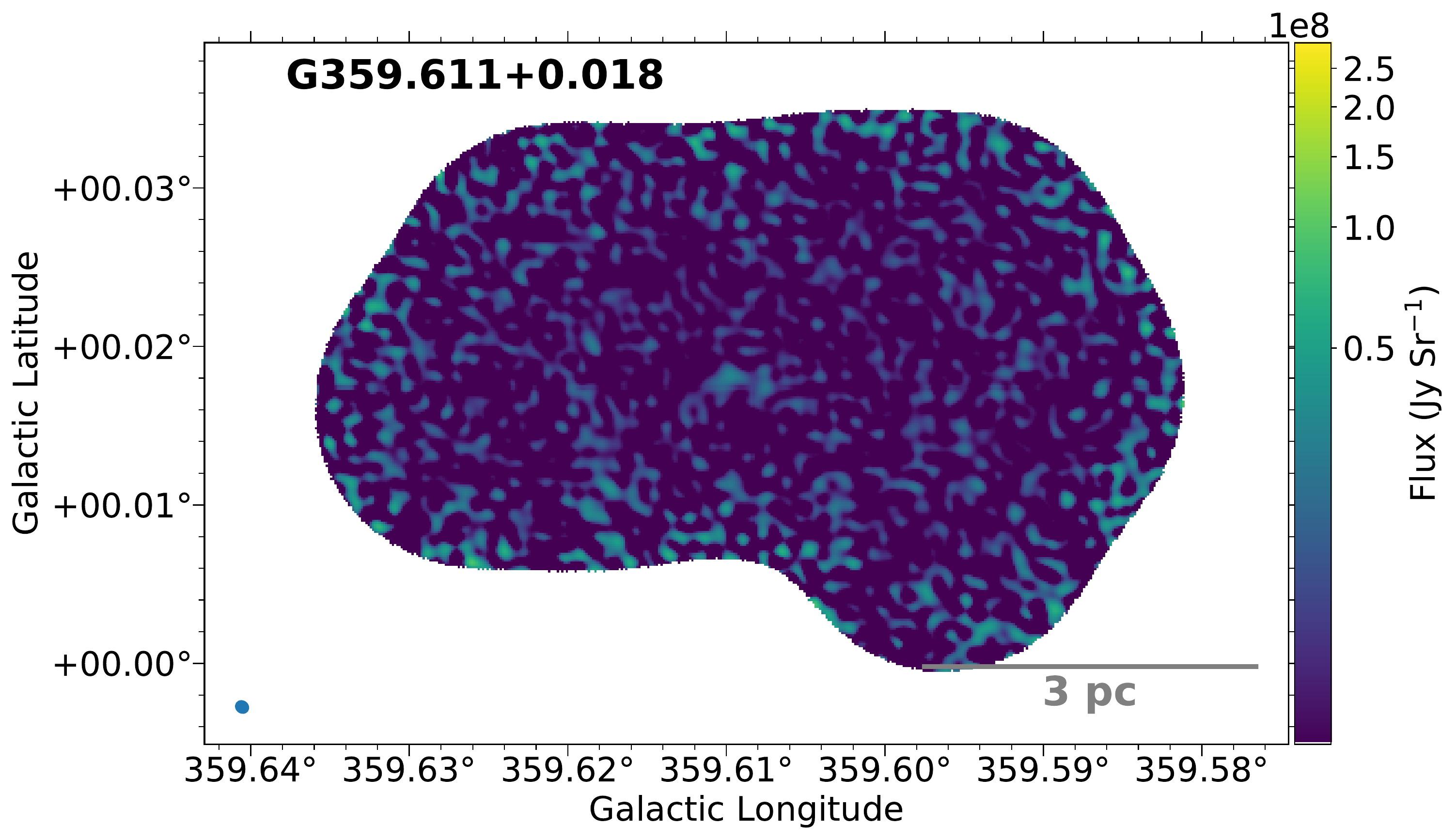}}\\
\subfigure{
\includegraphics[width=0.48\textwidth]{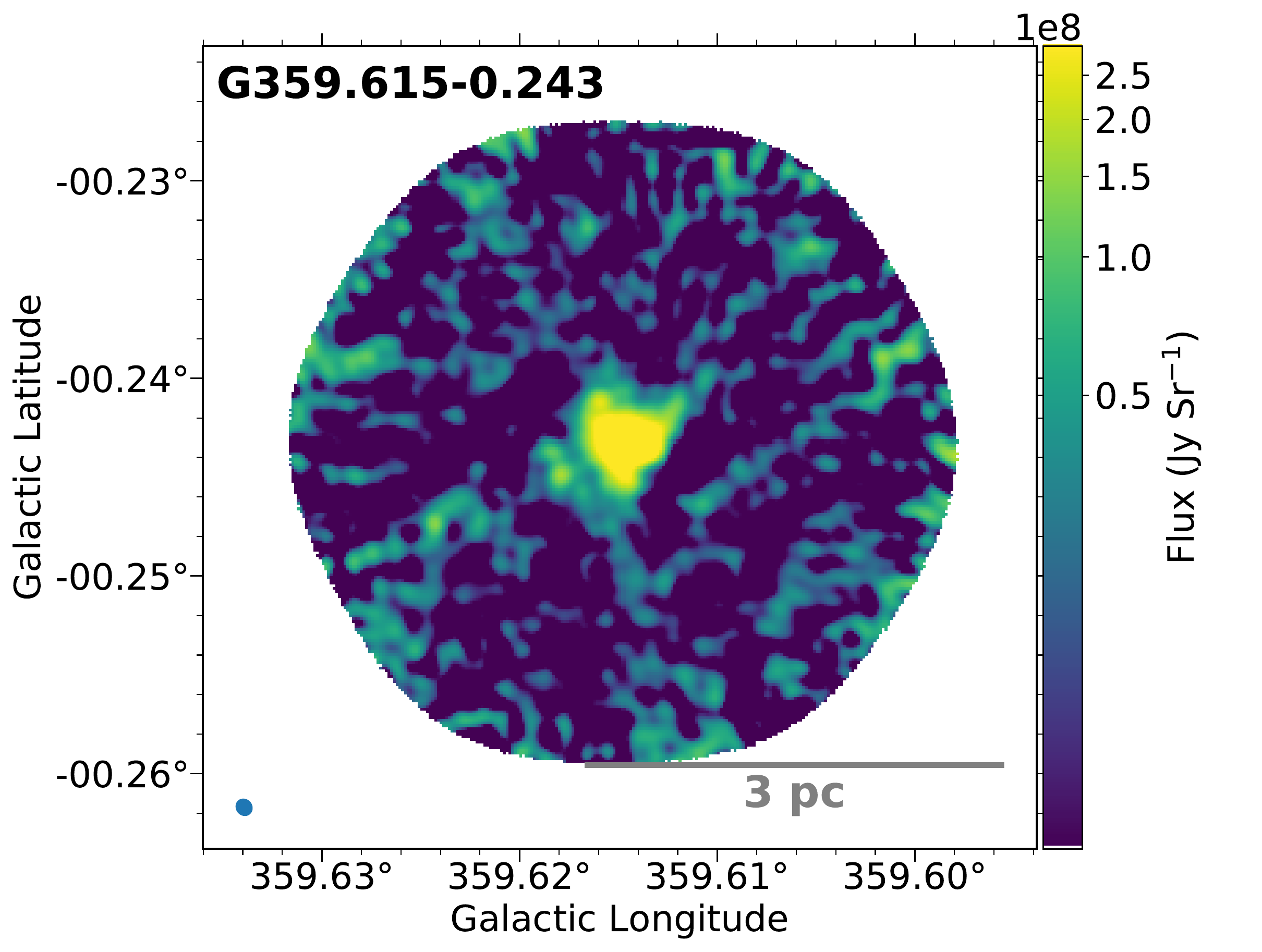}}
\subfigure{
\includegraphics[width=0.48\textwidth]{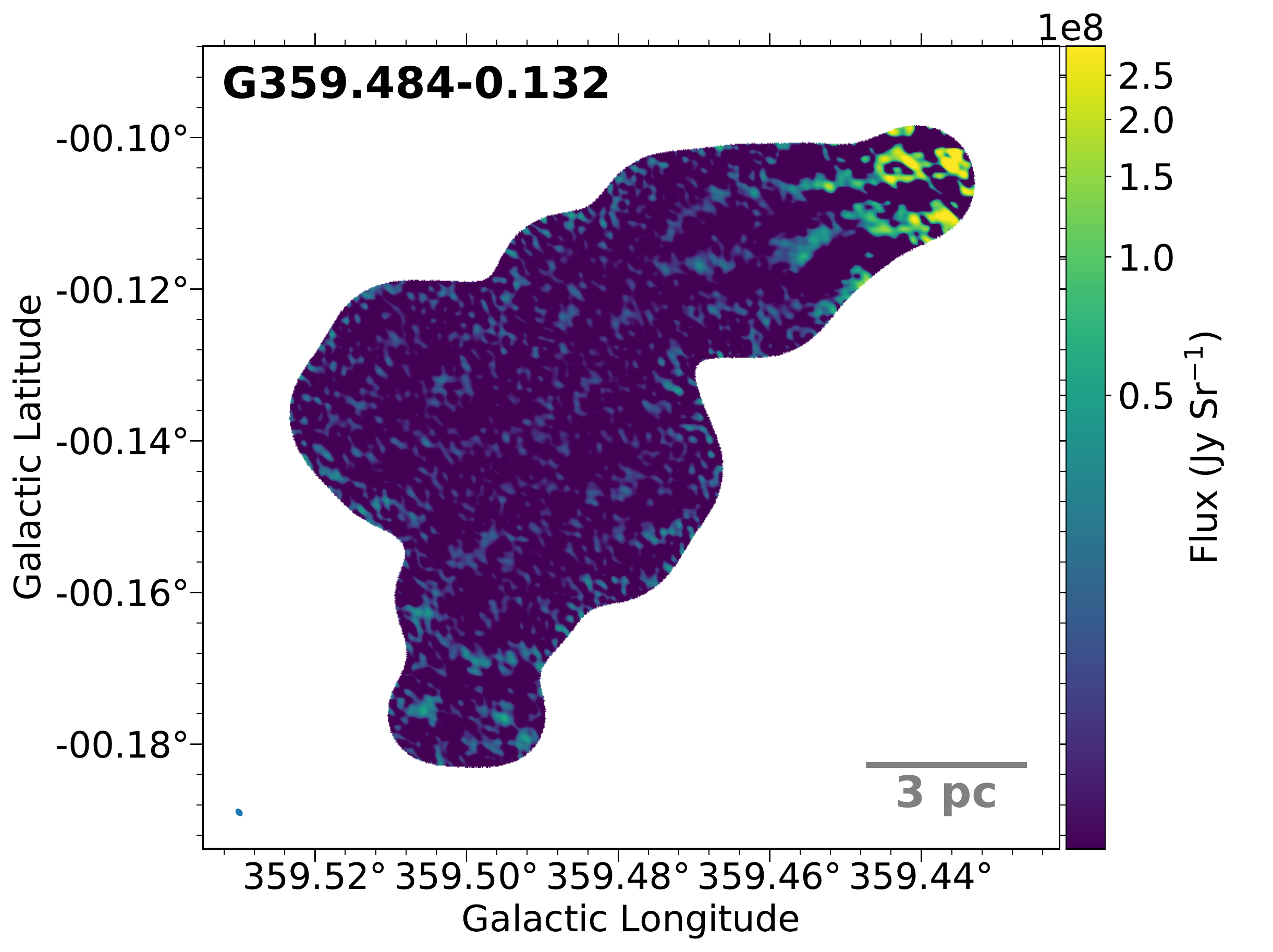}}\\
\subfigure{
\includegraphics[width=0.48\textwidth]{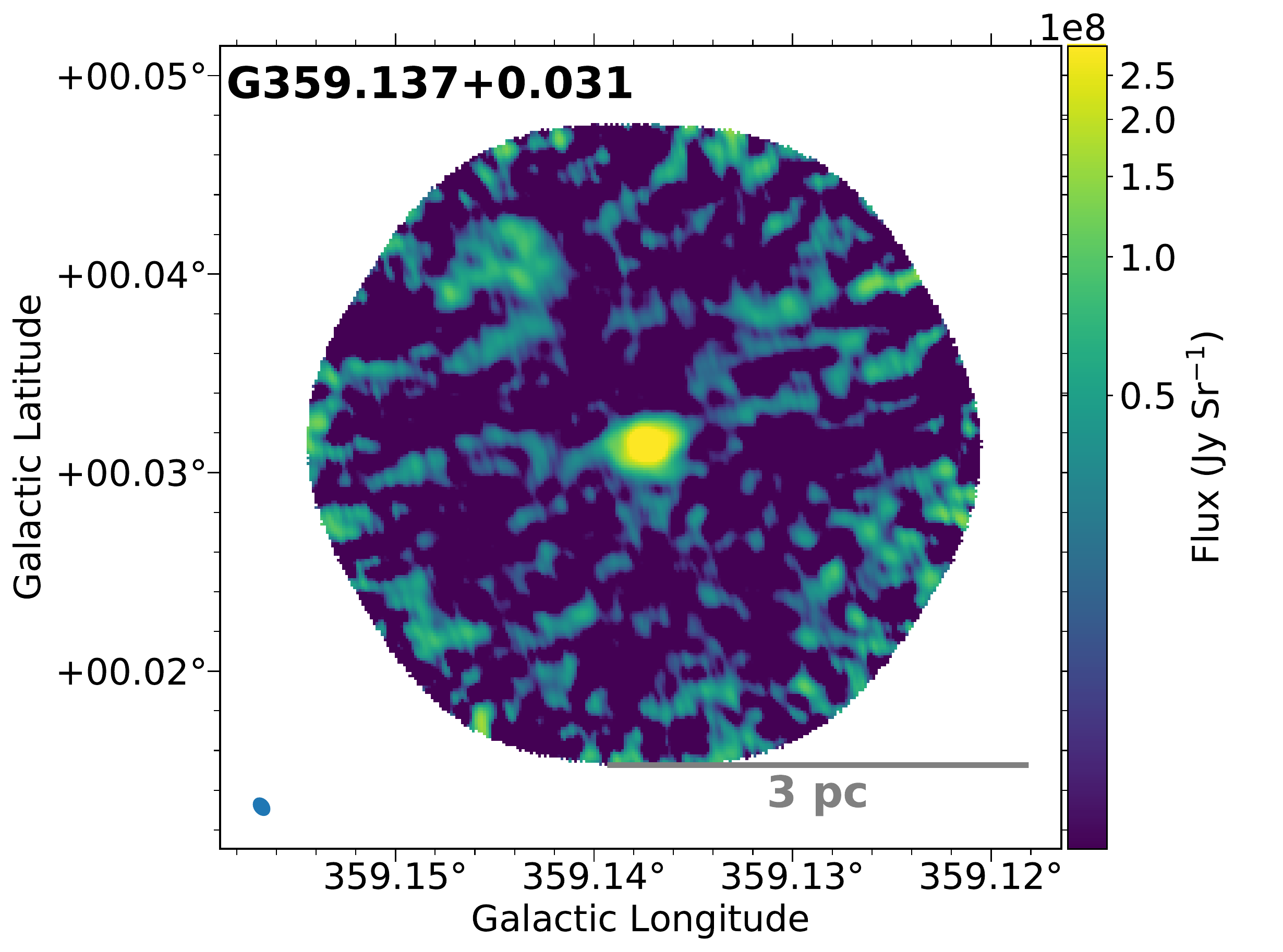}}
\singlespace\caption{Images show the 1.3 mm dust continuum from the \textit{CMZoom} Survey with a 3 pc scale-bar. The white contour shows the approximate 5-$\sigma$ level while the black contour shows the approximate 10-$\sigma$ level. All the images are displayed on same color scale from 2 - 290 MJy Sr$^{-1}$.} 
\label{fig:img_gallery}
\end{center}
\end{figure*}

\end{document}